\pdfoutput=1
\documentclass{article}
\usepackage{epic,eepic}
\usepackage{amsmath}
\usepackage{amssymb}
\usepackage{amsfonts}
\usepackage{latexsym}
\usepackage{graphicx}

\begin{document}

\newcommand{\abb}[1]{\mathrm{#1}}

\newtheorem{thm}{Theorem}[section]
\newtheorem{lem}[thm]{Lemma}
\newtheorem{cor}[thm]{Corollary}
\newtheorem{deff}[thm]{Definition}

\numberwithin{equation}{section}
\numberwithin{figure}{section}
\numberwithin{table}{section}
	
\begin{center}
{\Large The firing squad synchronization problem \\ 
for squares with holes}
\end{center}

\begin{center}
{\large Kojiro Kobayashi}
\end{center}
\vspace*{-2.0em}
\begin{center}
\texttt{kojiro@gol.com}
\end{center}

\begin{center}
{September 1, 2019}
\end{center}

\medskip

\noindent
{\bf Abstract}

\medskip

The firing squad synchronization problem (FSSP, for short) is 
a problem in automata theory introduced in 1957 by John Myhill.
Its goal is to design a finite automaton $A$ such that, if copies of $A$ 
are placed in a line and connected and are started at time $0$ 
with their leftmost copy in a special triggering state, 
then at some time (the ``firing time'') 
all copies enter a special ``firing state'' 
simultaneously for the first time.
FSSP has many variations and for many of them we know 
minimal-time solutions (solutions having shortest firing time).
One of such variations is the FSSP for squares (denoted by $\abb{SQ}$) 
in which copies are placed in a square.
In this paper we introduce a variation which we call 
the FSSP for squares with $k$ holes and denote by 
$\abb{SH}[k]$ by slightly modifying $\abb{SQ}$ ($k \geq 1$).  
In the variation, copies of a finite automaton are placed in a square 
but there are $k$ positions (``holes'') in the square 
where no copies are placed.
We show that $\abb{SH}[1]$ has a minimal-time solution.  
Moreover, for each problem instance (a placement of copies in a square) $C$ 
of $\abb{SH}[2]$, we determine the minimum firing time of $C$ 
(the minimum value of firing times of $C$ by $A$ 
where $A$ ranges over all solutions of $\abb{SH}[2]$).
The variation $\abb{SQ}$ was introduced and its minimal-time solutions were 
found in 1970's.
However, to find minimal-time solutions of $\abb{SH}[k]$, 
a very simple modification of $\abb{SQ}$, seems to be 
a very difficult and challenging problem for $k \geq 2$.

\medskip


\medskip

\section{Introduction}
\label{section:introduction}

The firing squad synchronization problem is a puzzle 
in automata theory.
According to \cite{Moore} it was devised in 1957 
as a problem of how to turn on all parts of 
a self-reproducing machine simultaneously.
The problem continues to attract interests of 
researchers even now.
The problem is formulated as follows.

In the formulation we use a finite automaton $A$ 
that has two inputs, one from the left and another from the right 
and two outputs, one to the left and another to the right.
The state of $A$ at a time $t + 1$ is completely 
determined by the state $s$ of $A$, the value $x$ of its left input 
and the value $y$ of its right input at the time $t$ by 
a state transition function $\delta(s, x, y)$.
The values of the two outputs of $A$ at a time $t$ 
are the state $s$ of $A$ at the time $t$.
The set of states of $A$ contains three special states 
$\abb{G}$, $\abb{Q}$, $\abb{F}$ called 
the {\it general state}\/, the {\it quiescent state}\/ and 
the {\it firing state}\/ respectively.

\begin{figure}[htbp]
\centering
\includegraphics[scale=1.0]{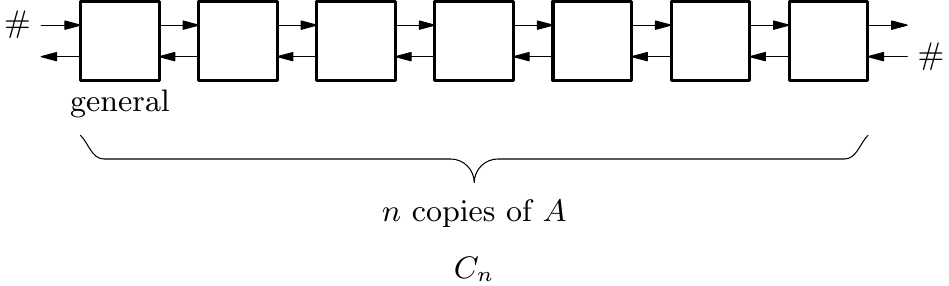}
\caption{A configuration $C_{n}$ of FSSP 
($n = 7$, in this case).}
\label{figure:fig048}
\end{figure}

We place copies of $A$ in a line and connect their inputs and outputs 
as shown in Fig. \ref{figure:fig048}.
When the number of copies is $n$, we call this line 
{\it the configuration of size $n$} and denote it by $C_{n}$.
We call each copy of $A$ in $C_{n}$ a {\it node} of $C_{n}$.
The values of the left input of the leftmost node and the right input 
of the rightmost node are special value $\#$ meaning that 
there are no nodes there.

We call the leftmost node of $C_{n}$ the {\it general} of $C_{n}$.
At time $0$, the general of $C_{n}$ is in the general state $\abb{G}$ 
and other nodes are in the quiescent state $\abb{Q}$.
Then the state of a node $v$ in $C_{n}$ at a time $t$ is completely determined 
by the state transition function $\delta$ of $A$.
We denote this state by $\abb{st}(v, t, C_{n}, A)$.
The transition function $\delta$ must satisfy the following 
condition: if both of $x$, $y$ are either $\abb{Q}$ or $\#$ then 
$\delta(\abb{Q}, x, y)$ must be $\abb{Q}$.
Intuitively, a node in the quiescent state $\abb{Q}$ remains 
in the state until at least one of its adjacent nodes is in a non-quiescent state. 

The firing squad synchronization problem, or FSSP for short, is the problem 
to design $A$ so that for any number $n$ of nodes, all nodes in $C_{n}$ 
enter the firing state $\abb{F}$ for the first time simultaneously 
at a time.
Formally stated, for any $n$ there must exist a time $t_{n}$ such that 
the following statement is true:
\begin{equation}
(\forall t < t_{n}) (\forall v \in C_{n}) 
[\abb{st}(v, t, C_{n}, A) \not= \abb{F}] \wedge
(\forall v \in C_{n})[\abb{st}(v, t_{n}, C_{n}, A) = \abb{F}].
\label{equation:eq007}
\end{equation}
We call a finite automaton $A$ that satisfies this condition a {\it solution} 
of FSSP.  
We call the time $t_{n}$ mentioned in \eqref{equation:eq007} 
the {\it firing time} of the solution $A$ for the configuration $C_{n}$ 
and denote it by $\abb{ft}(C_{n}, A)$.
Intuitively, $C_{n}$ is a firing squad and each node of $C_{n}$ is a soldier.
The special state $\abb{G}$ represents the general (the leftmost soldier) 
giving the order ``Fire'' and the special state $\abb{F}$ represents 
a firing soldier.

This is an interesting puzzle and we can easily find a solution 
using the well-known ``Divide and Rule'' strategy in the design of algorithms.
We usually find a solution $A$ with the firing time 
$\abb{ft}(C_{n}, A) = 3n + O(\log n)$.

One of the most interesting problems is to find fast solutions (that is, 
solutions with small values of $\abb{ft}(C_{n}, A)$).
We call a solution $\tilde{A}$ a \textit{minimal-time solution} 
if it is fastest among all solutions, or more precisely, 
if the following statement is true:
\begin{equation}
(\forall A) (\forall C_{n})[\abb{ft}(C_{n}, \tilde{A}) \leq 
\abb{ft}(C_{n}, A)].
\label{equation:eq005}
\end{equation}
Here $A$ ranges over all solutions.

It is not obvious that such solutions exist.
However, one such solution was found by E. Goto 
(\cite{Goto_1962}, reconstructed by H. Umeo 
\cite{Umeo_2018_Goto_reconstruction}) 
and later by A. Waksman (\cite{Waksman_1966}) and 
R. Balzer (\cite{Balzer_1967}) using ideas that are different from Goto's.
By these minimal-time solutions we know that 
the firing time $\abb{ft}(C_{n}, \tilde{A})$ of 
a minimal-time solution $\tilde{A}$ is $2n - 2$ for $n \geq 2$.

Many variations of FSSP have been also studied.
The following is a list of some of the variations: 
(1) lines (the general is the leftmost node) (the original FSSP,
\cite{Balzer_1967, Goto_1962, Waksman_1966}), 
(2) lines (the general may be an arbitrary node, \cite{Moore_Langdon_1968}), 
(3) two-way rings (\cite{Culik_1987, Gruska_et_al_2004, Kobayashi_Res_Rep_1976}), 
(4) one-way rings (\cite{Culik_1987, Gruska_et_al_2004, Kobayashi_Res_Rep_1976, Nishitani_Honda}), 
(5) squares (the general is the left down corner node, \cite{Shinahr_1974}), 
(6) cubes (the general is the left down bottom corner node, \cite{Shinahr_1974}), 
(7) rectangles (the general is the left down corner node, \cite{Shinahr_1974}), 
(8) rectangles (the general may be an arbitrary node, \cite{Szwerinski_1982}), 
(9) cuboids (that is, rectangular parallelepiped) 
(the general may be an arbitrary node, \cite{Szwerinski_1982}), 
(10) two-way tori constructed from squares (\cite{Gruska_et_al_2004}), 
(11) one-way tori constructed from squares (\cite{LaTorre_1996}), 
(12) two-way tori constructed from rectangles (\cite{Umeo_Kubo_2015_CANDAR}), 
(13) undirected networks (\cite{Nishitani_Honda, Rosenstiehl_1966, Rosenstiehl_1972}), 
(14) directed networks (\cite{Even_Litman_Winkler, Kobayashi_JCSS_1978, Ostrovsky_Wilkerson}).

For all of these variations except (13), (14) we know minimal-time 
solutions (see also \cite{Goldstein_Kobayashi_SIAM_2012}).
Variations of FSSP for Cayley graphs have been also studied 
(\cite{Roka_2000}).

The main motivation for studying variations of FSSP is that 
their solutions can be used to synchronize networks 
composed of large numbers of identical computing devices.
Another motivation is that it is a mathematical formulation of 
one case of the general problem: 
how to control the global behavior of a large network 
using only local information exchanges.
This general problem is one of the most fundamental problems 
in the theory of distributed computing.

Concerning FSSP, mainly three research themes have been studied.
First is to find fast solutions.
Especially interesting is to determine whether a variation 
has minimal-time solutions or not, and to find one if they exist.
If we cannot find them, it is desirable to formally prove that 
they do not exist.
The second is to find small solutions (that is, solutions with 
small numbers of states).  
For example, for the original FSSP a six state 
minimal-time solution is known (\cite{Mazoyer_1987}) 
and it is known that four state minimal-time solutions 
do not exist (see, for example, \cite{Umeo_2017_PaCT}).
The third is to find ``good'' solutions.
There are many criteria for ``goodness.''
The followings are among what we mean by ``good'' solutions:
easy to understand solutions, 
solutions that are easy to prove correctness, 
solutions using interesting or useful ideas, 
solutions having interesting or useful features, 
and solutions especially suited for some specific hardware implementation.

There are many surveys on FSSP and we refer the reader to them.
Mazoyer \cite{Mazoyer_Survey} provides a survey of the problem up to 
1986 and Napoli and Parente \cite{Napoli_Parente} give a survey of recent developments.
Goldstein and Kobayashi 
\cite{Goldstein_Kobayashi_SIAM_2005, Goldstein_Kobayashi_SIAM_2012} 
give surveys concentrating on the problem of existence/nonexistence of 
minimal-time solutions.
Umeo, Hisaoka and Sogabe \cite{Umeo_2005_Survey} give a survey 
of the minimal-time solutions of the original FSSP.

Before proceeding to explain the main results of this paper 
we give one more definition.
Let $\Gamma$ be a variation of FSSP that has at least one solution.
For this $\Gamma$ too we can define ``minimal-time solutions'' 
of $\Gamma$ as solutions $\tilde{A}$ of $\Gamma$ that satisfy 
the condition \eqref{equation:eq005}.
However there is another way to define minimal-time solutions of 
$\Gamma$.
For each configuration (a problem instance) $C$ of $\Gamma$, 
we define the \textit{minimum firing time} of $C$ 
(denoted by $\abb{mft}_{\Gamma}(C)$) by the following formula:
\[
\abb{mft}_{\Gamma}(C) = \min \{ \abb{ft}(C, A) ~|~ \text{$A$ is 
a solution of $\Gamma$} \}.
\]
This value is well-defined because we assume that $\Gamma$ has at least 
one solution.
We say that a solution $\tilde{A}$ of $\Gamma$ is 
a \textit{minimal-time solution} of $\Gamma$ if 
it satisfies the following condition:
\begin{equation}
(\forall C) [ \abb{ft}(C, \tilde{A}) = \abb{mft}_{\Gamma}(C) ].
\label{equation:eq006}
\end{equation}

Now we have two definitions of minimal-time solutions, 
one using \eqref{equation:eq005} and another using 
\eqref{equation:eq006}.
However we can easily show that these two definitions 
are equivalent.
The first definition has a clear intuitive meaning.
The second is technical but is useful for the study of 
minimal-time solutions as we see in the following.
Moreover, we can determine the value $\abb{mft}_{\Gamma}(C)$ 
even when we do not know whether $\Gamma$ has 
minimal-time solutions or not.
In such cases the problem to find minimal-time solutions 
is reduced to the problem to find solutions $\tilde{A}$ 
that satisfy \eqref{equation:eq006}.

\begin{figure}[htbp]
\centering
\includegraphics[scale=1.0]{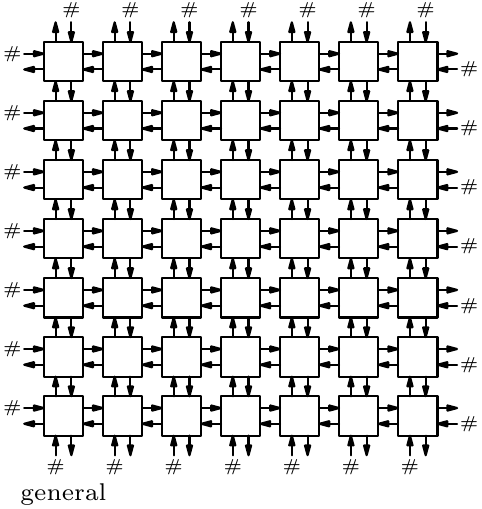}
\caption{A configuration $C_{w}$ of $\abb{SQ}$ ($w = 6$).}
\label{figure:fig079}
\end{figure}

We are ready to explain the main results of this paper.
One of the most basic variations of FSSP is the FSSP {\it for squares} 
(the variation (5) in our previous list).
We denote it by $\abb{SQ}$.
In the variation, for each $w$ ($\geq 0$) we have a configuration 
$C_{w}$ consisting of $(w + 1)^{2}$ nodes that are placed 
as a square of $w + 1$ rows and $w + 1$ columns.
Each node is a copy of a finite automaton $A$ 
that has four inputs and four outputs 
corresponding to the four 
directions the east, the north, the west and the south.
(From now on, we use these four directions 
instead of ``right,'' ``up,'' ``left'' and ``down.'')
The nodes are connected as shown in Fig. \ref{figure:fig079}.
(The value of $w$ is $6$ for this example.)
The general is at the southwest corner of the square.
A minimal-time solution was constructed by 
Shinahr (\cite{Shinahr_1974}).
The minimum firing time $\abb{mft}_{\abb{SQ}}(C_{w})$ is $2w$ 
for $w \geq 1$.

\begin{figure}[htbp]
\centering
\includegraphics[scale=1.0]{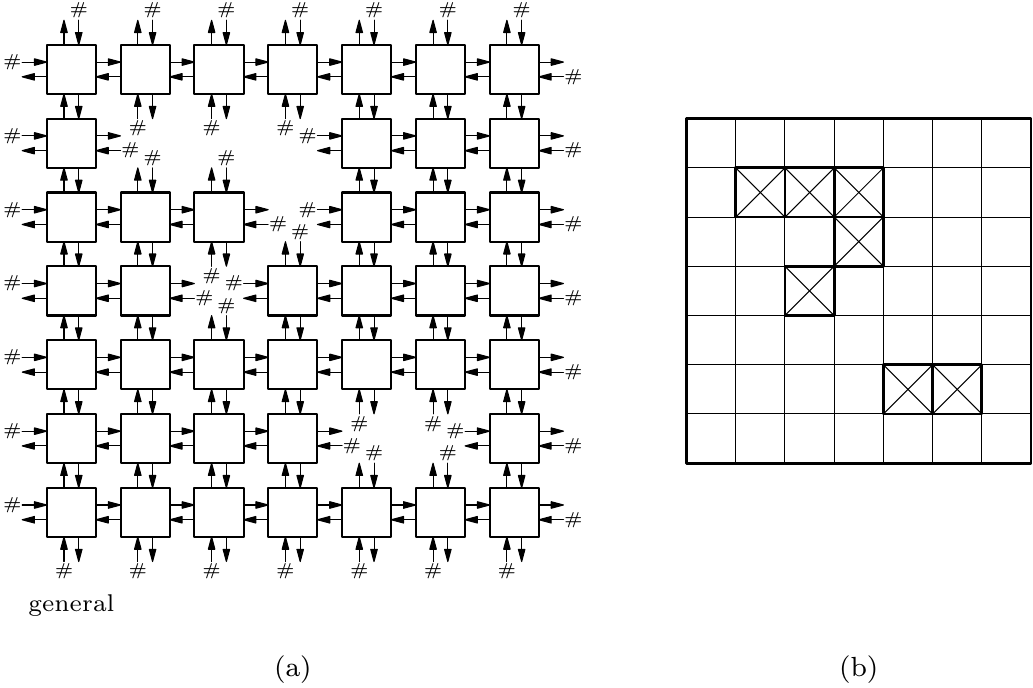}
\caption{An example of configurations of size $6$ of 
$\abb{SH}[7]$.}
\label{figure:fig059}
\end{figure}

In this paper we consider a variation of FSSP which 
we call the FSSP \textit{for squares with $k$ holes}.
Here, $k$ ($\geq 0$) is an integer and is a parameter of 
the variation.
We denote this variation by $\abb{SH}[k]$.
A \textit{configuration $C$ of size} $w$ of $\abb{SH}[k]$ 
is obtained by removing $k$ nodes from the configuration $C_{w}$ of $\abb{SQ}$ 
so that 
\begin{enumerate}
\item[$\bullet$] nodes on the boundary of the square are not removed, and 
\item[$\bullet$] the set of the remaining $(w + 1)^{2} - k$ nodes must be connected.
\end{enumerate}
We call the position of a removed node of a configuration 
a \textit{hole} of the configuration.
Each configuration of $\abb{SH}[k]$ has $k$ holes.
In Fig. \ref{figure:fig059} (a) we show an example 
of configurations of $\abb{SH}[k]$. 
The size $w$ of this configuration is $6$ and 
the number of holes $k$ of the configuration is $7$.
(Note that by a hole we mean a position that is not occupied 
by a node, not a set (a region) of adjacent nonoccupied positions 
such as $\{(1, 5), (2, 5), (3, 4), (3, 5)\}$ or 
$\{(4, 1), (5, 1)\}$.)
To save space we represent the configuration 
shown in Fig. \ref{figure:fig059} (a) by 
the figure shown in Fig. \ref{figure:fig059} (b).

We explain our motivations to study this variation $\abb{SH}[k]$ later 
and here we show the results concerning $\abb{SH}[k]$ obtained in this paper.
They are as follows:
\begin{enumerate}
\item[(1)] ${\abb{SH}}[1]$ has a minimal-time solution.
The minimum firing time of a configuration $C \in {\abb{SH}}[1]$ 
of size $w$ is 
$\abb{mft}_{\abb{SH}[1]}(C) = 2w$. 
(Theorem \ref{theorem:thm000}.)
\item[(2)] For each $k \geq 0$ there is a constant $c_{k}$ 
such that 
\begin{enumerate}
\item[$\bullet$] $2w \leq \abb{mft}_{{\abb{SH}}[k]}(C) \leq 2w + c_{k}$ 
for any $C \in {\abb{SH}}[k]$ of size $w$, and 
\item[$\bullet$] for all sufficiently large $w$, there are 
$C, C' \in \abb{SH}[k]$ of size $w$ such that 
$\abb{mft}_{\abb{SH}[k]}(C) = 2w$, 
$\abb{mft}_{\abb{SH}[k]}(C') = 2w + c_{k}$.
\end{enumerate}
(Theorem \ref{theorem:thm001}, Corollary \ref{corollary:cor001}.)
\item[(3)] $c_{0} = c_{1} = 0$, $c_{2} = 1$ and 
$k - 2 \leq c_{k} \leq k^{2} + 4k$ 
for all $k \geq 3$.
(Theorem \ref{theorem:thm005}.)
\item[(4)] The definition of $c_{k}$ itself gives an algorithm 
to compute the values $c_{k}$.
Using this algorithm we determined the values of $c_{k}$ 
for $2 \leq k \leq 9$ with computers. 
The results are $c_{2} = 1$ and 
$c_{k} = k - 2$ for $3 \leq k \leq 9$.
(In Table \ref{table:tab001} we summarize known values of 
$c_{k}$.)
(Subsection \ref{subsection:algorithm})
\item[(5)] We give a characterization of 
the minimum firing time $\abb{mft}_{\abb{SH}[2]}(C)$ of $\abb{SH}[2]$ 
(Theorem \ref{theorem:thm012}).
Although the characterization itself is simple, its derivation is very lengthy 
and tedious. (All of 
Section \ref{section:determination} and Appendix \ref{section:tedious_proof} 
are devoted to the derivation.)
\end{enumerate}

\begin{table}[htbp]
\centering
\begin{tabular}{|c|c|}
\hline
$k$ & $c_{k}$ 
\\ \hline
$0$ & $0$ \\
$1$ & $0$ \\
$2$ & $1$ \\
$3$ & $1$ \\
$4$ & $2$ \\
$5$ & $3$ \\
$6$ & $4$ \\
$7$ & $5$ \\
$8$ & $6$ \\
$9$ & $7$ \\ \hline
\end{tabular}
\caption{Known values of $c_{k}$.}
\label{table:tab001}
\end{table}

The following is a summary of the results we have 
at present on $\abb{SH}[k]$.
\begin{enumerate}
\item[$\bullet$] For $\abb{SH}[1]$ we know a minimal-time solution but 
for $k \geq 2$ we do not know whether $\abb{SH}[k]$ has 
minimal-time solutions or not.
\item[$\bullet$] For $\abb{SH}[2]$ we know the minimum firing time 
$\abb{mft}_{\abb{SH}[2]}(C)$ 
but for $k \geq 3$ we do not know it.
\end{enumerate}

As we mentioned above, the characterization of the minimum firing time 
$\abb{mft}_{\abb{SH}[2]}(C)$ of a configuration $C$ of $\abb{SH}[2]$ 
is simple.
We show it using the case where the size $w$ of $C$ is $12$ as an example.
(The characterization is slightly different for even $w$ and odd $w$.)
Suppose that $C$ is a configuration of size $12$.

\begin{figure}[htbp]
\centering
\includegraphics[scale=1.0]{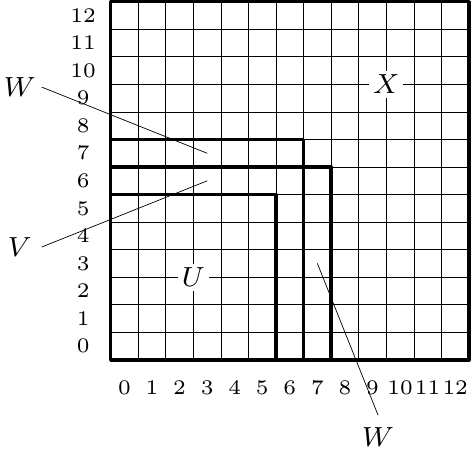}
\caption{Four sets $U$, $V$, $W$, $X$.}
\label{figure:fig081}
\end{figure}

There are $13^{2} = 169$ positions in the square of $C$ and two of them are holes.
We classify these $169$ positions into the following four disjoint sets $U$, $V$, $W$, $X$ 
(see Fig. \ref{figure:fig081}).
\begin{align*}
U & = \{(x, y) \mid 0 \leq x \leq 5, 0 \leq  y \leq 5 \}, \\
V & = \{(x, y) \mid \text{$0 \leq x \leq 5$, $y = 6$ or
                          $x = 6$, $0 \leq y \leq 5$ or 
                          $x = 6$, $y = 6$} \}, \\
W & = \{(x, y) \mid \text{$0 \leq x \leq 6$, $y = 7$ or 
                          $x = 7$, $0 \leq y \leq 6$} \}, \\
X & = \{(x, y) \mid 0 \leq x \leq 12, 0 \leq y \leq 12\} - U \cup V \cup W.
\end{align*}
Then, for the following three cases we have $\abb{mft}_{\abb{SH}[2]}(C) = 25$ ($= 2w + 1$) 
and for other cases we have $\abb{mft}_{\abb{SH}[2]}(C) = 24$ ($= 2w$).
\begin{enumerate}
\item[$\bullet$] $C$ has no holes in $U \cup V \cup W$.
\item[$\bullet$] $C$ has no holes in $U \cup V$, 
has one hole in $W$, and 
the position $v = (x, y)$ of the hole in $W$ satisfies $|x - y| = 2$.
\item[$\bullet$] $C$ has two holes in $U \cup V$ and their positions $v = (x, y)$, $v'$ 
satisfy $|x - y| = 2$ and $v' = v + (1, 1)$.
\end{enumerate}
In Fig. \ref{figure:fig047} we show five examples of $C$ of size $12$ 
such that $\abb{mft}_{\abb{SH}[2]}(C)$ is $25$ ($= 2w + 1$).
The example (a) is for the first case, the examples (b), (c) are 
for the second case and the examples (d), (e) are for the third case.

Now we explain our main motivation for studying $\abb{SH}[k]$.
In these two decades, some pairs of variations of $\abb{FSSP}$ $(\Gamma, \Gamma')$ having the following 
properties have been discovered.
\begin{enumerate}
\item[$\bullet$] $\Gamma$ is one of the basic variations that were 
extensively studied in the early days of the research of $\abb{FSSP}$ 
(that is, in 1960s and 1970s) and their minimal-time solutions were 
obtained at that time. 
\item[$\bullet$] $\Gamma'$ is a natural and simple modification of $\Gamma$.
\item[$\bullet$] At present we do not know whether $\Gamma'$ has 
minimal-time solutions or not and moreover the problem to know it seems to be 
very difficult. 
\end{enumerate}

\begin{figure}[htbp]
\centering
\includegraphics[scale=1.0]{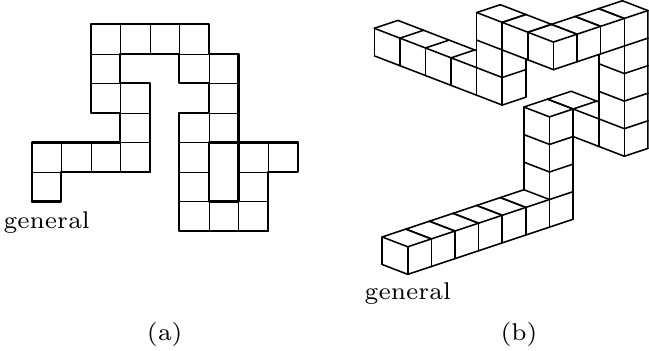}
\caption{(a) A configuration of $\abb{2PATH}$.
(b) A configuration of $\abb{3PATH}$.}
\label{figure:fig080}
\end{figure}

The first example is the pairs 
$(\abb{ORG}$, $\abb{2PATH})$ and 
$(\abb{ORG}$, $\abb{3PATH})$.
Here $\abb{ORG}$ denotes the original $\abb{FSSP}$ (the variation (1) in our previous list), 
$\abb{2PATH}$ denotes the FSSP of paths in the two-dimensional grid space 
and $\abb{3PATH}$ denotes the same problem for the three-dimensional grid space.
In both of $\abb{2PATH}$ and $\abb{3PATH}$ 
the general of a path is one of the two terminal nodes of the path.
In Fig. \ref{figure:fig080} (a) and (b) we show examples of configurations of $\abb{2PATH}$ and 
$\abb{3PATH}$ respectively.

In all of $\abb{ORG}$, $\abb{2PATH}$, $\abb{3PATH}$, 
configurations are lines of nodes.
The difference is that in $\abb{ORG}$ they must be straight but in $\abb{2PATH}$, $\abb{3PATH}$ 
they may be bent in the grid spaces.
Therefore, $\abb{2PATH}$, $\abb{3PATH}$ are natural and simple modifications of $\abb{ORG}$.
Minimal-time solutions of $\abb{ORG}$ were obtained by 
\cite{Balzer_1967, Goto_1962, Waksman_1966}.

Both of $\abb{2PATH}$ and $\abb{3PATH}$ were studied in 
\cite{Kobayashi_TCS_2001} and 
\cite{Goldstein_Kobayashi_SIAM_2005} respectively.
For them we know the minimum firing times 
$\abb{mft}_{\abb{2PATH}}(C)$, 
$\abb{mft}_{\abb{3PATH}}(C)$ 
but at present we do not know 
whether they have minimal-time solutions or not.
Nevertheless, we have the following circumstantial evidences that 
they have no minimal-time solutions.
\begin{enumerate}
\item[$\bullet$] If the problem $\abb{2PEP}$ has no polynomial-time algorithms 
then $\abb{2PATH}$ has no minimal-time solutions 
(\cite{Kobayashi_TCS_2001}).
\item[$\bullet$] If $\abb{P} \not= \abb{NP}$ then $\abb{3PATH}$ has 
no minimal-time solutions 
(\cite{Goldstein_Kobayashi_SIAM_2005}).
\end{enumerate}
Here, $\abb{2PEP}$ (the {\it two-dimensional path extension problem}) is 
a purely combinatorial problem on paths in the two-dimensional grid space 
such that (1) it is in $\abb{NP}$, (2) at present we know only exponential-time 
algorithms for it, but (3) at present we cannot prove that it is $\abb{NP}$-complete 
(\cite{Kobayashi_TCS_2001}).
The first of the above two results implies that to find a minimal-time solution of $\abb{2PATH}$ 
is at least as difficult as finding a polynomial-time algorithm for $\abb{2PEP}$.
The second result is a sufficiently convincing evidence that $\abb{3PATH}$ has no 
minimal-time solutions.

The second example is the pair $(\abb{SQ}, \abb{gSQ})$.
Here $\abb{gSQ}$ denotes the variation obtained from $\abb{SQ}$ by modifying 
so that the general of a configuration may be an arbitrary node in it.
We call $\abb{gSQ}$ the ``{\it generalized} FSSP {\it for squares}.''
As we mentioned previously a minimal-time solution of $\abb{SQ}$ was 
obtained by Shinahr (\cite{Shinahr_1974}).
In \cite{Umeo_2012_ACRI}, Umeo and Kubo noted that 
we do not know whether $\abb{gSQ}$ has minimal-time solutions or not.
By \cite{Kobayashi_TCS_2014}, at least we know the minimum firing time 
$\abb{mft}_{\abb{gSQ}}(C)$ of $\abb{gSQ}$.

The pairs $(\abb{SQ}, \abb{SH}[k])$ for $k \geq 2$ 
are the third example of such pairs.
As we mentioned above, we do not know whether $\abb{SH}[k]$ has minimal-time solutions 
or not for $k \geq 2$.
For $\abb{SH}[2]$ we know the minimum firing time.
However, its lengthy and tedious derivation given in 
Section \ref{section:determination} and 
Appendix \ref{section:tedious_proof} suggests that 
the problem to determine the minimum firing time of $\abb{SH}[k]$ 
seems to be very difficult for $k \geq 3$.

These examples of $(\Gamma, \Gamma')$ give us the following impression 
concerning variations of FSSP.
\begin{enumerate}
\item[$\bullet$] Most variations of FSSP are very difficult.
\item[$\bullet$] In our previous list (1) -- (14) of variations of FSSP, 
the variations (1) -- (12) for which we know 
minimal-time solutions are exceptionally simple and are isolated 
in difficult variations.
\end{enumerate}
If this impression is correct, to construct a general theory of FSSP 
that includes a much broader class of variations of FSSP 
is a very interesting open problem.
In that case, $\abb{2PATH}$, $\abb{3PATH}$, $\abb{gSQ}$, 
$\abb{SH}[k]$ are good variations to start with for the study of such general theory.
This is our main motivation for introducing and studying $\abb{SH}[k]$.

Another motivation for studying $\abb{SH}[k]$ is that 
it is one of the formulations of the problem to synchronize 
networks of computing devices in situations where the networks may have 
faulty devices 
(\cite{
Dimitriadis_Kutrib_Sirakoulis_2018, 
Kutrib_Vollmar_1995_Faulty, 
Umeo_2004_Faulty, 
Umeo_Kamikawa_Maeda_Fujita_2018_ACRI, 
Yunes_2006_Faulty}).

This paper is organized as follows.
In Section \ref{section:preliminaries} 
we explain basic notions and notations.
In Section \ref{section:sq_hole_1} we prove that 
$\abb{SH}[1]$ has a minimal-time solution (Result (1)).
In Section \ref{section:general_results} 
we show many results on 
the function $\abb{mft}_{\abb{SH}[k]}(C)$ for general 
values of $k$ (Results (2), (3), (4)).
In Section \ref{section:determination} 
we concentrate on $\abb{SH}[2]$ and give 
a characterization of the value $\abb{mft}_{\abb{SH}[2]}(C)$ 
(Result (5)).
Section \ref{section:discussions_and_conclusion} is for 
discussions and conclusion.

\section{Preliminaries}
\label{section:preliminaries}

Let $v = (x, y)$ and $v' = (x', y')$ be positions 
in the two-dimensional grid space 
$\mathbb{Z}^{2}$ ($\mathbb{Z}$ denotes the 
set $\{\ldots, -2, -1, 0, 1, 2, \ldots \}$ of integers).
We say that $v, v'$ are \textit{adjacent} if 
$x = x'$ and $|y - y'| = 1$ or 
$|x - x'| = 1$ and $y = y'$, and 
$v, v'$ \textit{touch with corners} if  
$|x - x'| = 1$ and $|y - y'| = 1$.
By the \textit{Manhattan distance} 
(or the \textit{MH distance} for short) 
\textit{between} $v$ \textit{and} $v'$, 
we mean the value $|x - x'| + |y - y'|$ and denote it 
by $\abb{d}_\abb{MH}(v, v')$.

By a \textit{path} we mean a sequence of positions 
$P = v_{0}, \ldots, v_{n}$ in $\mathbb{Z}^{2}$ such that 
$v_{i}$, $v_{i+1}$ are adjacent for each $i$ ($0 \leq i \leq n-1$).
We call $P$ a path \textit{from} $v_{0}$ \textit{to} $v_{n}$ 
or a path \textit{between} $v_{0}$ \textit{and} $v_{n}$.
We call the value $n$ the \textit{length} of the path $P$ 
and denote it by $|P|$.
When $P$, $P'$ are paths such that the end position 
$\tilde{v}$ of $P$ 
and the start position $\tilde{\tilde{v}}$ of $P'$ 
are the same, by $P + P'$ 
we mean the path obtained from $P$ and $P'$ 
by concatenating them (but deleting one of the overlapping 
$\tilde{v}$, $\tilde{\tilde{v}}$).

For each $w$ ($\geq 1$), by $S_{w}$ we denote the square 
\begin{equation}
S_{w} = \{ (x, y) \mid 0 \leq x \leq w, 0 \leq y \leq w \}
\label{equation:eq013}
\end{equation}
in $\mathbb{Z}^{2}$.
By the \textit{main diagonal} of $S_{w}$ we mean the 
set of positions 
$\{ (u, u) \mid 0 \leq u \leq w\}$.
By the \textit{boundary} of $S_{w}$ we mean the set of 
positions $\{ (x, y) \in S_{w} \mid 
\text{either $x = 0$, $x = w$, $y = 0$, or $y = w$} \}$.

We give a formal definition of $\abb{SH}[k]$.
A configuration $C$ is obtained as follows.
First we select an integer $w$ such that $(w - 1)^{2} \geq k$.
Next we select $(w + 1)^{2} - k$ positions from the 
$(w + 1)^{2}$ positions in $S_{w}$ so that two conditions 
are satisfied.
The first condition is that all positions in the boundary 
of $S_{w}$ are selected.  
We assume that $(w - 1)^{2} \geq k$ and hence 
there is at least one way to select positions so that 
this condition is satisfied.
The second condition is that, for any two selected positions 
$v$, $v'$ there is a path of selected positions 
between $v$ and $v'$.
Finally copies of a finite automaton $A$ are placed on 
the selected positions.
The placement of these copies of $A$ is 
a configuration $C$ of $\abb{SH}[k]$ of size $w$.
The general of $C$ is the copy placed at the position $(0, 0)$.
By $v_{\abb{gen}}$ we denote the general of a configuration.

We call each copy a \textit{node} of $C$.
When there is a node at a position $v$ 
we may say ``a node $v$'' instead of ``the node 
at a position $v$.''
For a position $v$, by the expression ``$v \in C$'' 
we mean that $v$ is a node of $C$ 
(not that $v$ is a position in $S_{w}$).
We call a position in $S_{w}$ that is not selected for $C$ 
a \textit{hole} of $C$.
There are $k$ holes of $C$.

We say that a path is a path in $C$ when all positions 
in it are nodes of $C$. 
For nodes $v$, $v'$ of $C$, by the \textit{distance} 
between $v$ and $v'$ we mean the minimum value of 
$|P|$ when $P$ ranges over all paths in $C$ between $v$ and $v'$, 
and denote it by $\abb{d}_{C}(v, v')$ or 
$\abb{d}(v, v')$ when $C$ is understood.
By $\abb{d}_\abb{MH}(v, v''; v')$ and 
$\abb{d}_{C}(v, v''; v')$ we mean 
$\abb{d}_\abb{MH}(v, v') + \abb{d}_\abb{MH}(v', v'')$ 
and $\abb{d}_{C}(v, v') + \abb{d}_{C}(v', v'')$ 
respectively.

When the length $|P|$ of a path $P$ from $v$ to $v'$ is 
$\abb{d}_{\abb{MH}}(v, v')$ we say that the path is 
{\it of the MH distance length}\/.
When an event occurs at a node $v$ at time 
$\abb{d}_{\abb{MH}}(v_{\abb{gen}}, v)$ we say that 
the event {\it occurs at the MH distance length time}\/.

We call the direction in $\mathbb{Z}^{2}$ in which 
the $x$-variable value increases the \textit{east}, 
and define the directions the \textit{north}, the \textit{west}  
and the \textit{south} similarly.

By the \textit{boundary condition} of a node $v = (x, y)$ 
in a configuration $C$ we mean the vector 
$(b_{0}, b_{1}, b_{2}, b_{3})$.
Here, $b_{0}$ is $1$ if the position $v' = (x + 1, y)$ 
east of $v$ is a node of $C$.
Otherwise (that is, either $v'$ is a hole of $C$ or 
$v'$ is out of the square $S_{w}$), $b_{0}$ is $0$.
We define $b_{1}$, $b_{2}$, $b_{3}$ similarly 
for the directions the north, the west, the south.
By $\abb{bc}_{C}(v)$ we denote the boundary condition of 
$v$ in $C$.

We defined ``solutions'' of the original FSSP in 
Section \ref{section:introduction} 
and this definition can be modified for any variation $\Gamma$ 
of FSSP by replacing configurations $C_{n}$ of 
the original FSSP with configurations $C$ of $\Gamma$ 
and replacing the times $t_{n}$ that may depend on $n$ 
with times $t_{C}$ that may depend on $C$.

For each variation $\Gamma$, we define a \textit{partial solution} 
\footnote{The term ``a partial solution of a variation of 
FSSP'' is also used for a different meaning (\cite{Umeo_2015_Survey}).}
of $\Gamma$ as a finite automaton $A$ such that 
for any configuration $C$ of $\Gamma$, 
either (1) each node of $C$ never fires 
(that is, the statement $(\forall t) (\forall v \in C) [\abb{st}(v, t, C, A) 
\not= \abb{F}]$ is true) or 
(2) there is a time $t_{C}$ such that 
all nodes in $C$ fire for the first time simultaneously at 
the time (that is, the statement \eqref{equation:eq007} 
is true with $C_{n}$ and $t_{n}$ replaced with $C$ 
and $t_{C}$ respectively).

When $A$ is a partial solution of $\Gamma$, 
by the \textit{domain} of $\Gamma$ we mean the set of 
configurations $C$ of $\Gamma$ for which the case (2) 
of the definition holds true.
Moreover, for each $C$ in the domain of $A$, 
by $\abb{ft}(C, A)$ we denote the time $t_{C}$ mentioned 
in (2).

Suppose that $\Gamma$ is a variation that has a solution.
Then we have 
\begin{equation}
\abb{mft}_{\Gamma}(C) \leq \abb{ft}(C, A)
\label{equation:eq012}
\end{equation}
for any partial solution $A$ and any configuration $C$ 
in the domain of $A$.  
The proof is as follows.

Let $A_{0}$ be a solution of $\Gamma$ and $A_{1}$ be any 
partial solution of $\Gamma$.
Let $A_{2}$ be the finite automaton that simulates 
both of $A_{0}$, $A_{1}$ and fires as soon as at least 
one of $A_{0}$, $A_{1}$ fires.
Then $A_{2}$ is a solution of $\Gamma$ and 
\begin{align*}
\abb{mft}_{\Gamma}(C) & \leq \abb{ft}(C, A_{2}) \\
 & = 
\begin{cases}
\min \{ \abb{ft}(C, A_{0}), \abb{ft}(C, A_{1}) \}& 
\text{$C$ is in the domain of $A_{1}$}, \\
\abb{ft}(C, A_{0})& \text{otherwise}
\end{cases}
\end{align*}
for any configuration $C$ of $\Gamma$.
Therefore, if $C$ is in the domain of $A_{1}$ we have 
$\abb{mft}_{\Gamma}(C) \leq \abb{ft}(C, A_{1})$.

\section{The variation $\abb{SH}[1]$}
\label{section:sq_hole_1}

In this section we show that 
$\abb{SH}[1]$ has a minimal-time solution.

\begin{thm}
\label{theorem:thm000}
{\rm (1)} For a configuration $C$ of size $w$ of 
$\abb{SH}[1]$, $\abb{mft}_{\abb{SH}[1]}(C) = 2w$.
{\rm (2)} $\abb{SH}[1]$ has a minimal time solution.
\end{thm}

\noindent
{\it Proof}. 
We construct a solution of $\abb{SH}[1]$ 
that fires a configuration $C$ of size $w$ 
at time $2w$.  
This shows both of (1), (2) of the theorem 
because we have $\abb{d}_{C}(v_\abb{gen}, (w, w)) = 2w$ 
and hence the firing time of $C$ of any solution cannot be 
smaller than $2w$.
We modify the idea by Shinahr (\cite{Shinahr_1974}) 
used to construct a minimal-time solution of $\abb{SQ}$.

We construct four finite automata $A_{0}$, $A_{1}$, $A_{2}$, $A_{3}$.
The last one $A_{3}$ is the desired solution.
First we explain the finite automaton $A_{0}$.

\begin{figure}[htbp]
\centering
\includegraphics[scale=1.0]{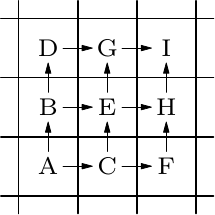}
\caption{The rule of generation of signals 
$\abb{A}$, $\abb{B}$, \ldots, $\abb{I}$.}
\label{figure:fig060}
\end{figure}

\begin{figure}[htbp]
\centering
\includegraphics[scale=1.0]{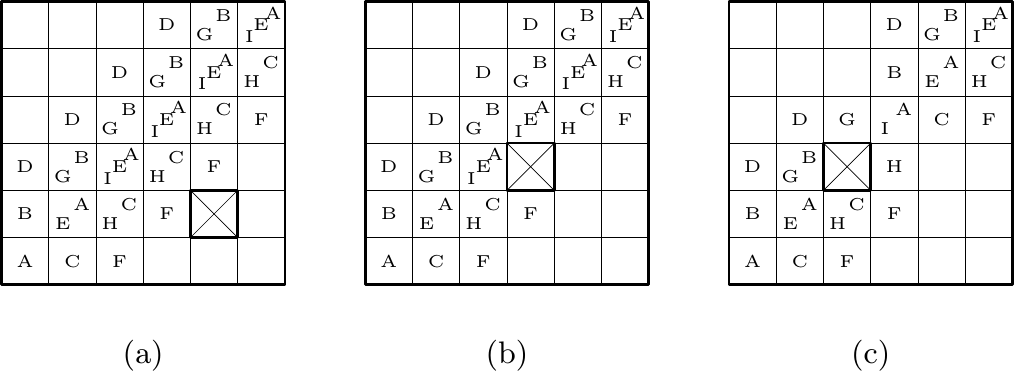}
\caption{Examples of the generation of signals.}
\label{figure:fig061}
\end{figure}

To locate nodes $(i, i)$ in the main diagonal $A_{0}$ 
uses nine signals $\abb{A}$, $\abb{B}$, $\ldots$, 
$\abb{I}$.
These signals are generated by the following rules.
\begin{enumerate}
\item[$\bullet$] A signal $\abb{A}$ is generated at $v_\abb{gen}$ at 
time $0$.
\item[$\bullet$] A signal $\abb{A}$ generates other signals 
as shown in Fig. \ref{figure:fig060} (if the generation 
is not blocked by holes).
\item[$\bullet$] When a signal $\abb{E}$ or $\abb{I}$ 
is generated at a node, 
a signal $\abb{A}$ is simultaneously generated at the node.
\end{enumerate}
For example, the arrow from $\abb{B}$ to $\abb{E}$ in Fig. \ref{figure:fig060} 
means that if a signal $\abb{B}$ is generated at a node $(x, y)$ at time $t$ 
and the position $(x + 1, y)$ is a node then a signal $\abb{E}$ 
is generated at the node $(x + 1, y)$ at time $t + 1$.
In Fig. \ref{figure:fig061} we show three examples 
of the generation of signals.

For any node $(i, i)$ ($0 \leq i \leq w$) in the main diagonal 
the signal $\abb{A}$ is generated at the node 
at time $2i$.
We can prove this by the induction on $i$ 
using our assumption that there is exactly one hole.

When a signal $\abb{A}$ is generated at a node $(i, i)$ at time $2i$ 
($0 \leq i \leq w - 1$), 
it activates the original FSSP for 
the horizontal sequence of positions 
$(i + 1, i)$, $(i+2, i)$, $\ldots$ at time $2i + 1$ 
assuming that the general is at the left end position 
$(i + 1, i)$ of the sequence.
We use a minimal-time solution of the original FSSP 
(\cite{Balzer_1967,Waksman_1966}) that fires a sequence of $n$ nodes 
at time $2n - 2$.
There are three cases.

\medskip

\noindent
Case 1. The position $(i + 1, i)$ is a hole. 
In this case there is no node at the position supposed to 
be the general.  Hence no nodes 
$(i + 2, i)$, $(i + 3, i)$, \ldots, $(w, i)$ fire.

\medskip

\noindent
Case 2.  A position $(i + s, i)$ for some $2 \leq s \leq w - i - 1$ is 
a hole.
In this case the activated original FSSP fires the $s - 1$ 
nodes $(i + 1, i)$, $(i + 2, i)$, \ldots, $(i + s - 1, i)$ 
at time $(2i + 1) + \{2(s - 1) - 2\} = 2i + 2s - 3 \leq 2w - 5$.
The remaining $w - i - s$ nodes $(i + s + 1, i)$, \ldots, $(w, i)$ 
do not fire because the hole $(i + s, i)$ blocks 
the signals from the node $(i + s - 1, i)$.

\medskip

\noindent
Case 3. There is no hole in the $w - i$ positions 
$(i + 1, i)$, $(i + 2, i)$, \ldots, $(w, i)$. 
In this case the activated original FSSP fires 
all of these nodes at time $(2i + 1) + \{2(w - i) - 2\} = 2w - 1$. 

\medskip

For the vertical sequence of positions 
$(i, i + 1)$, $(i, i + 2)$, $\ldots$ too, 
the signal $\abb{A}$ activates the original FSSP at time $2i + 1$ similarly.
This completes the definition of $A_{0}$.

\begin{figure}[htbp]
\centering
\includegraphics[scale=1.0]{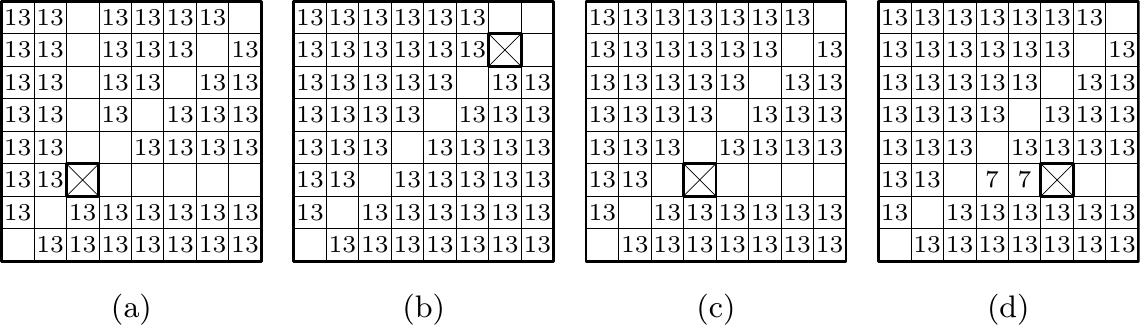}
\caption{Examples of configurations and firing times with $A_{0}$.}
\label{figure:fig062}
\end{figure}

In Fig. \ref{figure:fig062} we show four examples of 
configurations of size $w = 7$.
At each node we write the firing time if 
the node fires and left the node blank if the node does not fire.

\begin{figure}[htbp]
\centering
\includegraphics[scale=1.0]{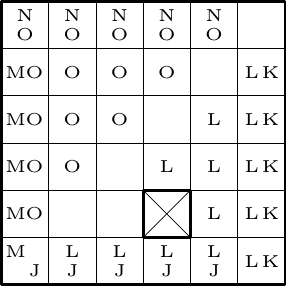}
\caption{An example of the generation of signals.}
\label{figure:fig063}
\end{figure}

We modify $A_{0}$ to another finite automaton $A_{1}$.
By the modification, each node that fires with $A_{0}$ 
before or at time $2w - 5$ by Case 2 never fires with $A_{1}$.
For this modification $A_{1}$ uses six signals 
$\abb{J}$, $\abb{K}$, $\abb{L}$, $\abb{M}$, 
$\abb{N}$, $\abb{O}$.
The three signals $\abb{J}$, $\abb{K}$, $\abb{L}$ 
are generated and travel as follows.
\begin{enumerate}
\item[$\bullet$] A signal $\abb{J}$ is generated at 
$v_{\abb{gen}}$ at time $0$ and proceeds to the east 
to the node $(w, 0)$.
\item[$\bullet$] When the signal $\abb{J}$ arrives at the node 
$(w, 0)$ it changes to a signal $\abb{K}$.
The signal $\abb{K}$ proceeds to the north to the node 
$(w, w - 1)$.
\item[$\bullet$] At each node $(w, j)$ ($0 \leq j \leq w - 1$), 
the signal $\abb{K}$ generates a signal $\abb{L}$.
The signal $\abb{L}$ proceeds to the west to the node 
$(j + 1, j)$ unless it is blocked by a hole.
\end{enumerate}
The generation and the travel of the three signals 
$\abb{M}$, $\abb{N}$, $\abb{O}$ are similar, 
replacing the directions the east, the north and the west 
with the north, the east and the south respectively.
In Fig. \ref{figure:fig063} we show all the generated signals 
in an example configuration.  

We define the firing rule of $A_{1}$ as follows: 
a node fires at a time $t$ with $A_{1}$ if and only if 
the node fires at the time $t$ with $A_{0}$ and 
it has received the signal $\abb{L}$ or $\abb{O}$ 
before or at the time $t$.

It is obvious that any node $(i, i)$ on the main diagonal 
never fires.
Consider a node $(i, j)$ such that $i \geq j + 1$.
If the node fires at time $2w - 1$ by Case 3 with $A_{0}$ 
then the signal $\abb{L}$ arrives at the node 
at time $w + j + (w - i) = 2w - (i - j) \leq 2w - 1$ 
and the node fires at time $2w - 1$ with $A_{1}$.
If the node fires before or at time $2w - 5$ by Case 2 with $A_{0}$ 
then the signal $\abb{L}$ does not arrive at the node 
blocked by a hole, and hence the node never fires with $A_{1}$.
The same is true also for a node $(i, j)$ such that $i \leq j - 1$.
Hence, for any node $v = (i, j)$ the following statement is true with $A_{1}$: 
either (1) $v$ fires at time $2w - 1$ or (2) $v$ never fires.
\begin{figure}[htbp]
\centering
\includegraphics[scale=1.0]{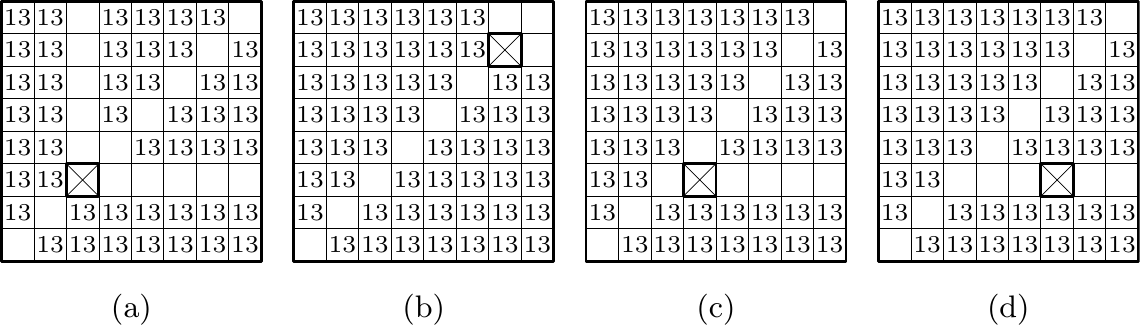}
\caption{The firing times of nodes in the four configurations 
shown in Fig. \ref{figure:fig062} with $A_{1}$.}
\label{figure:fig064}
\end{figure}
In Fig. \ref{figure:fig064} we show the firing times of nodes 
in the four configurations shown in Fig. \ref{figure:fig062}.

We will prove that the following stronger statement is true with $A_{1}$ 
except the special case where $v = (w, w)$ and the position $(w - 1, w - 1)$ is a hole: 
either (1) $v$ fires at time $2w - 1$ or 
(2) $v$ never fires but $v$ is adjacent to a node $v'$ that fires 
at time $2w - 1$.  We assume that $i \geq j$.
There are three cases.

\medskip

\noindent
(Case 1) $j \leq w - 1$ and there are no holes at positions 
$(j, j), (j + 1, j), \ldots, (w, j)$.
In this case all nodes $(j + 1, j), \ldots, (w, j)$ fire at time $2w - 1$.
Although the node $(j, j)$ never fires it is adjacent to the node $(j + 1, j)$ 
that fires at time $2w - 1$.
(Note that we assume $j \leq w - 1$ and hence there is really a node 
at $(j + 1, j)$.)
Therefore, if $i \geq j + 1$ then $v$ fires at time $2w - 1$ and 
if $i = j$ then $v = (j, j)$ is adjacent to a node $v' = (j + 1, j)$ 
that fires at time $2w - 1$.

\medskip

\noindent
(Case 2) $j \leq w - 1$ and there is a hole at one of the positions 
$(j, j), (j + 1, j), \ldots, (w, j)$.
In this case any of the nodes in these positions (including the node $v = (i, j)$) 
never fires.
We have $j \geq 1$ because the boundary of $C$ has no holes in it.
There are no holes in positions $(j - 1, j - 1), (j, j - 1), \ldots, (w, j - 1)$ 
because there is only one hole. 
All nodes $(j, j - 1), (j + 1, j - 1), \ldots, (w, j - 1)$ 
(including the node $v - (0, 1) = (i, j - 1)$) fire at time $2w - 1$.
Therefore, $v = (i, j)$ never fires and it is adjacent to a node $v' = (i, j - 1)$ 
that fires at time $2w - 1$.

\medskip

\noindent
(Case 3) $j = w$.  In this case we have $v = (w, w)$ and the position $(w - 1, w - 1)$ 
is a node because we exclude the case where $v = (w, w)$ and 
the position $(w - 1, w - 1)$ is a hole.
By the definition of $A_{1}$ the node $v$ never fires.
At least one of the two positions $(w - 1, w)$, $(w, w - 1)$ is a node 
and it fires at time $2w - 1$.
Therefore, $v$ never fires and $v$ is adjacent to a node that fires at time 
$2w - 1$.

\medskip

In Fig. \ref{figure:fig064} (b) we show an example of the exceptional case.
In this case the node $v = (w, w)$ never fires and 
the two nodes adjacent to $v$ 
(that is, nodes $(w - 1, w)$ and $(w, w - 1)$) also do not fire at $2w - 1$.

We define the third finite automaton $A_{2}$.
$A_{2}$ simulates the behavior of $A_{1}$.
A node $v$ in $C$ fires at a time $t$ with $A_{2}$ 
if and only if either $v$ fires at the time $t - 1$ with $A_{1}$ 
or $v$ is adjacent to a node $v'$ that fires at the time 
$t - 1$ with $A_{1}$.
Then, by what we have proved above, 
any node $v$ of $C$ fires at time $2w$ except the case 
where $v = (w, w)$ and the position $(w - 1, w - 1)$ is a hole.

We modify $A_{2}$ by adding the following ad hoc rule 
to let the node $(w, w)$ fire at $2w$:  
if the signal $\abb{A}$ (see Fig. \ref{figure:fig060}) 
arrives at a node $v$ having the boundary condition 
$(0, 0, 1, 1)$ (that is, the boundary condition of 
the node $(w, w)$) at a time $t$ the node $v$ fires at the 
time $t$.
Let $A_{3}$ be the finite automaton obtained by this modification.
Then all nodes in $C$ fire at time $2w$ with $A_{3}$ and 
$A_{3}$ is the desired solution.
\hfill $\Box$

\medskip

\section{Some results on minimum firing times \\
$\abb{mft}_{\abb{SH}[k]}(C)$ of $\abb{SH}[k]$}
\label{section:general_results}

\medskip

For $k \geq 2$, we do not know whether 
$\abb{SH}[k]$ has minimal-time solutions or not.
However the author has the conjecture that $\abb{SH}[k]$ has 
minimal-time solutions for all values of $k$.
A first step to prove this is to know the exact value of 
$\abb{mft}_{\abb{SH}[k]}(C)$.
In this section we consider the minimum value and the maximum value 
of $\abb{mft}_{\abb{SH}[k]}(C)$ 
when $k$, $w$ are fixed and 
$C$ ranges over all configurations of ${\abb{SH}}[k]$ of size $w$.
Our results were summarized in 
Section \ref{section:introduction}.
Here we give a more detailed outline of this section.

In Subsection \ref{subsection:minimum_value} we show that 
$2w$ is the smallest value of $\abb{mft}_{\abb{SH}[k]}(C)$ 
for all $k$ and all sufficiently large $w$.
In Subsection \ref{subsection:a_characterization} 
we define a value $H_{k,w}$ and show that 
$H_{k,w}$ is the maximum value of 
$\abb{mft}_{\abb{SH}[k]}(C)$ for all $k$ and all $w$.
In Subsection \ref{subsection:maximal_barriers} 
we introduce  a notion ``maximal barriers of configurations'' 
and show one property of this notion.
In Subsection \ref{subsection:maximal_barriers_and_h} 
we define a value $c_{k}$ using this notion 
and show that $H_{k,w} = 2w + c_{k}$ for all $k$ 
and all sufficiently large $w$.
Therefore $2w + c_{k}$ is the maximum value of 
$\abb{mft}_{\abb{SH}[k]}(C)$ for all $k$ and 
all sufficiently large $w$.
The definition of $c_{k}$ itself gives an algorithm 
for computing $c_{k}$.
Using this algorithm we determined the value of 
$c_{k}$ for $2 \leq k \leq 9$.
In Subsection \ref{subsection:algorithm} we show 
the result.

\subsection{The minimum value of 
$\abb{mft}_{\abb{SH}[k]}(C)$}
\label{subsection:minimum_value}

By $\mathcal{C}_{k,w}$ we denote the set of 
all configurations $C$ of size $w$ of ${\abb{SH}}[k]$.
First we show a result on the minimum value 
of $\abb{mft}_{\abb{SH}[k]}(C)$ for $C \in \mathcal{C}_{k,w}$.

\begin{thm}
\label{theorem:thm001}
{\rm (1)} For any $k$ and any $w$, 
$2w \leq \abb{mft}_{\abb{SH}[k]}(C)$ 
for any $C \in \mathcal{C}_{k,w}$.
{\rm (2)} For any $k$ and any $w \geq 2k+1$, 
there exists $C \in \mathcal{C}_{k,w}$ 
such that $\abb{mft}_{\abb{SH}[k]}(C) 
= 2w$.
\end{thm}

\noindent
{\it Proof}. 
(1) The lower bound is obvious because 
$2w = \abb{d}_{C}(v_\abb{gen}, (w, w)) 
\leq \abb{mft}(C)$.

\medskip

\begin{figure}[htbp]
\centering
\includegraphics[scale=1.0]{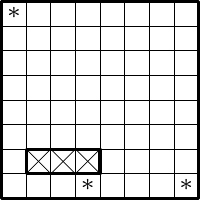}
\caption{A configuration $\tilde{C}$ 
in $\mathcal{C}_{3,7}$ 
that has the minimum firing time $14$.}
\label{figure:fig002}
\end{figure}

\noindent
(2) As an example we consider the case $k = 3$ 
and show a configuration $\tilde{C}$ of size $\tilde{w} = 7$ 
of $\abb{SH}[3]$ such that 
$\abb{mft}_{\abb{SH}}[3](\tilde{C}) \leq 14$ 
(and hence $\abb{mft}_{\abb{SH}}[3](\tilde{C}) = 14$ by 
(1) of this theorem).
This $\tilde{C}$ is shown in Fig. \ref{figure:fig002}.
We will construct a partial solution $A$ 
of $\abb{SH}[3]$ that fires $\tilde{C}$ at time $14$.
This shows $\abb{mft}(\tilde{C}) \leq 14$ 
by \eqref{equation:eq012}.
Suppose that copies of $A$ are placed 
in a configuration $C$ 
of $\abb{SH}[3]$ of size $w$.

$A$ uses two signals to check the condition ``$w = 7$.''
The first signal starts at $v_\abb{gen}$ at time $0$ 
and proceeds to the node $(0, 7)$ by going $7$ steps 
to the north.
If the signal falls off the north boundary of $C$ 
before arriving at the node $(0, 7)$ (and hence $w < 7$) 
the signal vanishes.
If the signal arrives at the node $(0, 7)$ but its 
boundary condition is not $(1, 0, 0, 1)$ 
(the boundary condition of the northwest corner) 
(and hence $w > 7$) the signal vanishes.
If the signal arrives at the node $(0, 7)$ and 
its boundary condition is $(1, 0, 0, 1)$ then 
it knows that the condition ``$w = 7$'' is true and it generates 
a message $\abb{W}_{0}$ at the node $(0, 7)$ at time $7$.

Similarly the second signal starts at $v_\abb{gen}$ at time $0$ 
and proceeds to the node $(7, 0)$ by going $7$ steps to the east.
If $w < 7$ or $w > 7$ then the signal vanishes.
If $w = 7$ then the signal generates a message $\abb{W}_{1}$ 
at the node $(7, 0)$ at time $7$.

$A$ also uses a signal that checks the condition 
``there are holes at the positions 
$(1, 1)$, $(2, 1)$, $(3, 1)$.''
It starts at $v_\abb{gen}$ at time $0$, proceeds 
to the east, and checks the above condition 
by checking the boundary conditions of the 
three nodes $(1, 0)$, $(2, 0)$, $(3, 0)$.
If the condition is satisfied the signal knows it 
at the node $(3, 0)$ at time $3$, 
and the signal generates a message $\abb{M}$ 
at the node $(3, 0)$ at time $3$.
If the condition is not satisfied the signal vanishes.

Messages $\abb{W}_{0}$, $\abb{W}_{1}$, $\abb{M}$ 
propagate to all nodes in $C$ as soon as they are generated.

$A$ uses the following rule to fire: 
a node fires at a time $t$ if and only if $t = 14$ and 
the node has received at least one of the two messages 
$\abb{W}_{0}$, $\abb{W}_{1}$ and 
also the message $\abb{M}$ 
before or at the time $t$.
(We may assume that each message keeps the current time up to $14$ 
and hence each node knows the current time when it receives 
a message before or at time $14$.)
We show that $A$ is a partial solution 
that has the set $\{\tilde{C}\}$ as its domain 
and that fires configurations in the domain at time $14$.

Suppose that $C = \tilde{C}$. 
Then all of the three messages are generated, 
$\abb{W}_{0}$ at $(0, 7)$ at time $7$, 
$\abb{W}_{1}$ at $(7, 0)$ at time $7$, and 
$\abb{M}$ at $(3, 0)$ at time $3$.
(In Fig. \ref{figure:fig002} the three ``*'' 
denote the nodes where these messages 
are generated.)
Therefore, any node $v = (x, y)$ in $C$ receives at least 
one of $\abb{W}_{0}$, $\abb{W}_{1}$ and also $\abb{M}$ 
before or at time $14$ and hence fires at $14$. 
This follows from the following observation.
As for $\abb{W}_{0}$, $\abb{W}_{1}$, 
we have 
$\min\{ 7 + \abb{d}_{C}((0, 7), v), 
7 + \abb{d}_{C}((7, 0), v)\} \leq 14$ 
(the equality is true for $v = (0, 0), (2, 2), (3, 3), \ldots, (7, 7)$).
As for $\abb{M}$, we have 
$3 + \abb{d}_{C}((3, 0), v) \leq 14$ 
(the equality is true for $v = (1, 7), (7, 7)$).

Conversely suppose that a node $v$ in $C$ fires at some 
time.
Then at least one of $\abb{W}_{0}$ and $\abb{W}_{1}$ 
was generated.
Hence $w = 7$ is true.
Moreover $\abb{M}$ was generated.
Hence there are holes at $(1, 1)$, $(2, 1)$, $(3, 1)$.
Therefore $C = \tilde{C}$.

Thus we have proved that $A$ is a partial solution 
that has the domain $\{ \tilde{C} \}$ and that fires 
configurations in the domain at time $14$.
\hfill $\Box$

\medskip

In the proof of Theorem \ref{theorem:thm001} 
we used two messages $\abb{W}_{0}$, $\abb{W}_{1}$ that 
imply $w = \tilde{w}$.  
We call these two messages the \textit{size check messages} 
and from now on we use them repeatedly.
We summarize the situations where these messages are used 
as follows.

$\tilde{w}$ is some fixed value and $C$ is an arbitrary configuration 
of $\abb{SH}[k]$ of an arbitrary size $w$.  
Nodes of $C$ are copies of a finite automaton $A$.
The two size check messages $\abb{W}_{0}$, $\abb{W}_{1}$ are 
used for nodes in $C$ to know whether $w = \tilde{w}$ or not.

$\abb{W}_{0}$, $\abb{W}_{1}$ are generated and propagate to 
all nodes in $C$ as explained in the proof of 
Theorem \ref{theorem:thm001}.
If $w = \tilde{w}$ then $\abb{W}_{0}$ and $\abb{W}_{1}$ 
are generated at time $w$ at $(0, w)$ and 
$(w, 0)$ respectively and propagate to all nodes 
in $C$.
A node $v$ in $C$ receives at least one of 
$\abb{W}_{0}$, $\abb{W}_{1}$ (and hence knows that 
$w = \tilde{w}$) at time 
$w + \min \{ \abb{d}_{C}((0, w), v), \abb{d}_{C}((w, 0), v) \}$.
If $w \not= \tilde{w}$ then $\abb{W}_{0}$, $\abb{W}_{1}$ 
are not generated and nodes in $C$ never receive them.

\subsection{A characterization of the maximum value of 
$\abb{mft}_{\abb{SH}[k]}(w)$}
\label{subsection:a_characterization}

In this subsection we concentrate our attention on the maximum value of 
$\abb{mft}(C)$ for $C \in \mathcal{C}_{k, w}$.
First we show a result (Corollary \ref{corollary:cor000}) 
that is used repeatedly to show 
lower bounds of $\abb{mft}(C)$.

Suppose that $C$, $C'$ are configurations of $\abb{SH}[k]$, 
$t$ ($\geq 0$) is a number, and $v$ is a node in both of $C$, $C'$ 
(or more precisely, $v$ is a position in $\mathbb{Z}^{2}$ 
that is a node in both of $C$, $C'$).
By $C \equiv_{t, v}' C'$ we mean that the following two statements are true.
\begin{enumerate}
\item[$\bullet$] If $P$ is a path in $C$ from $v_\abb{gen}$ 
to $v$ of length at most $t$ then $P$ is also a path 
in $C'$ and $\abb{bc}_{C}(u) = \abb{bc}_{C'}(u)$ 
for any node $u$ in $P$.
\item[$\bullet$] The same statement with $C$, $C'$ 
exchanged.
\end{enumerate}

\begin{thm}
\label{theorem:thm008}
If $C \equiv_{t, v}' C'$ then\/ 
$\abb{mft}_{\abb{SH}[k]}(C) \geq t + 1$ if and only if\/ 
$\abb{mft}_{\abb{SH}[k]}(C')$ $\geq t + 1$.
\end{thm}

\noindent
{\it Proof}. 
Suppose that $C \equiv_{t, v}' C'$.
Let $A$ be any solution of $\abb{SH}[k]$.
First we prove that $\abb{st}(v, s, C, A) = \abb{st}(v, s, C', A)$ 
for any $s$ such that $s \leq t$.
To prove it we assume $\abb{st}(v, s, C, A) \not= \abb{st}(v, s, C', A)$ 
for a value $s$ ($\leq t$) and derive a contradiction.

If $\abb{st}(u, r, C, A) \not= \abb{st}(u, r, C', A)$ 
for a node $u$ in both of $C$, $C'$ and a time $r$, then 
one of the following is true.
\begin{enumerate}
\item[(1)] $\abb{bc}_{C}(u) \not= \abb{bc}_{C'}(u)$.
\item[(2)] $\abb{bc}_{C}(u) = \abb{bc}_{C'}(u)$ and $r = 0$.
\item[(3)] $\abb{bc}_{C}(u) = \abb{bc}_{C'}(u)$, $0 < r$ and 
$\abb{st}(u, r - 1, C, A) \not= \abb{st}(u, r - 1, C', A)$.
\item[(4)] $\abb{bc}_{C}(u) = \abb{bc}_{C'}(u)$, $0 < r$ and 
$\abb{st}(w, r - 1, C, A) \not= \abb{st}(w, r - 1, C', A)$ 
for a node $w$ that is in both of $C$, $C'$ and is 
adjacent to $u$.
\end{enumerate}

Repeatedly using this property starting with our assumption 
$\abb{st}(v, s, C, A) \not= \abb{st}(v, s, C', A)$ 
we know that there is a sequence $u_{r_{0}}$, $u_{r_{0} + 1}$, 
\ldots, $u_{s-1}$, $u_{s}$ of nodes in both of $C$, $C'$ 
such that 
\begin{enumerate}
\item[$\bullet$] $u_{s} = v$, 
\item[$\bullet$] $\abb{st}(u_{r}, r, C, A) \not= 
\abb{st}(u_{r}, r, C', A)$ for all $r_{0} \leq r \leq s$, 
\item[$\bullet$] $\abb{bc}_{C}(u_{r}) = \abb{bc}_{C'}(u_{r})$ 
for all $r_{0} < r \leq s$,
\item[$\bullet$] either $u_{r-1} = u_{r}$ or $u_{r-1}$ and 
$u_{r}$ are adjacent for all $r_{0} < r \leq s$,
\end{enumerate}
and moreover one of the following is true:
\begin{enumerate}
\item[(5)] $\abb{bc}_{C}(u_{r_{0}}) \not= 
\abb{bc}_{C'}(u_{r_{0}})$, 
\item[(6)] $\abb{bc}_{C}(u_{r_{0}}) = 
\abb{bc}_{C'}(u_{r_{0}})$ and $r_{0} = 0$.
\end{enumerate}
However, from each of (5), (6) we can derive a contradiction.

Suppose the case (5).
One of $\abb{st}(u_{r_{0}}, r_{0}, C, A)$, 
$\abb{st}(u_{r_{0}}, r_{0}, C', A)$ is not the quiescent state 
$\abb{Q}$.
We consider the case where the former is not $\abb{Q}$.
Then $\abb{d}_{C}(v_\abb{gen}, u_{r_{0}}) \leq r_{0}$, 
there is a path $P$ in $C$ from $v_\abb{gen}$ to $v$ of 
length at most $s$ ($\leq t$), and there is a node $u_{r_{0}}$ 
in it such that $\abb{bc}_{C}(u_{r_{0}}) \not= 
\abb{bc}_{C'}(u_{r_{0}})$.
This contradicts our assumption that $C \equiv_{t, v}' C'$.

Suppose the case (6).
In this case, 
$\abb{st}(u_{r_{0}}, r_{0}, C, A) = 
\abb{st}(u_{r_{0}}, r_{0}, C', A) = \abb{G}$ 
if $u_{r_{0}} = v_\abb{gen}$ and
$\abb{st}(u_{r_{0}}, r_{0}, C, A) = 
\abb{st}(u_{r_{0}}, r_{0}, C', A) = \abb{Q}$ 
otherwise.
Therefore, 
$\abb{st}(u_{r_{0}}, r_{0}, C, A) \not= 
\abb{st}(u_{r_{0}}, r_{0}, C', A)$ 
cannot be true.

Thus we have proved that $\abb{st}(v, s, C, A) = \abb{st}(v, s, C', A)$ 
for any $s$ ($\leq t$).
Next we assume that $\abb{mft}(C) \geq t + 1$, $\abb{mft}(C') \leq t$ and 
derive a contradiction.
(The derivation of a contradiction for the case where 
$\abb{mft}(C) \leq t$, $\abb{mft}(C') \geq t + 1$ is similar.)
We select a solution that fires $C'$ at time $\abb{mft}(C')$ ($\leq t$) as $A$.
Then we have $\abb{F} = \abb{st}(v,  \abb{mft}(C'), C', A) 
= \abb{st}(v, \abb{mft}(C'), C, A)$.
This means that $A$ is a solution that fires $C$ at the time $\abb{mft}(C')$ ($\leq t$).
This contradicts our assumption that $\abb{mft}(C) \geq t + 1$.
\hfill $\Box$

\medskip

For $C \in \mathcal{C}_{k,w}$, $v \in C$, 
let $T(v, C)$ be defined by 
\begin{equation}
T(v, C) = \min \{ \abb{d}_{C}(v_\abb{gen}, v; (0, w)), 
                  \abb{d}_{C}(v_\abb{gen}, v; (w, 0)) \}. 
\label{equation:eq003} 
\end{equation}

\begin{thm}
\label{theorem:thm002}
For any configuration $C$ of $\abb{SH}[k]$, 
$\max_{v \in C} T(v, C) \leq \abb{mft}_{\abb{SH}[k]}(C)$.
\end{thm}

\noindent
{\it Proof}. 
It is sufficient to prove $T(v, C) \leq \abb{mft}(C)$ for any $v \in C$.
Suppose that $v$ is a fixed node of $C$.

\begin{figure}[htbp]
\centering
\includegraphics[scale=1.0]{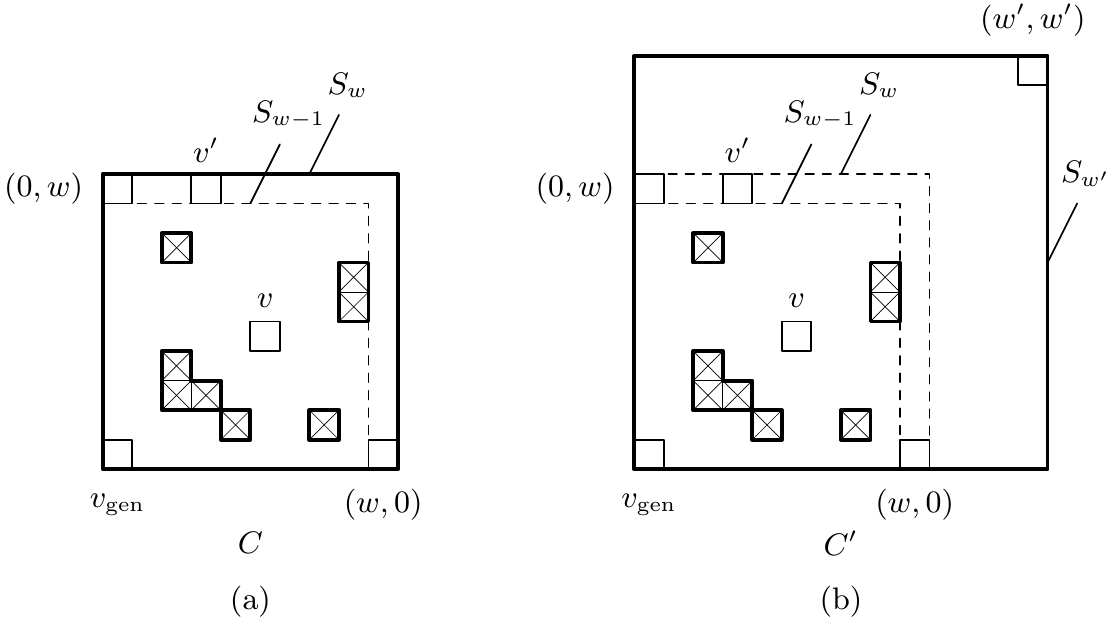}
\caption{Two configurations $C$, $C'$ used in the proof of 
Theorem \ref{theorem:thm002}.}
\label{figure:fig003}  
\end{figure}

Let $w$ be the size of $C$. 
We define another configuration $C'$ of size $w'$ of 
$\abb{SH}[k]$ (see Fig. \ref{figure:fig003}) 
such that 
\begin{enumerate}
\item[$\bullet$] $w < w'$ and $T(v, C) \leq 2w'$.
\item[$\bullet$] The distribution of holes in the square 
$S_{w}$ is the same in $C$ and in $C'$ 
(and hence $C'$ has no holes in $S_{w'} - S_{w}$). 
(For the definition of $S_{w}$ and so on, see 
\eqref{equation:eq013} in Section \ref{section:preliminaries}.)
\end{enumerate}

We assume that $P$ is a path in $C$ from $v_\abb{gen}$ to $v$ 
of length at most $T(v, C) - 1$ 
and $P$ passes through the set $S_{w} - S_{w - 1}$ 
(the north and the east boundaries of $C$) 
and derive a contradiction.

Let $v'$ be a node in $P$ that is in $S_{w} - S_{w - 1}$.
We consider the case where $v'$ is 
in the horizontal part of $S_{w} - S_{w - 1}$ 
(the north boundary of $C$).
Then we have the following contradiction:

\begin{align*}
T(v, C) - 1 
& \geq |P| \\
& \geq \abb{d}_{C}(v_\abb{gen}, v; v') \\
& = \abb{d}_{C}(v_\abb{gen}, v') + \abb{d}_{C}(v', v) \\
& \geq \abb{d}_\abb{MH}(v_\abb{gen}, v') + \abb{d}_{C}(v', v) \\
& = \abb{d}_{C}(v_\abb{gen}, (0, w)) + 
      \abb{d}_{C}((0, w), v') + \abb{d}_{C}(v', v) \\
& \geq \abb{d}_{C}(v_\abb{gen}, (0, w)) + \abb{d}_{C}((0, w), v) \\
& = \abb{d}_{C}(v_\abb{gen}, v; (0, w)) \\
& \geq T(v, C).
\end{align*}

This means that if $P$ is a path from $v_\abb{gen}$ 
to $v$ in $C$ of length at most $T(v, C) - 1$ 
then $P$ is in $S_{w - 1}$.
Therefore, $P$ is also in $C'$ and $\abb{bc}_{C}(u) = 
\abb{bc}_{C'}(u)$ for any $u$ in $P$.

Similarly, we can show that if 
$P$ is a path from $v_\abb{gen}$ 
to $v$ in $C'$ of length at most $T(v, C) - 1$ 
then $P$ is also in $C$ and 
$\abb{bc}_{C}(u) = \abb{bc}_{C'}(u)$ for any $u$ in $P$.
In the proof we use the fact that 
$S_{w'} - S_{w - 1}$ has no holes of $C'$ in it and hence 
for any nodes $v'$, $v''$ of $C'$ in $S_{w} - S_{w - 1}$ there is 
a path that is from $v'$ to $v''$, 
is a path in $C'$, is a path in $S_{w} - S_{w - 1}$, 
and is of length $\abb{d}_{\abb{MH}}(v', v'')$.

Thus we have proved that $C \equiv_{t, v}' C'$ with $t = T(v, C) - 1$.
By Theorem \ref{theorem:thm008}, 
$T(v, C) \leq \abb{mft}(C)$ is true if and only if 
$T(v, C) \leq \abb{mft}(C')$ is true.
However the latter is true because 
$T(v, C) \leq 2w' = \abb{d}_{C'}(v_{\abb{gen}}, (w', w')) 
\leq \abb{mft}(C')$.
Therefore we have $T(v, C) \leq \abb{mft}(C)$.
\hfill $\Box$

\medskip

We simply write $C \equiv_{t}' C'$ if there exists $v$ such that 
$C \equiv_{t, v}' C'$.
Let ``$\equiv_{t}$'' be the reflexive and transitive closure 
of the relation ``$\equiv_{t}'$.''
More precisely, $C \equiv_{t} C'$ is true if and only if 
there exists a sequence $C_{0}, \ldots, C_{n}$ of configurations 
($0 \leq n$) such that $C = C_{0}$, $C' = C_{n}$, 
and $C_{i} \equiv_{t}' C_{i+1}$ for any $0 \leq i \leq n - 1$.

\begin{cor}
\label{corollary:cor000}
For any configuration $C$ of $\abb{SH}[k]$ and any $t$, 
if there exists a configuration $C'$ of $\abb{SH}[k]$ such that 
$C \equiv_{t} C'$ and $t + 1 \leq \max_{v \in C'} T(v, C')$ 
then $t + 1 \leq \abb{mft}(C)$.
\end{cor}

\noindent
{\it Proof}. 
By Theorem \ref{theorem:thm008}, the definition of the relation ``$\equiv_{t}$'' 
and $C \equiv_{t} C'$, 
$t + 1 \leq \abb{mft}(C)$ is true if and only if 
$t + 1 \leq \abb{mft}(C')$ is true.
However the latter is true because 
$t + 1 \leq \max_{v \in C'} T(v, C') \leq \abb{mft}(C')$ by 
Theorem \ref{theorem:thm002}.
\hfill $\Box$

\medskip

Corollary \ref{corollary:cor000} can be used to prove a lower bound 
$t + 1 \leq \abb{mft}(C)$.
If we try to prove this lower bound using the corollary 
it is necessary to find configurations $C_{0}$, $C_{1}$, \ldots, $C_{n}$ 
($n \geq 0$) such that 
$C = C_{0} \equiv_{t}' C_{1} \equiv_{t}' \ldots \equiv_{t}' C_{n}$ and 
$t + 1 \leq \max_{v \in C_{n}} T(v, C_{n})$.
Here we have a problem.
The corollary gives us no hint about the sizes of 
$C_{1}$, \ldots, $C_{n}$.
However, by the following theorem and its corollary we may assume that 
the sizes of $C_{1}$, \ldots, $C_{n}$ are the same as the size of $C$.

\begin{thm}
\label{theorem:thm015}
Let $C$, $C'$ be configurations of\/ $\abb{SH}[k]$ and let $t$ {\rm (}$\geq 0${\rm )} 
be a number.
If $C \equiv_{t}' C'$ and $\max_{v \in C} T(v, C) \leq t$ then 
$C$ and $C'$ have the same size.
\end{thm}

\noindent
{\it Proof}. 
Let $v$ be a node in both of $C$, $C'$ such that 
$C \equiv_{t, v}' C'$ and $w$ be the size of $C$.
Then we have either $\abb{d}_{C}(v_{\abb{gen}}, v; (0, w)) \leq t$ or 
$\abb{d}_{C}(v_{\abb{gen}}, v; (w, 0)) \leq t$ 
because $\max_{v \in C} T(v, C) \leq t$.
We assume the former is true.
Then there is a path in $C$ from $v_{\abb{gen}}$ to $v$ 
via $(0, w)$ of length at most $t$.
By $C \equiv_{t, v}' C'$, this path is also a path in $C'$ 
and $\abb{bc}_{C}(u) = \abb{bc}_{C'}(u)$ for any node $u$ on the path.
This means that the size of $C'$ is $w$.
\hfill 
$\Box$

\medskip

\begin{cor}
\label{corollary:cor004}
Let $C$ be a configuration of $\abb{SH}[k]$ and $t$ {\rm (}$\geq 0${\rm )} be 
a number.
If there exists a configuration $C'$ of $\abb{SH}[k]$ such that 
$C \equiv_{t} C'$ and $t + 1 \leq \max_{v \in C'} T(v, C')$ 
then there exist configurations $C_{0}$, $C_{1}$, \ldots, $C_{n}$ {\rm (}$n \geq 0${\rm )} 
of $\abb{SH}[k]$ of the same sizes as $C$ such that 
$C = C_{0} \equiv_{t}' C_{1} \equiv_{t}' \ldots \equiv_{t}' C_{n}$
and $t + 1 \leq \max_{v \in C_{n}} T(v, C_{n})$.
\end{cor}

\noindent
{\it Proof}. 
Let $C_{0}$, $C_{1}$, \ldots, $C_{m}$ be configurations of $\abb{SH}[k]$ 
such that $C = C_{0} \equiv_{t}' C_{1} \equiv_{t}' \ldots \equiv_{t}' C_{m} = C'$.
Let $i_{0}$ be the smallest value of $i$ (possibly $0$) such that 
$t + 1 \leq \max_{v \in C_{i}} T(v, C_{i})$.  
Then if we set $n = i_{0}$ then the configurations 
$C_{0}, C_{1}, \ldots, C_{n}$ satisfy the condition stated in the corollary 
by Theorem \ref{theorem:thm015}
\hfill $\Box$

\medskip

Although 
$\max_{v \in C} T(v, C) \leq \abb{mft}(C)$ is true, 
$\abb{mft}(C) \leq \max_{v \in C} T(v, C)$ 
is not necessarily true.
However we have a weaker result.
Let $H_{k,w}$ be defined by
\begin{equation}
H_{k,w} = \max_{C \in \mathcal{C}_{k, w}, v \in C} T(v, C).
\label{equation:eq010}
\end{equation}

\begin{thm}
\label{theorem:thm003}

\noindent
\begin{enumerate}
\item[$(1)$] For any $k$, any $w$ and 
any $C \in \mathcal{C}_{k,w}$, 
$\abb{mft}_{\abb{SH}[k]}(C) \leq H_{k,w}$.
\item[$(2)$] For any $k$ and any $w$ 
there exists $C \in \mathcal{C}_{k,w}$ 
such that 
$\abb{mft}_{\abb{SH}[k]}(C) = H_{k,w}$.
\end{enumerate}
\end{thm}

\noindent
{\it Proof}. 
(1) Let $\tilde{w}$ be some fixed value and 
let $\tilde{C}$ be some fixed configuration of size $\tilde{w}$ 
of $\abb{SH}[k]$.
We construct a partial solution $A$ 
that has $\mathcal{C}_{k,\tilde{w}}$ as its domain 
and that fires configurations in the domain 
at time $H_{k,\tilde{w}}$.
Suppose that $C$ is a configuration of size $w$ of $\abb{SH}[k]$ 
and that copies of $A$ are placed on $C$.

$A$ uses the size check messages $\abb{W}_{0}$, $\abb{W}_{1}$ 
(see the comment after Theorem \ref{theorem:thm001}).
$\abb{W}_{0}$, $\abb{W}_{1}$ are generated if and only if $w = \tilde{w}$ 
and if they are generated it is at the nodes $(0, \tilde{w})$, 
$(\tilde{w}, 0)$ and at time $\tilde{w}$.
They propagate in $C$ as soon as they are generated.

A node in $C$ fires at a time $t$ if and only if 
$t = H_{k, \tilde{w}}$ and 
it has received at least one of $\abb{W}_{0}$, $\abb{W}_{1}$ 
before or at the time $t$.
We show that $A$ is a desired partial solution.

Suppose that $C \in \mathcal{C}_{k, \tilde{w}}$.
Then $w = \tilde{w}$ is true and 
$\abb{W}_{0}$, $\abb{W}_{1}$ are generated.
Therefore any node $v$ in $C$ receives 
at least one of $\abb{W}_{0}$, $\abb{W}_{1}$ at time 
$\tilde{w} + \min \{ \abb{d}_{C}((0, \tilde{w}), v), 
\abb{d}_{C}((\tilde{w}, 0), v) \} 
= T(v, C) \leq H_{k, \tilde{w}}$.
Hence $v$ fires at time $H_{k, \tilde{w}}$.
Therefore, $A$ fires $C$ at time $H_{k, \tilde{w}}$.

Conversely, suppose that a node in $C$ fires at some time.
Then at least one of $\abb{W}_{0}$, $\abb{W}_{1}$ was generated, 
$w = \tilde{w}$ is true, and hence 
$C \in \mathcal{C}_{k, \tilde{w}}$.

Thus we constructed a partial solution $A$ that has 
$\mathcal{C}_{k, \tilde{w}}$ as its domain and 
that fires configurations in the domain 
(including $\tilde{C}$ itself) at time 
$H_{k, \tilde{w}}$.
This shows $\abb{mft}(\tilde{C}) \leq H_{k, \tilde{w}}$.

(2) Let $C \in \mathcal{C}_{k,w}$ and $v \in C$ be such that 
$T(v, C) = H_{k,w}$.
Then by Theorem \ref{theorem:thm002}, 
$H_{k, w} = T(v, C) \leq \abb{mft}(C)$.
However, by (1) of the present theorem, $\abb{mft}(C) \leq H_{k,w}$.
Hence $\abb{mft}(C) = H_{k,w}$.
\hfill$\Box$

\medskip

\subsection{Maximal barriers of configurations}
\label{subsection:maximal_barriers}

We have proved that $H_{k, w}$ is the largest value 
of $\abb{mft}_{\abb{SH}[k]}(C)$ 
for $C \in \mathcal{C}_{k, w}$.
To determine the exact value of $H_{k, w}$, 
in this subsection we introduce a notion 
``maximal barriers'' and study its properties.
Suppose that a configuration $C$ of size $w$ 
of $\abb{SH}[k]$ is given and is fixed.
Let the square $S_{w}$ defined by \eqref{equation:eq013} 
be the set of positions in $C$.

\begin{figure}[htbp]
\centering
\includegraphics[scale=1.0]{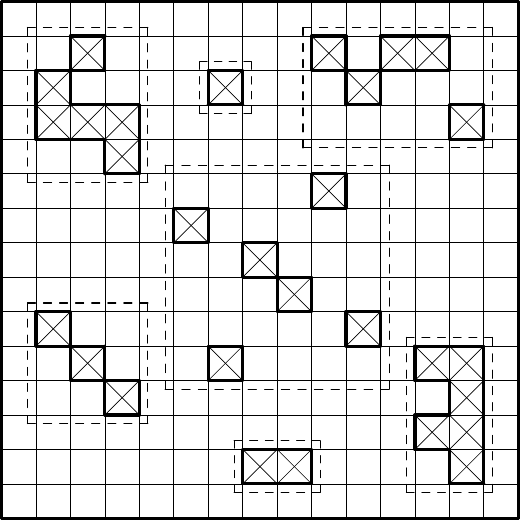}
\caption{An example of a configuration of size $14$ of 
$\abb{SH}[29]$ and its maximal barriers.}
\label{figure:fig006}  
\end{figure}

Let $R$ be a nonempty subset of $S_{w}$ of the form of 
a rectangle 
$\{(x, y) ~|~ 
x_{0} \leq x \leq x_{1}, 
y_{0} \leq y \leq y_{1}\}$ 
($0 \leq x_{0} \leq x_{1} \leq w$, 
$0 \leq y_{0} \leq y_{1} \leq w$).
We call $R$ a \textit{barrier} of $C$ if 
each column and each row of $R$ contain at least 
one hole.
Moreover, we say that a barrier $R$ is \textit{maximal} 
if it is not properly contained in another barrier.
As an example, in Fig. \ref{figure:fig006} 
we show a configuration of size $14$ of $\abb{SH}[29]$.
It has $7$ maximal barriers (regions shown by dotted lines) 
and $70$ barriers.
For example, the northwest maximal barrier (with three columns and 
four rows) has $20$ barriers in it.

\begin{figure}[htbp]
\centering
\includegraphics[scale=1.0]{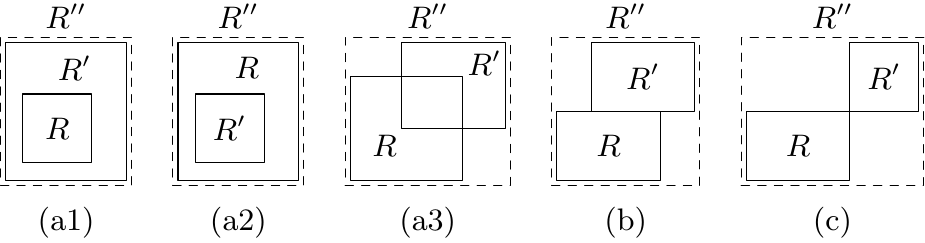}
\caption{Three cases mentioned in (2) of 
Theorem \ref{theorem:thm009}.}
\label{figure:fig005}
\end{figure}

\begin{thm}
\label{theorem:thm009}
\begin{enumerate}
\item[{\rm (1)}] Suppose that $R$, $R'$ are different 
barriers of $C$ and one of the following three 
statements is true.
\begin{enumerate}
\item[$\bullet$] $R \cap R' \not= \emptyset$ 
{\rm (}Fig. $\ref{figure:fig005}$ {\rm (a1)}, 
{\rm (a2)}, {\rm (a3))}.
\item[$\bullet$] $R \cap R' = \emptyset$ but 
there are positions $v \in R$, $v' \in R'$ 
that are adjacent {\rm (}the figure {\rm (b))}.
\item[$\bullet$] $R \cap R' = \emptyset$, 
there are no positions $v \in R$, $v' \in R'$ 
that are adjacent but there are positions 
$v \in R$, $v' \in R'$ that touch with 
corners {\rm (}the figure {\rm (c))}.
\end{enumerate}
Then the smallest rectangle $R''$ that includes 
both of $R$, $R'$ is also a barrier.
\item[{\rm (2)}] If $R$, $R'$ are different 
maximal barriers of $C$ then none of the 
three statements in {\rm (1)} are true.
\item[{\rm (3)}] Any barrier $R$ of $C$ is included in 
exactly one maximal barrier of $C$.
Especially, any hole $v$ of $C$ is contained in 
exactly one maximal barrier of $C$ because 
$\{ v \}$ is a barrier.
\item[{\rm (4)}] A barrier of $C$ does not contain 
a position in the boundary of $S_{w}$.
\item[{\rm (5)}] Suppose that $R$ is a maximal barrier 
of $C$ and a position $v$ is out of $R$ but 
either $v$ is adjacent to a position in $R$ or 
$v$ touches a position in $R$ with corners.
Then $v$ is a node of $C$ and $v$ is not in maximal barriers of $C$.
\end{enumerate}
\end{thm}

\noindent
{\it Proof}. 
(1) Any column of $R''$ includes a column of $R$ or $R'$.
Therefore it contains at least one holes.
The same is also true for rows.

\medskip

\noindent
(2) Suppose that $R$, $R'$ are different maximal barriers 
of $C$ and one of the three statements is true.
Then the smallest rectangle $R''$ that includes both of 
$R$, $R'$ is a barrier by (1).
Therefore we have $R = R''$, $R' = R''$ because $R$, $R'$ 
are maximal barriers.
However this contradicts our assumption that $R$, $R'$ 
are different.

\medskip

\noindent
(3) Let $\mathcal{R}$ be the set of all barriers 
that include $R$.
$\mathcal{R}$ is not empty because $R$ itself 
is a barrier.
If $R'$, $R''$ are two different barriers in 
$\mathcal{R}$ then the smallest rectangle $R'''$ 
that includes  both of $R'$, $R''$ is a barrier 
by $R' \cap R'' \not= \emptyset$ and (1).
Moreover $R'''$ includes $R$.
Hence $R'''$ is in $\mathcal{R}$.
This means that there is one barrier $\tilde{R}$ 
in $\mathcal{R}$ 
that is maximum in $\mathcal{R}$ 
with respect to the inclusion relation.
It is obvious that this $\tilde{R}$ is a maximal 
barrier and it includes $R$.
Moreover we can show that there is at most one 
maximal barrier that includes $R$ using (2).

\medskip

\noindent
(4) Suppose that $v$ is a position in the boundary of $S_{w}$ 
and that $v$ is in a rectangle $R$ in $S_{w}$.
Then either the column of $R$ containing $v$ or 
the row of $R$ containing $v$ has no holes in it.
Therefore $R$ cannot be a barrier.

\medskip

\noindent
(5) Suppose that $v$ is in a maximal barrier $R'$.
$R$ and $R'$ are different because $R$ does not 
contains $v$ but $R'$ contains $v$.
Moreover it is obvious that one of the three statements 
in (1) is true.  This is a contradiction by (2).
Therefore $v$ is not in maximal barriers of $C$.
This implies that $v$ is a node of $C$ because 
a hole is in a maximal barrier by (4).
\hfill $\Box$

\medskip

The definition of maximal barriers itself gives the following 
algorithm to enumerate all maximal barriers of $C$: 
enumerate all rectangles in $S_{w}$ having at most 
$k$ columns and at most $k$ rows, 
delete all rectangles that 
are not barriers, and select maximal rectangles.
However this algorithm is not efficient.
We show a more efficient algorithm.

Let $\mathcal{R}$ be the set consisting of 
one large rectangle 
$\{(x, y) ~|~ 1 \leq x \leq w-1, 1 \leq y \leq w-1\}$.
Starting with this $\mathcal{R}$, repeat the following.
If all rectangles in $\mathcal{R}$ are barriers then 
finish the algorithm with $\mathcal{R}$ as its result.
Otherwise, select a rectangle $R$ that is not a barrier from 
$\mathcal{R}$.  
Suppose that a column of $R$ contains no holes.
If there is such a column that is also a side boundary column 
of $R$ 
then replace $R$ in $\mathcal{R}$ with the rectangle $R'$ 
that is obtained from $R$ be deleting that side boundary column.
Otherwise the column must be an inner column.
In this case replace $R$ with the two rectangles $R'$, $R''$ 
that are obtained from $R$ by deleting that inner column.
Similarly for the case when a row of $R$ contains no holes.
It is evident that at some step 
all rectangles in $\mathcal{R}$ are barriers and 
the algorithm finishes.
Let $\tilde{\mathcal{R}}$ denote the resulting set $\mathcal{R}$.

\begin{thm}
\label{theorem:thm010}
$\tilde{\mathcal{R}}$ is the set of all maximal barriers of $C$.
\end{thm}

\noindent
{\it Proof}. 
First we show that any barrier $R$ is included in a barrier in 
$\tilde{\mathcal{R}}$.
At the start of the algorithm $R$ is included in the unique large 
rectangle in $\mathcal{R}$ by Theorem \ref{theorem:thm009} (4).
Moreover, as the algorithm is executed 
$R$ continues to be in a set in $\mathcal{R}$.
This is because $R$ is a barrier and hence any deleted column 
or row does not include columns or rows of $R$.
Therefore, $R$ must be in a barrier in $\tilde{\mathcal{R}}$.

Suppose that $R$ is a maximal barrier but 
$R$ is not in $\tilde{\mathcal{R}}$. 
Then $R$ is included properly in a barrier $R'$ in 
$\tilde{\mathcal{R}}$.
But this contradicts the assumption that $R$ is maximal.
Therefore any maximal barrier is in $\tilde{\mathcal{R}}$.

Suppose that $R$ is a barrier in $\tilde{\mathcal{R}}$ but 
$R$ is not maximal.
Then there is a barrier $R'$ that includes $R$ 
properly.
But then there is another barrier $R''$ in 
$\tilde{\mathcal{R}}$ that includes $R'$.
This contradicts our assumption that $R$ is 
in $\tilde{\mathcal{R}}$.
Therefore, any barrier in $\tilde{\mathcal{R}}$ is 
maximal.
\hfill $\Box$

\medskip

\begin{figure}[htbp]
\centering
\includegraphics[scale=1.0]{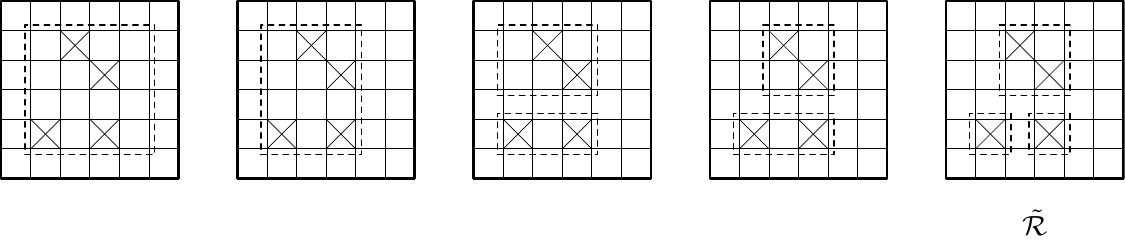}
\caption{An example of application of the algorithm 
to obtain all maximal barriers.}
\label{figure:fig051}
\end{figure}

Fig. \ref{figure:fig051} shows an example of 
application of the above algorithm.
We show the change of $\mathcal{R}$ from left to 
right.  
We obtain the final $\tilde{\mathcal{R}}$ 
in $4$ steps.

The usefulness of maximal barriers in the analysis 
of $H_{k, w}$ comes from the following theorem.

\begin{thm}
\label{theorem:thm004}
Let $v$ be one of the four corners 
$(0, 0)$, $(0, w)$, $(w, 0)$, $(w, w)$ of $C$  
and $v'$ be any node of $C$ that is not in maximal barriers of $C$.
Then $\abb{d}_{C}(v, v') = \abb{d}_\abb{MH}(v, v')$, 
and hence there is a path in $C$ from $v$ to $v'$ of the MH distance length.
\end{thm}

\noindent
{\it Proof}. 
We prove this only for the case $v = (0, 0)$.
Proofs for other cases are similar.

Suppose that a node $(x, y)$ is not in maximal barriers and 
$0 < x$, $0 < y$.
We assume that both of the two positions 
$(x - 1, y)$, $(x, y - 1)$ are 
in maximal barriers and derive a contradiction.
Let $R$ and $R'$ respectively be the maximal barriers 
that contain $(x - 1, y)$ and $(x, y - 1)$ respectively.

It is not possible that $R = R'$ because 
if two positions $(x - 1, y)$, $(x, y - 1)$ 
are in one maximal barrier then the position $(x, y)$ 
is also in the maximal barrier, contradicting our assumption 
that $(x, y)$ is not in maximal barriers.
Hence $R$, $R'$ are different maximal barriers.

Both of $R$ and $R'$ do not contain $(x, y)$ and hence 
$(x - 1, y)$ is in the east boundary of $R$ and 
$(x, y - 1)$ is in the north boundary of $R'$.
Therefore, one of the three statements 
mentioned in Theorem \ref{theorem:thm009} (1) 
must be true and we have a contradiction 
by Theorem \ref{theorem:thm009} (2).

We have proved that if a node $(x, y)$ is not 
in maximal barriers and $x > 0$, $y > 0$ then 
at least one of the positions 
$(x - 1, y)$, $(x, y - 1)$ is 
not in maximal barriers (and hence is a node).
This, together with our assumption that 
$v'$ is not in maximal barriers,  
means that there is a path $P$ in $C$ from $v'$ to a node $v''$ 
that is either in the west boundary or in the south boundary of $C$ 
satisfying the condition: the path proceeds only to the west or 
to the south.
It is obvious that there is a similar path $P'$ in $C$ from $v''$ to $v$.
Therefore, there is a similar path $P + P'$ in $C$ from $v'$ to $v$.
Hence we have $\abb{d}_{C}(v, v') = \abb{d}_\abb{MH}(v, v')$.
\hfill
$\Box$

\medskip

\subsection{Maximal barriers and $H_{k,w}$}
\label{subsection:maximal_barriers_and_h}

The following lemma is used repeatedly.

\begin{lem}
\label{lemma:lem000}
Let $C$ be a configuration of size $w$ of 
$\abb{SH}[k]$.  
Then
\[
\max \{ \min \{ \abb{d}_{\abb{MH}}(v_\abb{gen}, v; (0, w)), 
                \abb{d}_{\abb{MH}}(v_\abb{gen}, v; (w, 0)) \} ~|~ 
      v \in C \} = 2w.
\]
\end{lem}

\noindent
{\it Proof}. 
Let $v$ be $(x, y)$ and let $\delta$ denote $x - y$.
Then $-w \leq \delta \leq w$ and $\delta = 0$ 
for at least two nodes $v = (0, 0), (w, w)$ in $C$.
Therefore, 
\begin{align*}
\lefteqn{ 
  \max \{ \min \{ \abb{d}_\abb{MH}(v_\abb{gen}, v; (0, w)), 
                \abb{d}_\abb{MH}(v_\abb{gen}, v; (w, 0)) \} 
        ~|~ v \in C \} } \\
& =  
  w + \max \{ \min \{ x + (w - y), (w - x) + y \} 
        ~|~ v \in C \} \\
& = w + \max \{ \min \{ w + \delta, w - \delta \} ~|~ v \in C \} \\
& = 2w.
\end{align*}
\hfill $\Box$

\medskip

Suppose that a node $v$ in $C \in \mathcal{C}_{k, w}$ 
is given and is fixed.
We derive a formula for the value $T(v, C)$.
If $v$ is not in maximal barriers of $C$, 
we have 
\begin{align}
T(v, C) & = \min \{ \abb{d}_{C}(v_\abb{gen}, v; (0, w)), 
                      \abb{d}_{C}(v_\abb{gen}, v; (w, 0)) \} \nonumber \\
        & = \min \{ \abb{d}_\abb{MH}(v_\abb{gen}, v; (0, w)), 
                      \abb{d}_\abb{MH}(v_\abb{gen}, v; (w, 0)) \} 
              \nonumber \\
        & \leq 2w.
\label{equation:eq016}
\end{align}
by \eqref{equation:eq003}, 
Theorem \ref{theorem:thm004} and Lemma \ref{lemma:lem000}.
From now on, we consider the case when $v$ is in 
a maximal barrier $R$ of $C$.
We will show that $T(v, C)$ is expressed as $T(v, C) = 2w + E(S, p, \delta)$ 
(see \eqref{equation:eq014}).
This equation implies that $T(v, C)$ is 
determined by the three factors: 
(1) the form of $R$ (represented by $S$), 
(2) the position of $v$ in $R$ (represented by $p$), 
(3) the position of $R$ in $C$ relative to the main diagonal (represented by $\delta$).

When $R$ is a rectangle (a maximal barrier, for example) 
$\{ (x, y) \mid x_{0} \leq x \leq x_{1}, 
y_{0} \leq y \leq y_{1} \}$ and 
$1 \leq x_{0}$, $x_{1} \leq w - 1$, $1 \leq y_{0}$, 
$y_{1} \leq w - 1$, 
by the \textit{enlarged rectangle of} $R$ we mean the 
rectangle $X = \{ (x, y) \mid x_{0} - 1 \leq x \leq x_{1} + 1, 
y_{0} - 1 \leq y \leq y_{1} + 1 \}$.

\begin{figure}[htbp]
\centering
\includegraphics[scale=1.0]{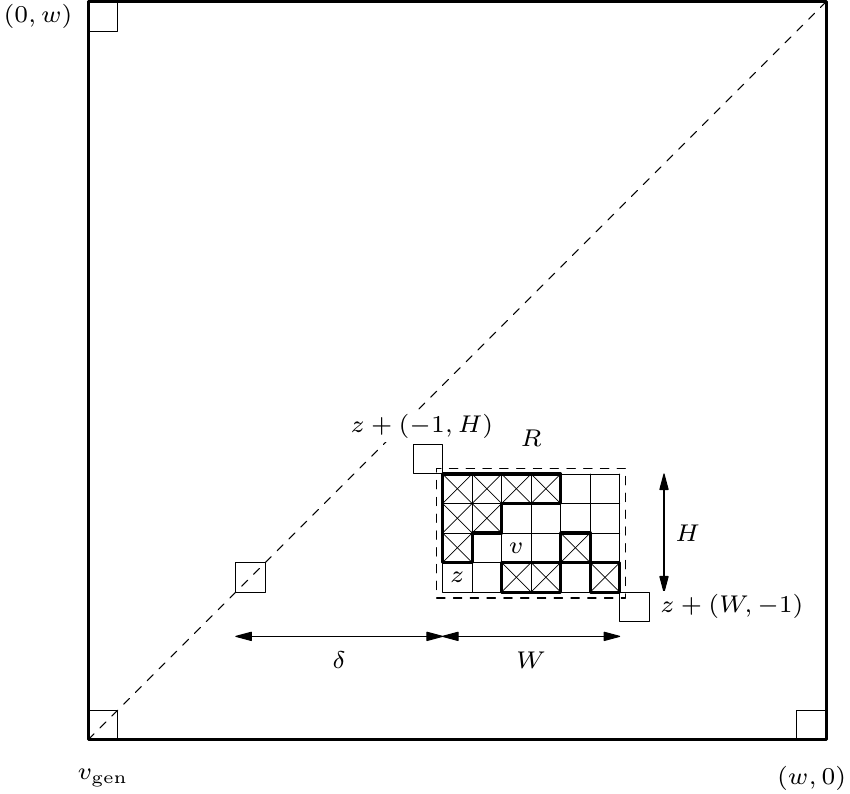}
\caption{A configuration $C$ in $\mathcal{C}_{k,w}$ 
and node $v$ that is in a maximal barrier $R$ of $C$.}
\label{figure:fig007}  
\end{figure}

\begin{thm}
\label{theorem:thm014}
Suppose that a node $v$ of $C \in \mathcal{C}_{k, w}$ is 
in a maximal barrier $R$ of $C$.
Let $W$, $H$, $z$, $\delta$, $d_{0}$, $d_{1}$ be defined 
as follows {\rm (}see Fig. $\ref{figure:fig007}$\;{\rm )}{\rm :}
\begin{enumerate}
\item[$\bullet$] $W$ is the number of columns of $R$.
\item[$\bullet$] $H$ is the number of rows of $R$.
\item[$\bullet$] $z = (z_{x}, z_{y})$ is the position of the southwest 
corner of $R$.
\item[$\bullet$] $\delta = z_{x} - z_{y}$.
\item[$\bullet$] $d_{0} = \abb{d}_{C}(z + (-1, H), v)$, 
$d_{1} = \abb{d}_{C}(z + (W, - 1), v)$.
\end{enumerate}
Then 
\begin{equation}
T(v, C) = 2w + \min \{ \delta - H - 1 + d_{0}, 
                       - \delta - W - 1 + d_{1} \}.
\label{equation:eq001}
\end{equation}
\end{thm}

\begin{figure}[htbp]
\centering
\includegraphics[scale=1.0]{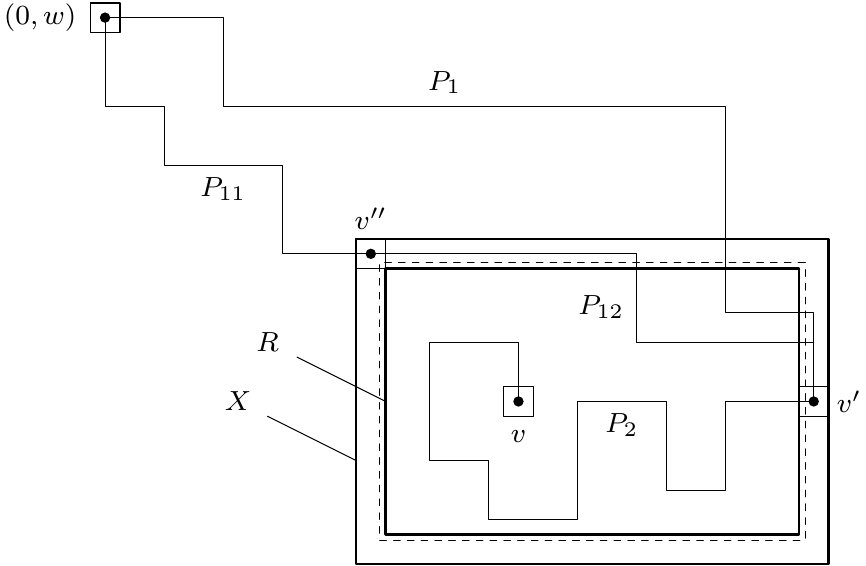}
\caption{Paths and nodes used in the estimation 
of $\abb{d}_{C}(v_\abb{gen}, v; (0, w))$.}
\label{figure:fig052}
\end{figure}

\noindent
{\it Proof}. 
First we represent the value $\abb{d}_{C}(v_\abb{gen}, v; (0, w))$ 
with $W$, $H$, $z = (z_{x}, z_{y})$, $\delta$, $d_{0}$, $d_{1}$.
Let $X$ be the enlarged rectangle of $R$, 
\[
X = \{ (x, y) \mid z_{x} - 1 \leq x \leq z_{x} + W, 
                     z_{y} - 1 \leq y \leq z_{y} + H \}
\]
(see Fig. \ref{figure:fig052}).

$R$ is a maximal barrier.
Hence by Theorem \ref{theorem:thm009} (5), $X - R$ 
has no holes in it and any node in $X - R$ is not 
in maximal barriers.

Let $P$ be a shortest path in $C$ from $(0, w)$ to $v$, 
$v'$ be the last node in $P$ that is not in $R$, and 
$P_{1}$ and $P_{2}$ be the parts of $P$ 
from $(0, w)$ to $v'$ and from $v'$ to $v$ respectively.
Let $v''$ be the node $z + (- 1, H)$ and 
let $P_{11}$ and $P_{12}$ be shortest paths from 
$(0, w)$ to $v''$ and from $v''$ to $v'$ respectively.

Then we have 
$|P_{1}| = \abb{d}_\abb{MH}((0, w), v')$, 
$|P_{11}| = \abb{d}_\abb{MH}((0, w), v'')$ and 
$|P_{12}| = \abb{d}_\abb{MH}(v'', v)$.
The first and the second equalities are true 
because $v', v''$ are in $X - R$ and hence are 
not in maximal barriers (Theorem \ref{theorem:thm004}).
The third equality is true because both of $v'$, $v''$ 
are in $X - R$, $X - R$ has no holes, and $v''$ is 
the northwest corner of $X - R$.
From this we have $|P_{1}| = |P_{11}| + |P_{12}|$.
This implies that the path $P_{11} + P_{12} + P_{2}$ 
is a shortest path in $C$ from $(0, w)$ to $v$.
Therefore, $P_{12} + P_{2}$ is a shortest path 
from $v''$ to $v$ in $C$ and hence 
$|P_{12} + P_{2}| = d_{0}$ by the definition of $d_{0}$.

Hence we have 
\begin{align*}
\abb{d}_{C}(v_\abb{gen}, v; (0, w)) & = w + |P| \\
  & = w + |P_{11}| + |P_{12} + P_{2}| \\
  & = w + \abb{d}_\abb{MH}((0, w), z + (- 1, H)) + d_{0} \\
  & = w + (z_{x} - 1) + (w - (z_{y} + H)) + d_{0} \\
  & = 2w + \delta - H - 1 + d_{0}.
\end{align*}
Similarly we have $\abb{d}_{C}(v_\abb{gen}, v; (w, 0)) 
= 2w - \delta - W - 1 + d_{1}$.
Hence we have 
\[
T(v, C) = 2w + \min \{\delta - H - 1 + d_{0}, 
                      - \delta - W - 1 + d_{1} \}.
\]
\hfill $\Box$

\medskip

In the above proof we proved that $P_{12} + P_{2}$ is a 
shortest path in $C$ from $v''$ to $v$.
Moreover, this path is in $X$.
Therefore, $d_{0}$ is the length of a shortest of all paths 
from $v''$ to $v$ that are in $X$ and 
$d_{0}$ is completely determined by the distribution of holes of $C$ 
in $X$ irrespective of the distribution out of $X$.
The same is also true for $d_{1}$.

The value $\delta$ represents the relative position of $R$ 
with respect to the main diagonal of $C$.
If $\delta \geq 0$, the southwest corner of $R$ is 
to the east of the main diagonal by $\delta$ positions.
If $\delta < 0$, the position is to the west 
by $- \delta$ positions. 

Using the equation \eqref{equation:eq001} we can 
determine the value $H_{k, w}$ 
for all $k$ and for all sufficiently large $w$.
Let $E$ denote the value 
$E = \min \{ \delta - H - 1 + d_{0}, 
- \delta - W - 1 + d_{1}\}$. 
Then $T(v, C) = 2w + E$.
This value $E$ is a function of $C$ and $v$.
However, $E$ can be determined by the following 
three factors:
\begin{enumerate}
\item[$\bullet$] The shape $S$ of the maximal 
barrier $R$.  
More precisely, it is the triple:
\begin{enumerate}
\item[$\diamond$] $W$, 
\item[$\diamond$] $H$, 
\item[$\diamond$] whether the position $z + (i, j)$ 
in $C$ is a hole or not for each 
$(i, j)$ such that $0 \leq i \leq W - 1$, 
$0 \leq j \leq H - 1$.
\end{enumerate}
We regard $S$ as a barrier.
\item[$\bullet$] $p = v - z$.
\item[$\bullet$] $\delta = z_{x} - z_{y}$.
\end{enumerate}
Although the two values $d_{0}$, $d_{1}$ are not 
included, we can determine them from these three 
factors.
For example, as we mentioned above, $d_{0}$ is 
the length of a shortest of all paths in $C$ 
from the northwest corner of the 
enlarged rectangle $X$ of $R$ to $v$ that are 
in $X$.
This can be determined from $S$ and $p$.

Therefore, we will denote $E$ as a function 
$E(S, p, \delta)$ of these three factors 
$S$, $p$, $\delta$.
Then the equation \eqref{equation:eq001} is 
written as
\begin{equation}
T(v, C) = 2w + E(S, p, \delta).
\label{equation:eq014}
\end{equation}
We continue to call $S$ a ``barrier'' of 
$\abb{SH}[k]$ and write ``$p \in S$'' 
to mean that a position $p$ in $S$ is a node.

Let $\mathcal{S}_{k}$ denote the set of 
all barriers having at most $k$ holes.
Then, for any maximal barrier $R$ 
in any configuration $C \in \mathcal{C}_{k, w}$ 
the corresponding barrier $S$ is in $\mathcal{S}_{k}$.
Conversely, for any barrier $S$ in $\mathcal{S}_{k}$ 
and any sufficiently large $w$, 
there are $C \in \mathcal{C}_{k, w}$ and 
a maximal barrier $R$ in $C$ such that $S$ corresponds to $R$.
When a maximal barrier $R$ in $C \in \mathcal{C}_{k, w}$ and 
a barrier $S \in \mathcal{S}_{k}$ correspond, 
there is also a one-to-one correspondence between 
nodes $v$ in $R$ and nodes $p$ in $S$.

\begin{figure}[htbp]
\centering
\includegraphics[scale=1.0]{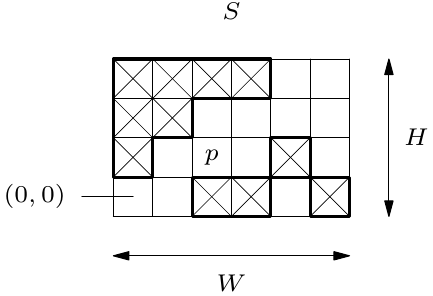}
\caption{The barrier $S$ and the node $p \in S$ corresponding 
to the maximal barrier $R$ and the node $v \in R$ 
in Fig. \ref{figure:fig007}.}
\label{figure:fig008}
\end{figure}

In Fig. \ref{figure:fig008} we show the barrier $S$ and 
the node $p$ in $S$ that correspond to the maximal barrier $R$ 
and the node $v$ in $R$ in Fig. \ref{figure:fig007}.

We define several values.  
We defined $E(S, p, \delta)$ 
by 
\begin{equation}
E(S, p, \delta) = 
\min \{ \delta - H - 1 + d_{0}, - \delta - W - 1 + d_{1} \}
\label{equation:eq008}
\end{equation}
for $S \in \mathcal{S}_{k}$, $p \in S$, 
$\delta \in \mathbb{Z}$.
We define $E_\abb{max}(S, p)$ by 
\begin{equation}
E_\abb{max}(S, p) = 
\max_{\delta \in \mathbb{Z}} E(S, p, \delta)
\label{equation:eq009}
\end{equation}
for $S \in \mathcal{S}_{k}$, $p \in S$.
Finally, we define $c_{k}$ by 
\begin{equation}
c_{k} = \max_{S \in \mathcal{S}_{k}, p \in S} 
E_\abb{max}(S, p)
\label{equation:eq002} 
\end{equation}
for $k \geq 2$.
By the assumption $k \geq 2$ 
there is at least one pair of $S$ and $p$ 
such that $p \in S$ 
(see Fig. \ref{figure:fig010}).

Let $X = \{ (i, j) \mid -1 \leq i \leq W, 
-1 \leq j \leq H \}$ be the enlarged rectangle 
of $S$ and let $v_{0} = (-1, H)$, $v_{1} = (W, -1)$ 
be its northwest and southeast corners respectively.
Then we have 
$W + H + 2  = \abb{d}_{\abb{HM}}(v_{0}, v_{1})$, 
$d_{0} + d_{1} = \abb{d}_{X}(v_{0}, p) + \abb{d}_{X}(p, v_{1})$.
Hence both of $W + H + 2$ and $d_{0} + d_{1}$ are lengths of paths 
in $X$ from $v_{0}$ to $v_{1}$.
This means that $W + H + 2 \equiv d_{0} + d_{1} \pmod{2}$.
(Regard $X$ as a checkerboard.)

Using this we obtain the following 
simplified expression for $E_{\abb{max}}(S, p)$.
\begin{align}
E_\abb{max}(S, p) & = \max \{ \min \{ \delta - H - 1 + d_{0}, 
                              -\delta - W - 1 + d_{1} \} ~|~ 
    \delta \in \mathbb{Z} \} \nonumber \\
        & = (- W - H - 2 + d_{0} + d_{1} ) \; / \; 2.
\label{equation:eq000} 
\end{align}
The value $\delta_\abb{opt}(S, p)$ such that 
$E_\abb{max}(S, p) = E(S, p, \delta_\abb{opt}(S, p))$ is given 
by 
\begin{equation}
\delta_\abb{opt}(S, p) = (- W + H - d_{0} + d_{1}) / 2.
\label{equation:eq011}
\end{equation}
We can also show $E_\abb{max}(S, p) \geq 0$ as follows.
\begin{align}
E_\abb{max}(S, p) & = (d_{0} + d_{1} - (W + H + 2)) / 2 
\nonumber \\
& = (\abb{d}_{X}(v_{0}, p) + \abb{d}_{X}(p, v_{1}) 
       - \abb{d}_\abb{MH}(v_{0}, v_{1})) / 2 \nonumber \\
& \geq (\abb{d}_{X}(v_{0}, v_{1}) - \abb{d}_\abb{MH}(v_{0}, v_{1})) / 2 
\nonumber \\
& \geq 0.
\label{equation:eq015}
\end{align}
This implies $c_{k} \geq 0$.

\begin{figure}[htbp]
\centering
\includegraphics[scale=1.0]{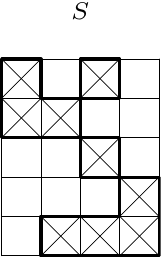}
\caption{A barrier $S$ having $9$ holes.}
\label{figure:fig056}
\end{figure}

Fig. \ref{figure:fig056} shows a barrier $S$ that has 
$9$ holes.
It is in $\mathcal{S}_{k}$ for $9 \leq k$.
For this $S$, we have $W = 4$, $H = 5$, 
$E(S, p, \delta) = \min \{ \delta - 6 + d_{0}, 
- \delta - 5 + d_{1} \}$, 
$E_\abb{max}(S, p) = (-11 + d_{0} + d_{1}) / 2$,  
and $\delta_\abb{opt}(S, p) = (1 - d_{0} + d_{1}) / 2$.
In Table \ref{table:tab000} we show values 
$d_{0}$, $d_{1}$, $E(S, p, \delta)$, $E_\abb{max}(S, p)$, 
$\delta_\abb{opt}(S, p)$ for each node $p$ of the $11$ nodes in $S$.
\begin{table}[htb]
\centering
\begin{tabular}{|c|c|c|c|c|c|}
\hline
$p$ & $d_{0}$ & $d_{1}$ & $E(S, p, \delta)$ & 
$E_\abb{max}(S, p)$ & $\delta_\abb{opt}(S, p)$ \\ \hline
$(0, 0)$ & $6$ & $5$  & $\min\{ \delta,     -\delta    \}$ & $0$ & $0$  \\
$(0, 1)$ & $5$ & $6$  & $\min\{ \delta - 1, -\delta + 1\}$ & $0$ & $1$  \\
$(0, 2)$ & $4$ & $7$  & $\min\{ \delta - 2, -\delta + 2\}$ & $0$ & $2$  \\
$(1, 1)$ & $6$ & $7$  & $\min\{ \delta,     -\delta + 2\}$ & $1$ & $1$  \\
$(1, 2)$ & $5$ & $8$  & $\min\{ \delta - 1, -\delta + 3\}$ & $1$ & $2$  \\
$(1, 4)$ & $3$ & $10$ & $\min\{ \delta - 3, -\delta + 5\}$ & $1$ & $4$  \\
$(2, 1)$ & $7$ & $8$  & $\min\{ \delta + 1, -\delta + 3\}$ & $2$ & $1$  \\
$(2, 3)$ & $7$ & $6$  & $\min\{ \delta + 1, -\delta + 1\}$ & $1$ & $0$  \\
$(3, 2)$ & $7$ & $4$  & $\min\{ \delta + 1, -\delta - 1\}$ & $0$ & $-1$ \\
$(3, 3)$ & $6$ & $5$  & $\min\{ \delta,     -\delta    \}$ & $0$ & $0$  \\
$(3, 4)$ & $5$ & $6$  & $\min\{ \delta - 1, -\delta + 1\}$ & $0$ & $1$  \\ \hline
\end{tabular}
\caption{Values of $d_{0}$, $d_{1}$ and so on 
for the barrier $S$ in Fig. \ref{figure:fig056} 
and $p \in S$.}
\label{table:tab000}
\end{table}

\begin{figure}[htbp]
\centering
\includegraphics[scale=1.0]{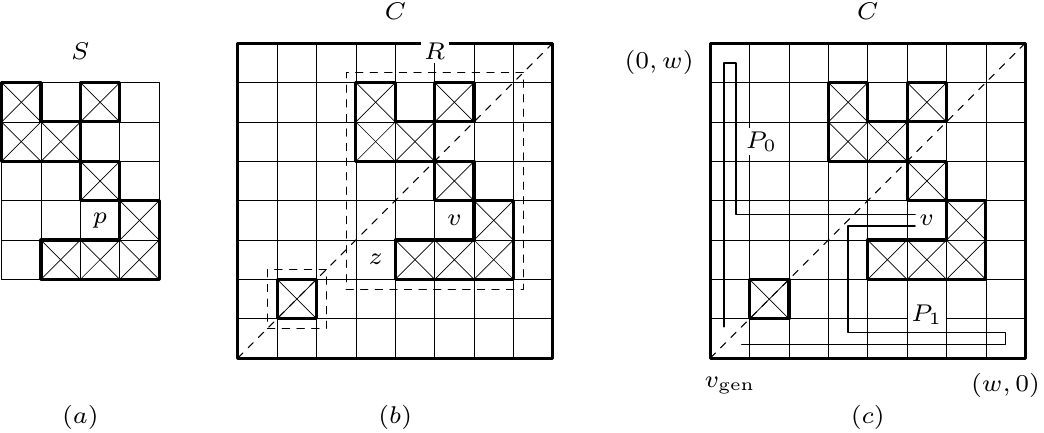}
\caption{(a) The position of $p = (2, 1)$ in $S$.
(b) A configuration $C$ in which $S$ appears as 
a maximal barrier $R$ so that 
$E(S, p, \delta) = E_\abb{max}(S, p) = 2$ 
with $p = (2, 1)$, $\delta = \delta_\abb{opt}(S, p) = 1$.
(C) Two paths $P_{0}$, $P_{1}$ from $v_\abb{gen}$ to $v$ via 
$(0, w)$ and via $(w, 0)$.
}
\label{figure:fig057}
\end{figure}

This table shows that the maximum value of 
$E_\abb{max}(S, p)$ for $p \in S$ is $2$ 
and this value $2$ is realized only by $p = (2, 1)$, $\delta = 1$.
In Fig. \ref{figure:fig057} (a) we show the position of 
$p = (2, 1)$ in $S$.
In Fig. \ref{figure:fig057} (b) we show a configuration 
$C$ of size $w = 7$ of $\abb{SH}[10]$ 
in which $S$ appears as a maximal barrier $R$ 
at a position $z = (z_{x}, z_{y}) = (3, 2) $ 
such that $\delta = z_{x} - z_{y} = 3 - 2 = 1$.
The node $v$ in $C$ that corresponds to $p = (2, 1)$ 
in $S$ is $v = z + p = (5, 3)$.  
This pair $C$, $v \in C$ realizes 
$T(v, C) = 2w + 2 = 16$.
In Fig. \ref{figure:fig057} (c) we show a shortest path $P_{0}$ from $v_{\abb{gen}}$ to $v$ 
via $(0, w)$ ($= (0, 9)$) and 
a shortest path $P_{1}$ from $v_{\abb{gen}}$ to $v$ 
via $(w, 0)$ ($= (9, 0)$).
We have $T(v, C) = \min \{ |P_{0}|, |P_{1}| \} = \min \{16, 16\} = 16$ 
confirming the above equation.
For this $C$ we have 
$\abb{mft}_{\abb{SH}[10]}(C) \geq 
\max_{v \in C} T(v, C) \geq 16$.

\begin{thm}
\label{theorem:thm006}

\noindent
\begin{enumerate}
\item[{\rm (1)}] For any $k \geq 2$ and any $w$, 
$H_{k,w} \leq 2w + c_{k}$.
\item[{\rm (2)}] For any $k \geq 2$ and any $w$ such that 
$(k^{2} + 7k + 5) / 2 \leq w$, $H_{k,w} = 2w + c_{k}$.
\end{enumerate}
\end{thm}

\noindent
{\it Proof}. 
(1) We assume that $C \in \mathcal{C}_{k,w}$, $v \in C$ 
and prove $T(v, C) \leq 2w + c_{k}$.  
This implies 
$H_{k, w} = \max_{C \in \mathcal{C}_{k, w}, v \in C} T(v, C)
\leq 2w + c_{k}$.

If $v$ is not in maximal barriers of $C$ then 
$T(v, C) \leq 2w \leq 2w + c_{k}$ by 
\eqref{equation:eq016}.
(Note that $c_{k} \geq 0$ by \eqref{equation:eq015}.)
Suppose that $v$ is in a maximal barrier $R$ of $C$.
Let $S \in \mathcal{S}_{k}$, $p \in S$, $\delta$ be the barrier and so on 
that are determined from $C$, $R$, $v$ by the correspondence explained 
in the proof of Theorem \ref{theorem:thm014}.
Then by \eqref{equation:eq014} we have 
$T(v, C) = 2w + E(S, p, \delta) 
\leq 2w + E_\abb{max}(S, p) \leq 2w + c_{k}$.

\medskip

\noindent
(2) We assume that $S \in \mathcal{S}_{k}$, 
$p \in S$, $(k^{2} + 7k + 5) / 2 \leq w$ and prove 
$2w + E_\abb{max}(S, p) \leq H_{k,w}$.
This implies that if $(k^{2} + 7k + 5) / 2 \leq w$ then 
$2w + c_{k} = 2w + 
\max_{S \in \mathcal{S}_{k}, v \in S} E_\abb{max}(S, p) 
\leq H_{k,w}$ 
and hence $H_{k,w} = 2w + c_{k}$.

\begin{figure}
\centering
\includegraphics[scale=1.0]{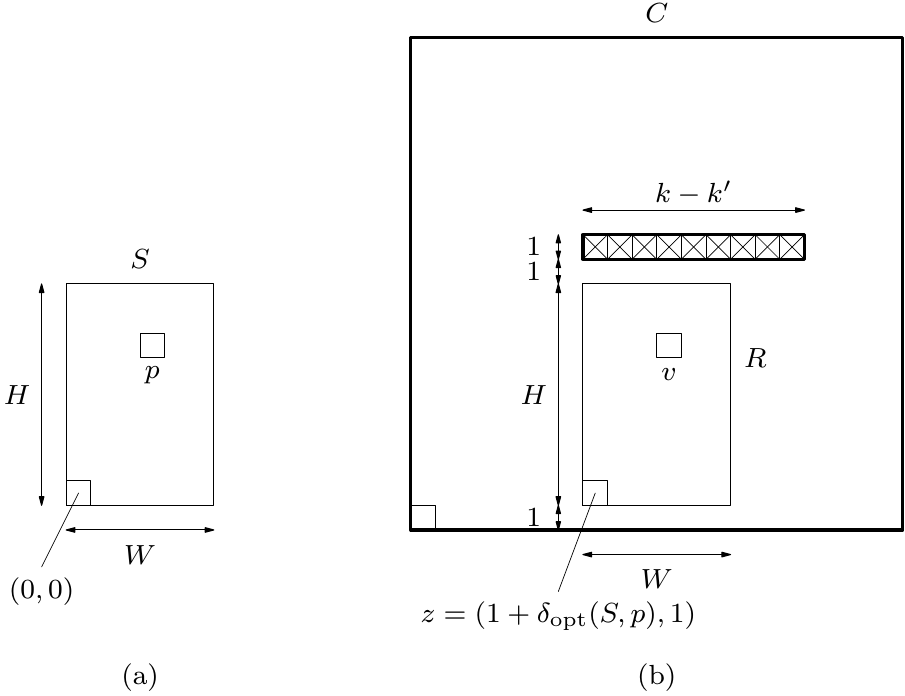}
\caption{(a) represents a barrier $S$ with $k'$ holes 
and a node $p$ in it. 
(b) represents a configuration $C$ in which $S$ 
appears as a maximal barrier $R$ at a position such that 
$\delta = \delta_\abb{opt}(S, p)$. 
Additional $k - k'$ holes are also included in $C$ 
so that $C$ is a configuration of $\abb{SH}[k]$.}
\label{figure:fig012}  
\end{figure}

We assume $\delta_\abb{opt}(S, p) \geq 0$.  
The proof for the other case is similar.
Let $S$ has $k'$ holes and let $W$ and $H$ 
be the number of columns and that of rows of $S$ respectively 
(see Fig. \ref{figure:fig012} (a)).
We have $W \leq k$, $H \leq k$ by the definition of 
barriers.

It is obvious that 
$d_{1} \leq (W + 2)(H + 2) - 1 \leq 
k^{2} + 4k + 3$ because the enlarged rectangle $X$ of $S$ 
has $(W + 2)(H + 2)$ positions and $d_{1}$ is the length 
of a shortest path from $(W, -1)$ to $p$ in $X$.
Using this we have an upper bound of  
$\delta_\abb{opt}(S, p)$:
\begin{align*}
\delta_\abb{opt}(S, p) & = (- W + H - d_{0} + d_{1}) / 2 \\
                 & \leq (H + d_{1}) / 2 \\
                 & \leq (k^{2} + 5k + 3) / 2.
\end{align*}

We define a configuration $C \in \mathcal{C}_{k,w}$ as 
shown in Fig. \ref{figure:fig012} (b).
In $C$, the barrier $S$ is placed at the position 
$z = (z_{x}, z_{y}) = 
(\delta_\abb{opt}(S, p) + 1, 1)$ 
as a barrier $R$ in $C$.
Moreover, $k - k'$ holes are placed in $C$ 
at positions 
$(1 + \delta_\abb{opt}(S, p), H + 2)$, 
\ldots, 
$(\delta_\abb{opt}(S, p) + k - k', H + 2)$ 
so that they constitute another barrier 
(if $k' < k$).
We can show that these two barriers are 
two different maximal barriers in $C$ 
using our assumption $(k^{2} + 7k + 5)/2 \leq w$, $k \geq 2$ 
and the above mentioned upper bound of $\delta_\abb{opt}(S, p)$.

For example, we can prove that 
the $x$-coordinate $(\delta_\abb{opt}(S, p) 
+ k - k')$ of the easternmost hole of the 
$k - k'$ holes (if $k' < k$) is at most $w - 1$ as follows:
\begin{align*}
\delta_\abb{opt}(S, p) + k - k' 
& \leq (k^{2} + 5k + 3) / 2 + k \\
& = (k^{2} + 7k + 3) / 2 \\
& \leq w - 1.
\end{align*}

Let $v = z + p$ be the node in $R$ 
that corresponds to $p$ in $S$ 
and $\delta$ be the value $\delta = z_{x} - z_{y} = \delta_\abb{opt}(S, p)$.
Then 
\begin{align*}
2w + E_\abb{max}(S, p) & = 2w + E(S, p, \delta_\abb{opt}(S, p)) \\
       & = 2w + E(S, p, \delta) \\
       & = T(v, C) \quad \text{(by \eqref{equation:eq014})} \\
       & \leq H_{k,w}.
\end{align*}
\hfill $\Box$

\medskip

\begin{cor}
\label{corollary:cor001}

\noindent
\begin{enumerate}
\item[{\rm (1)}] For any $k \geq 2$, any $w$ and any 
$C \in \mathcal{C}_{k,w}$, 
$\abb{mft}_{\abb{SH}[k]}(C) \leq 2w + c_{k}$.
\item[{\rm (2)}] For any $k \geq 2$ and any $w$ such that 
$(k^{2} + 7k + 5) / 2 \leq w$, there exists 
$C \in \mathcal{C}_{k,w}$ such that 
$\abb{mft}_{\abb{SH}[k]}(C) = 2w + c_{k}$.
\end{enumerate}
\end{cor}

\noindent
{\it Proof}. 
This corollary follows from Theorems \ref{theorem:thm003}, 
\ref{theorem:thm006}.
\hfill
$\Box$

\medskip

Below we show a lower bound and an upper bound of 
$c_{k}$.
As we explain later the lower bound $k - 2$ is 
the correct value of $c_{k}$ for $3 \leq k \leq 9$.
The upper bound $k^{2} + 4k$ was obtained by 
a very simple estimation and we expect 
to be able to improve it considerably.

\begin{thm}
\label{theorem:thm005}
For any $k \geq 3$, 
\[
k - 2 \leq c_{k} \leq k^{2} + 4k.
\]
\end{thm}

\noindent
{\it Proof}. 
Upper bound: In the proof of Theorem \ref{theorem:thm006} 
we showed $d_{1} \leq k^{2} + 4k + 3$ and 
we have a similar result also for $d_{0}$.
Hence, using $2 \leq W$, $2 \leq H$ 
(a barrier $S$ with either $W = 1$ or $H = 1$ has no nodes $p$ in it), 
we have 
\begin{align*}
E_\abb{max}(S, p) & = (- W - H - 2 + d_{0} + d_{1}) / 2 \\
& \leq (- 6 + 2(k^{2} + 4k + 3)) / 2 \\
& = k^{2} + 4k.
\end{align*}
Therefore, $c_{k} = 
\max_{S \in \mathcal{S}_{k}, p \in S} E_\abb{max}(S, p) 
\leq k^{2} + 4k$.

\medskip

\begin{figure}
\centering
\includegraphics[scale=1.0]{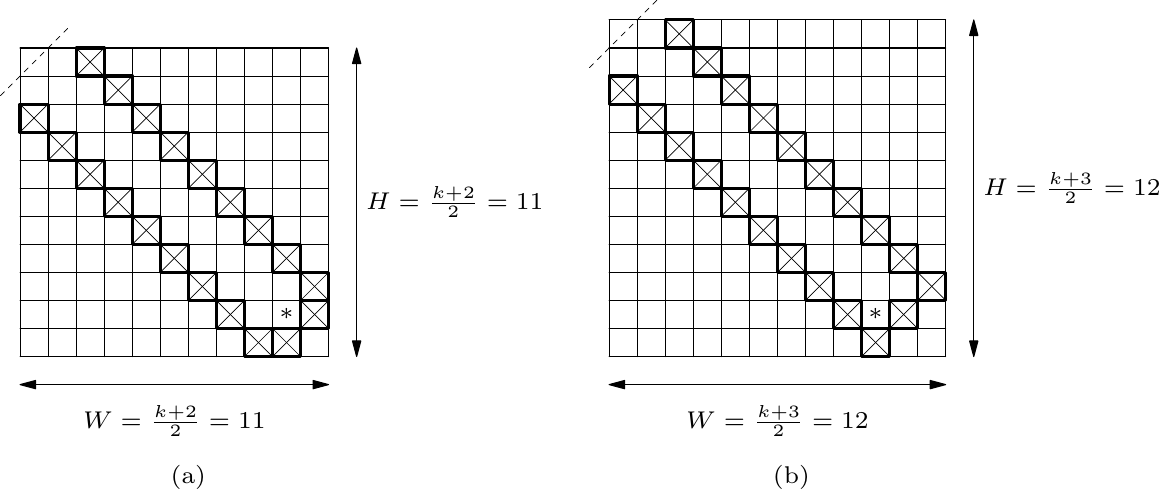}
\caption{Barriers and nodes that realize the lower 
bound $k - 2$ of $c_{k}$ for $k \geq 3$.}
\label{figure:fig011}  
\end{figure}

\noindent
Lower bound: In Fig. \ref{figure:fig011} we show 
barriers $S \in \mathcal{S}_{k}$ and $p \in S$ 
such that $E_\abb{max}(S, p) = k - 2$ 
implying $c_{k} \geq k - 2$.
The figure (a) is for even $k$ ($\geq 4$) and 
the figure (b) is for odd $k$ ($\geq 3$).  
The marks ``*'' denote the nodes $p$.
The dotted lines are main diagonals.

For (a), 
we have 
$k = 20$, $W = H = (k+2)/2 = 11$, 
$d_{0} = k = 20$, $d_{1} = 2k = 40$, 
$E_\abb{max}(S, p) = (- (k+2)/2 - (k+2)/2 + k + 2k - 2) / 2 
= k - 2 = 18$, 
$\delta_\abb{opt} = (- (k+2)/2 + (k+2)/2 - k + 2k) / 2 
= k/2 = 10$.

For (b), 
we have 
$k = 21$, $W = H = (k+3)/2 = 12$, 
$d_{0} = k = 21$, $d_{1} = 2k + 1 = 43$, 
$E_\abb{max}(S, p) = (- (k+3)/2 - (k+3)/2 + k + (2k + 1) - 2) / 2 
= k - 2 = 19$, 
$\delta_\abb{opt} = (- (k+3)/2 + (k+3)/2 - k + (2k + 1)) / 2 
= (k+1)/2 = 11$.
\hfill $\Box$

\medskip

\subsection{Determination of values $c_{k}$ by an algorithm}
\label{subsection:algorithm}

Definition \eqref{equation:eq002} of $c_{k}$ 
itself gives an algorithm for computing $c_{k}$.
We enumerate all $S$ in $\mathcal{S}_{k}$.
For each pair $(S, p)$ of $S \in \mathcal{S}_{k}$ and 
$p \in S$ we compute the value $E_\abb{max}(S, p)$.
Then the maximum value of this value $E_\abb{max}(S, p)$ 
over all pairs $(S, p)$ is the desired $c_{k}$.

\begin{figure}[htbp]
\centering
\includegraphics[scale=1.0]{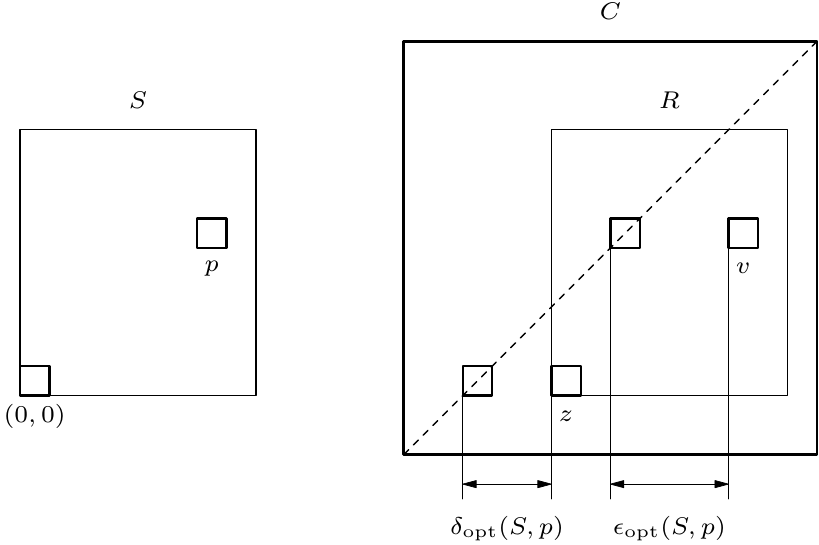}
\caption{An intuitive meaning of $\delta_{\abb{opt}}(S, p)$ and $\epsilon_{\abb{opt}}(S, p)$.}
\label{figure:fig083}
\end{figure}

Before carrying out this computation we define one value which 
we will denote by $\epsilon_{\abb{opt}}(S, p)$.
Let a barrier $S$ and a node $p = (p_{x}, p_{y})$ in $S$ 
correspond to a maximal barrier $R$ in a configuration $C$ 
and a node $v = (v_{x}, v_{y})$ in $R$ respectively.
Let $z = (z_{x}, z_{y})$ be the southwest corner of $R$ and 
let $\delta$ denote $z_{x} - z_{y}$.
Let $\epsilon$ denote $v_{x} - v_{y}$.
Then we have the relation 
$\epsilon = v_{x} - v_{y} = (z_{x} + p_{x}) - (z_{y} + p_{y}) 
= \delta + p_{x} - p_{y}$ between $\epsilon$ and $\delta$.
We defined the value $\delta_{\abb{opt}}(S, p)$ to be the value 
of $\delta$ such that $E(S, p, \delta)$ is maximum.
Let $\epsilon_{\abb{opt}}(S, p)$ be defined by 
$\epsilon_{\abb{opt}}(S, p) = \delta_{\abb{opt}}(S, p) 
+ p_{x} - p_{y}$.
Then $\epsilon_{\abb{opt}}(S, p)$ is the value of $\epsilon$ 
such that the value $E(S, p, \delta)$ is maximum as a function 
of $\epsilon$.
Intuitively, $\delta_{\abb{opt}}(S, p)$ and 
$\epsilon_{\abb{opt}}(S, p)$ represent the positions of $z$ and $v$ 
respectively relative to the main diagonal 
when $S$ is placed in $C$ so that the value $E(S, p, \delta)$ 
is maximum (Fig. \ref{figure:fig083}).

\begin{figure}[htb]
\centering
\includegraphics[scale=1.0]{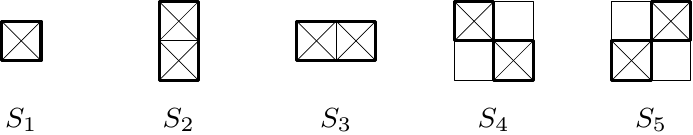}
\caption{Barriers in $\mathcal{S}_{2}$.}
\label{figure:fig010}
\end{figure}

Now we determine the value of $c_{2}$ using the above 
algorithm.
$\mathcal{S}_{2}$ has $5$ barriers $S_{1}$, \ldots, 
$S_{5}$ shown in Fig. \ref{figure:fig010}.
There are $4$ pairs $(S, p)$ such that 
$S \in \mathcal{S}_{2}$, $p \in S$.
In Table \ref{table:tab004} we show values 
$W$, $H$, $d_{0}$, $d_{1}$, $E_\abb{max}(S, p)$, 
$\delta_\abb{opt}(S, p)$, $\epsilon_\abb{opt}(S, p)$ 
for each of these four pairs.
\begin{table}[htb]
\centering
\begin{tabular}{|c|c|c|c|c|c|c|c|}
\hline
$(S, p)$ & $W$ & $H$ & $d_{0}$ & $d_{1}$ & 
    $E_\abb{max}(S, p)$ & $\delta_\abb{opt}(S, p)$ & 
    $\epsilon_\abb{opt}(S, p)$ \\ \hline 
$(S_{4}, (0, 0))$ & $2$ & $2$ & $3$ & $3$ & $0$ & $0$ & $0$ \\ 
$(S_{4}, (1, 1))$ & $2$ & $2$ & $3$ & $3$ & $0$ & $0$ & $0$ \\ 
$(S_{5}, (0, 1))$ & $2$ & $2$ & $2$ & $6$ & $1$ & $2$ & $1$ \\ 
$(S_{5}, (1, 0))$ & $2$ & $2$ & $6$ & $2$ & $1$ & $-2$ & $-1$ \\ \hline
\end{tabular}
\caption{Values of $W$, $H$ and so on 
for the four pairs $(S, p)$ such that 
$S \in \mathcal{S}_{2}$, $p \in S$.}
\label{table:tab004}
\end{table}
From this table we have 
$c_{2} = \max_{S \in \mathcal{S}_{2}, p \in S} E_\abb{max}(S, p) = 1$.

\begin{figure}[htb]
\centering
\includegraphics[scale=1.0]{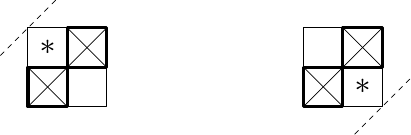}
\caption{Two pairs $(S, p)$ 
such that $E_\abb{max}(S, p) = 1$ ($= c_{2}$).}
\label{figure:fig014}  
\end{figure}

Two pairs $(S_{5}, (0, 1))$, $(S_{5}, (1, 0))$ 
realize the value $E_\abb{max}(S, p) = 1$ ($= c_{2}$).
In Fig. \ref{figure:fig014} we show these two pairs.
The left represents $(S_{5}, (0, 1))$ with the 
$\epsilon_{\abb{opt}}$ value $1$ and 
the right represents $(S_{5}, (1, 0))$ with the 
$\epsilon_{\abb{opt}}$ value $-1$.
Marks ``*'' represent positions of $p$.
The two pairs are symmetric with respect to the 
main diagonal represented by dotted lines.

\begin{figure}[htbp]
\centering
\includegraphics[scale=1.0]{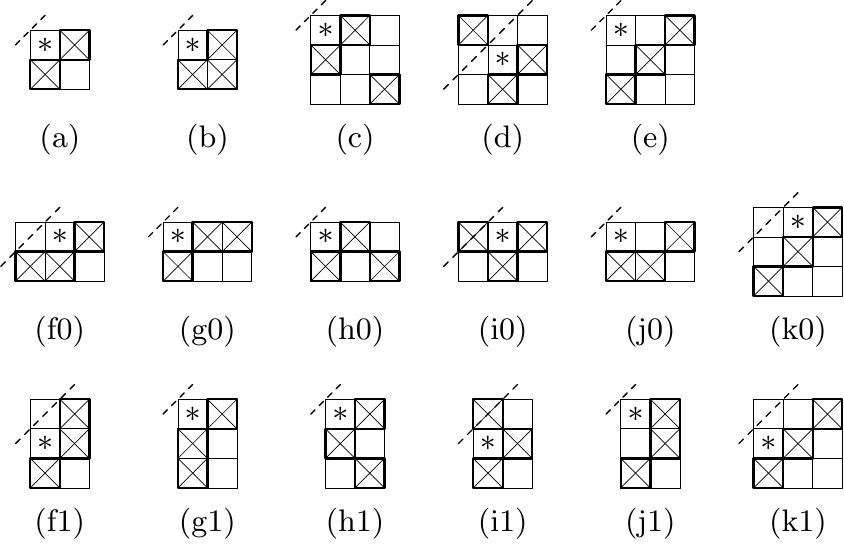}
\caption{The $17$ pairs $(S, p)$ for $k = 3$ 
such that $E_\abb{max}(S, p) = 1$ ($= c_{3}$) 
and $\epsilon_\abb{opt}(S, p) = 1$.}
\label{figure:fig013}
\end{figure}

Next we determine the value of $c_{3}$.
There are $29$ barriers in $\mathcal{S}_{3}$ and 
there are $80$ pairs $(S, p)$ such that 
$S \in \mathcal{S}_{3}$, $p \in S$.
Of these $80$ pairs, the value of 
$E_\abb{max}(S, p)$ is $0$ for $46$ pairs and 
$1$ for $34$ pairs.
Therefore $c_{3} = 1$.
Of the $34$ pairs with $E_\abb{max}(S, p) = 1$, 
the value of $\epsilon_\abb{opt}(S, p)$ 
is $1$ for $17$ pairs and $-1$ for $17$ pairs.
The former $17$ pairs and the latter $17$ pairs are 
symmetric with respect to the main diagonal.
In Fig. \ref{figure:fig013} we show the $17$ pairs 
having the $\epsilon_\abb{opt}$ value $1$.

These $17$ pairs have also symmetry with respect to the 
direction from the northwest to the southeast (the direction 
that is orthogonal to the direction of the main diagonal).
Each of the $5$ pairs (a), (b), \ldots, (e) is 
symmetric with itself.
The six pairs (f0), (g0), \ldots, (k0) are 
symmetric to the six pairs (f1), (g1), \ldots, (k1) 
respectively.

The pair (a) has only $2$ holes but it realizes 
$E_\abb{opt}(S, p) = 1 = c_{3}$.
The $13$ pairs (b), (c), (d), (f0), (g0), (h0), 
(i0), (k0), 
(f1), (g1), (h1), (i1), (k1) 
are obtained by adding 
one hole to the two hole pair (a) but 
realize the same $E_\abb{opt}(S, p)$ value as (a).
Hence they are essentially the two hole pair (a).
For the remaining three pairs (e), (j0), (j1), 
the three holes are essentially used.

We computed the value of $c_{k}$ for up to $k = 9$ 
by computer.
We show the result in Table \ref{table:tab002}.
In the table we also show the number of barriers 
$S \in \mathcal{S}_{k}$ 
(that is, the number of barriers $S$ having at most $k$ holes),  
the number of pairs $(S, p)$ 
such that $S \in \mathcal{S}_{k}$, $p \in S$, 
and the number of pairs $(S, p)$ such that 
$S \in \mathcal{S}_{k}$, $p \in S$, 
$E_\abb{max}(S, p) = c_{k}$.
\begin{table}[htb]
\centering
\begin{tabular}{|c|c|r|r|r|}
\hline
\begin{minipage}{2em}
\begin{center}
$k$ 
\end{center}
\end{minipage}
&
\begin{minipage}{2em}
\begin{center}
$c_{k}$ 
\end{center}
\end{minipage}
& 
\begin{minipage}{8em}
\begin{center}
The number of barriers $S$
\end{center}
\end{minipage}
&
\begin{minipage}{8em}
\begin{center}
The number of pairs $(S, p)$ 
\end{center}
\end{minipage}
&
\begin{minipage}{8em}
\begin{center}
\rule[0em]{0em}{1.5em} 
The number of pairs $(S, p)$ 
with $E_\abb{max}(S, p)$ $=$ $c_{k}$
\rule[-0.9em]{0em}{0em} 
\end{center}
\end{minipage}
\\ \hline
\rule{0em}{1.0em}
$2$ & $1$ & $5$ & $4$ & $2$ \\
\rule{0em}{1.0em}
$3$ & $1$ & $29$ & $80$ & $34$ \\
\rule{0em}{1.0em}
$4$ & $2$ & $224$ & $1,324$ & $16$ \\
\rule{0em}{1.0em}
$5$ & $3$ & $2,220$ & $22,588$ & $24$ \\
\rule{0em}{1.0em}
$6$ & $4$ & $26,898$ & $416,782$ & $14$ \\
\rule{0em}{1.0em}
$7$ & $5$ & $384,344$ & $8,397,762$ & $20$ \\ 
\rule{0em}{1.0em}
$8$ & $6$ & $6,314,747$ & $184,619,252$ & $26$ \\ 
\rule{0em}{1.0em}
$9$ & $7$ & $117,140,060$ & $4,411,162,884$ & $32$ \\ 
\hline
\end{tabular}
\caption{
The values of $c_{k}$ for $k = 2, \ldots, 9$ 
obtained by computer.
The table also shows 
the numbers of barriers $S$ in $S_{k}$, 
the numbers of pairs $(S, p)$, 
and the numbers of pairs $(S, p)$ such that 
$E_\abb{max}(S, p) = c_{k}$.}
\label{table:tab002}
\end{table}

\begin{figure}[htb]
\centering
\includegraphics[scale=1.0]{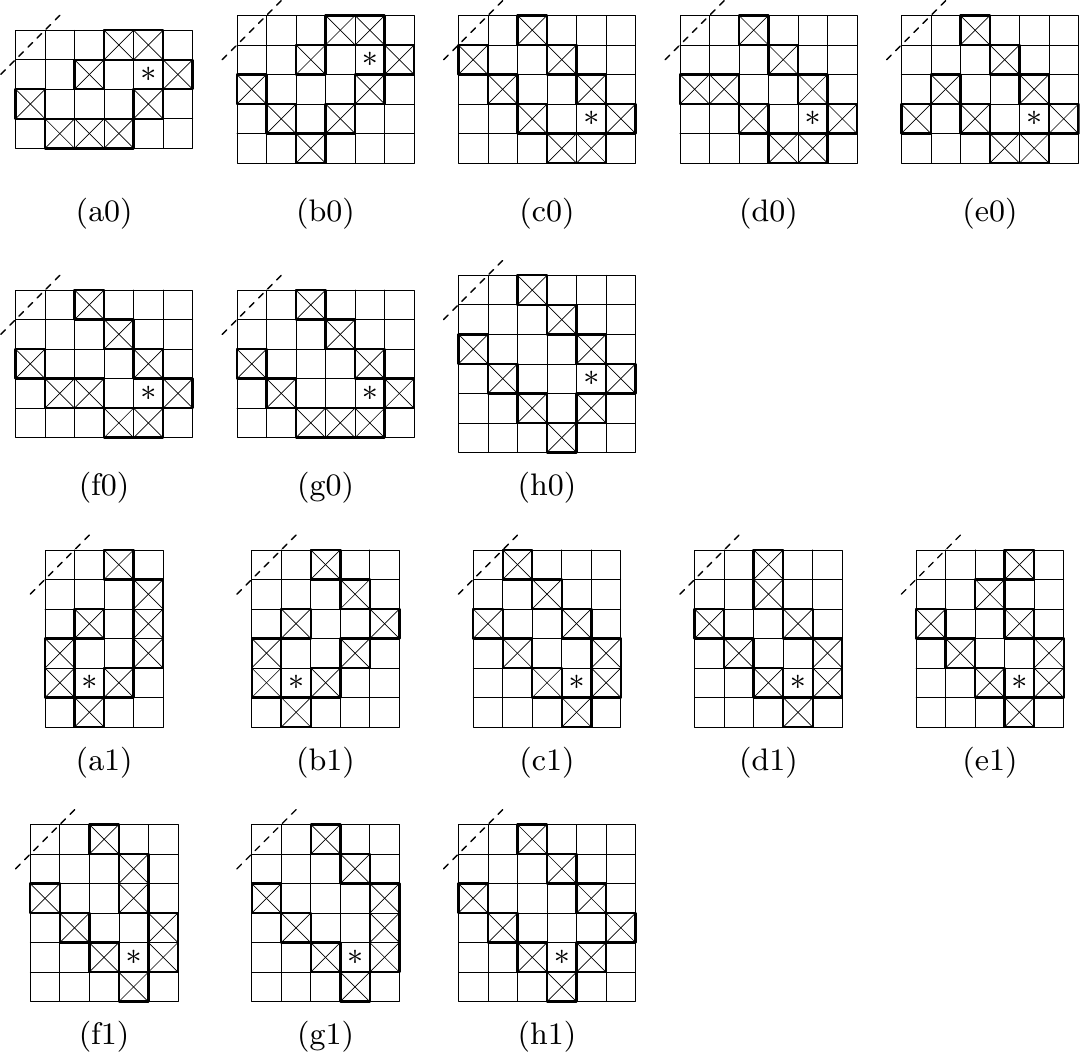}
\caption{The $16$ pairs $(S, p)$ for $k = 9$ 
such that $E_\abb{max}(S, p) = 7$ ($ = c_{9}$) and 
$\epsilon_\abb{opt}(S, p) \geq 0$.}
\label{figure:fig054}
\end{figure}

For $k = 9$ we have $c_{9} = 7$.
In Fig. \ref{figure:fig054} we show the $16$ pairs 
$(S, p)$ that realize the value $E_\abb{max}(S, p) = 7$ 
and that have nonnegative $\epsilon_\abb{opt}$ values.
The value $\epsilon_\abb{opt}(S, p)$ is 
$5$ for the $4$ pairs (a0), (b0), (a1), (b1) and 
$7$ for the remaining $12$ pairs (c0), (d0), \ldots, (h0), 
(c1), (d1), \ldots, (h1).
The forms of the barriers of the former $4$ pairs are 
curved caves and those of the latter $12$ pairs are 
straight caves, both with the positions of $p$ at their dead ends.
The $8$ pairs (a0), (b0), \ldots, (h0) and 
the $8$ pairs (a1), (b1), \ldots, (h1) are symmetric 
with respect to the direction from the northwest to the southeast.

\section{Determination of the minimum firing time \\
$\abb{mft}_{\abb{SH}[2]}(C)$ 
of $\abb{SH}[2]$}
\label{section:determination}

In this section we determine the value of 
the minimum firing time 
$\abb{mft}_{\abb{SH}[2]}(C)$ 
of $\abb{SH}[2]$.

Suppose that $C$ is a configuration of size $w$ of 
$\abb{SH}[2]$.
We know that $\abb{mft}(C)$ is either $2w$ or $2w + 1$.
Therefore to determine $\abb{mft}(C)$ it is only necessary 
to prove either the lower bound $\abb{mft}(C) \geq 2w + 1$ or 
the upper bound $\abb{mft}(C) \leq 2w$.
To show the lower bound $\abb{mft}(C) \geq 2w + 1$ we use 
Corollary \ref{corollary:cor000}.
In Subsection \ref{subsection:reformulation} we show some results 
that are used in proving lower bounds.
In Subsection \ref{subsection:four_subsets} we define 
a division of a square $S_{w}$ into four nonoverlapping subsets 
$U$, $V$, $W$, $X$ and show one theorem on this division.
To show the lower bound $\abb{mft}(C) \leq 2w$ we construct 
a partial solution that fires $C$ at time $2w$.
To construct such a partial solution we use one unified strategy.
In Subsection \ref{subsection:partial_solution_strategy} 
we explain it.
Finally, in Subsection \ref{subsection:statement_and_proof} 
we state the main result and prove it.

\subsection{Some results used for proving lower bounds of \\
$\abb{mft}_{\abb{SH}[2]}(C)$}
\label{subsection:reformulation}

We use Corollary \ref{corollary:cor000} to prove 
a lower bound $2w + 1 \leq \abb{mft}_{\abb{SH}[2]}(C)$ 
for a configuration $C$ of $\abb{SH}[2]$ of size $w$.
By Corollary \ref{corollary:cor004} we may assume that 
the size of $C'$ mentioned in Corollary \ref{corollary:cor000} 
is $w$.
In this subsection we show some results that are useful for finding 
such $C'$.

By a \textit{pattern} $\pi$ we mean a partial function from 
$\mathbb{Z}^{2}$ to the two element set $\{\abb{N}, \abb{H}\}$.
The letters ``$\abb{N}$'' and ``$\abb{H}$'' are abbreviations of 
\textit{nodes} and \textit{holes} respectively.
We say that a configuration $C$ \textit{has a pattern} $\pi$ if 
for any $v \in \mathbb{Z}^{2}$, 
if $\pi(v) = \abb{N}$ then the position $v$ is a node in $C$ and 
if $\pi(v) = \abb{H}$ then there is a hole at the position $v$ in $C$.
For a configuration $C$ of size $w$ and a set $X \subseteq \{0, 1, \ldots, w\}^{2}$, 
by $\pi(C, X)$ we denote the pattern having $X$ as its domain such that, 
for any $v \in X$, if $v$ is a node of $C$ then $\pi(C, X)(v) = \abb{N}$ and 
if there is a hole at $v$ in $C$ then $\pi(C, X)(v) = \abb{H}$.

\begin{figure}[htbp]
\centering
\includegraphics[scale=1.0]{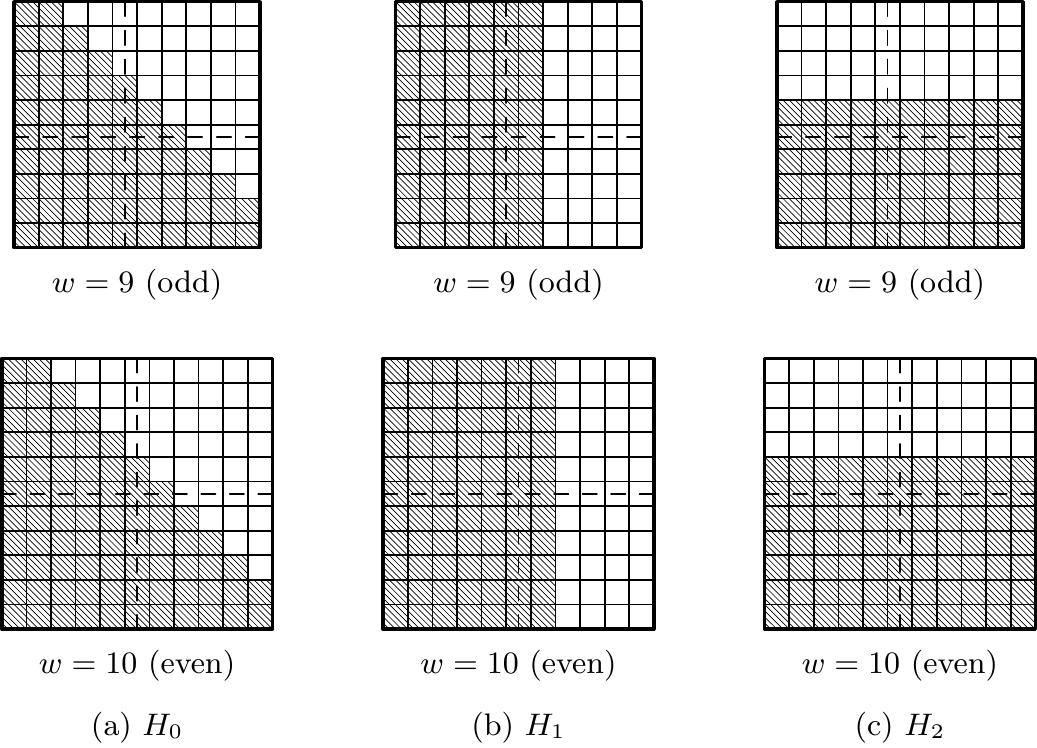}
\caption{Three sets $H_{0}$, $H_{1}$, $H_{2}$.}
\label{figure:fig018}
\end{figure}

Let $H_{0}$, $H_{1}$, $H_{2}$ be the following subsets of $S_{w}$ 
(see \eqref{equation:eq013} for the definition of $S_{w}$):
\begin{align*}
H_{0} & = \{ (x, y) ~|~ 0 \leq x \leq w, 0 \leq y \leq w, x + y \leq w + 1 \}, \\
H_{1} & = \{ (x, y) ~|~ 0 \leq x \leq \lfloor w / 2 \rfloor + 1, 0 \leq y \leq w \}, \\
H_{2} & = \{ (x, y) ~|~ 0 \leq x \leq w, 0 \leq y \leq \lfloor w / 2 \rfloor + 1 \}.
\end{align*}
In Fig. \ref{figure:fig018} (a), (b), (c) we show examples of these sets. 
For each of them we show an example for an even $w$ and for an odd $w$.
Dotted lines in the figures are vertical lines 
$x = \lfloor w / 2 \rfloor$ and horizontal lines 
$y = \lfloor w / 2 \rfloor$.
In the determination of $\abb{mft}(C)$ of $C$ of size $w$, 
these special vertical and horizontal lines 
play important roles.
Therefore, when we show $S_{w}$ by figures we write dotted lines 
in these lines as in Fig. \ref{figure:fig018}.

\begin{thm}
\label{theorem:thm007}
Suppose that $C$, $C'$ are configurations of size $w$ of $\abb{SH}[2]$ and 
$\pi(C, H) = \pi(C', H)$ for one of 
$H = H_{0}, H_{1}, H_{2}$.
Then we have $C \equiv_{2w}' C'$.
\end{thm}

\noindent
{\it Proof}. 
We consider the case $H = H_{0}$ and prove $C \equiv_{2w, (0, 0)}' C'$.
Let $H_{0}'$ be the set of positions
\[
H_{0}' = \{ (x, y) \in H_{0} ~|~ x + y \leq w \}.
\]
Then $H_{0}' \subseteq H_{0}$ and any position in $S_{w}$ adjacent 
to a position in $H_{0}'$ is in $H_{0}$.  

Let $P$ be an arbitrary path in $C$ of length 
at most $2w$ from $v_\abb{gen}$ ($= (0, 0)$) 
to the node $(0, 0)$.
Then for any node $u = (x, y)$ on this path we have 
$2(x + y) = \abb{d}_\abb{MH}(v_\abb{gen}, u) + 
\abb{d}_\abb{MH}(u, (0, 0)) 
\leq \abb{d}_{C}(v_\abb{gen}, u) + \abb{d}_{C}(u, (0, 0)) 
\leq |P| \leq 2w$ 
and hence $x + y \leq w$.
Therefore the node $u$ is in $H_{0}'$.
Then, by $\pi(C, H_{0}) = \pi(C', H_{0})$, $u$ is also 
a node in $C'$.
Let $u'$ be any position in $S_{w}$ adjacent to $u$.
Then $u'$ is in $H_{0}$ and 
by $\pi(C, H_{0}) = \pi(C', H_{0})$, $u'$ is a node in $C$ 
if and only if $u'$ is a node in $C'$.
This means that $\abb{bc}_{C}(u) = \abb{bc}_{C'}(u)$.

Therefore, any path in $C$ of length at most $2w$ 
from $v_\abb{gen}$ to $(0, 0)$ is 
also a path in $C'$ 
and the boundary condition of any node in the path is 
the same in $C$ and $C'$.
Similarly we can prove the same statement with $C$, $C'$ 
interchanged.
Hence we have $C \equiv_{2w, (0, 0)}' C'$.

For the cases $H = H_{1}$ and $H = H_{2}$ 
we show $C \equiv_{2w, (0, w)}' C'$ and 
$C \equiv_{2w, (w, 0)} C'$ respectively.
Instead of $2(x + y) \leq 2w$ we use 
$2x + w \leq 2w$ (and hence $x \leq \lfloor w / 2 \rfloor$) 
and $w + 2y \leq 2w$ (and hence $y \leq \lfloor w / 2 \rfloor$) 
respectively.
\hfill $\Box$

\medskip

This theorem is useful for finding $C'$ such that $C \equiv_{2w + 1} C'$ 
in applying Corollary \ref{corollary:cor000}.
Next we show a characterization of configurations $C'$ of size $w$ 
such that $2w + 1 \leq \max_{v \in C'} T(v, C')$.

We call a hole at $v = (x_{0}, y_{0})$ a \textit{critical hole} 
if $|x_{0} - y_{0}| = 2$ and by a \textit{critical pair} (\textit{of holes}) 
we mean a pair $v_{0}, v_{1}$ of two critical holes such that 
$v_{1} = v_{0} + (1, 1)$.
\begin{figure}[htbp]
\centering
\includegraphics[scale=1.0]{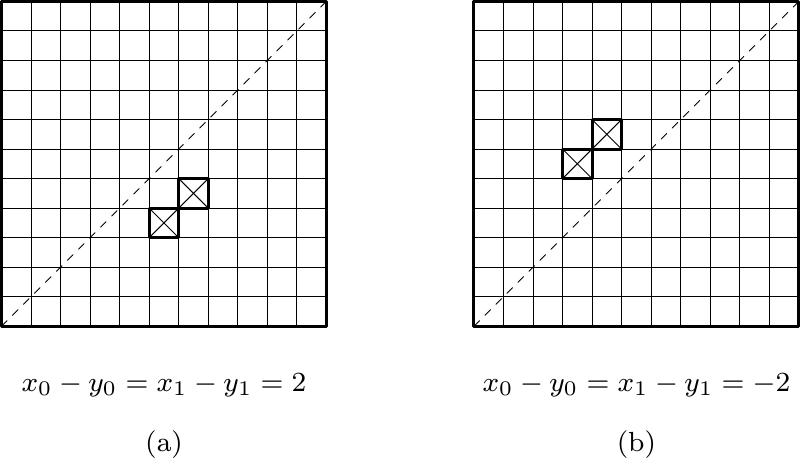}
\caption{Examples of critical pairs $v_{0} = (x_{0}, y_{0})$, $v_{1} = (x_{1}, y_{1})$.}
\label{figure:fig020}
\end{figure}
In Fig. \ref{figure:fig020} we show examples of critical pairs.

\begin{thm}
\label{theorem:thm013}
Let $C$ be a configuration of size $w$ of $\abb{SH}[2]$.
The following two statements are equivalent.
\begin{enumerate}
\item[{\rm (1)}] $2w + 1 \leq \max_{v \in C} T(v, C)$.
\item[{\rm (2)}] $C$ has a critical pair of holes.
\end{enumerate}
\end{thm}
%
\noindent
{\it Proof}. 
When a node $v$ of $C$ is in a maximal barrier $R$, 
by $S$, $p$, $\delta$, $z$ ($= (z_{x}, z_{y})$) 
we denote the items determined from $v$, $R$ by 
Figures \ref{figure:fig007}, \ref{figure:fig008}.

Suppose that the statement (1) is true.
Let $v$ be a node in $C$ such that $T(v, C) > 2w$.  
Then by \eqref{equation:eq016} 
$v$ must be in a maximal barrier $R$ of $C$.
By Fig. \ref{figure:fig010}, there are four pairs $(S, p)$ 
such that $p \in S$, that is, 
$(S_{4}, (0, 0))$, $(S_{4}, (1, 1))$, $(S_{5}, (0, 1))$ and 
$(S_{5}, (1, 0))$.
We have $E(S, p, \delta) = T(v, C) - 2w > 0$. 
Table \ref{table:tab004} and a simple calculation show that 
there are two triples $(S, p, \delta)$ such that $p \in S$ and 
$E(S, p, \delta) > 0$ and they are 
$(S_{5}, (0, 1), 2)$ and $(S_{5}, (1, 0), -2)$.

Suppose that $S = S_{5}$, $p = (0, 1)$, $\delta = 2$ correspond to 
$v$, $R$.
Then there are two holes at $v_{0} = z$, $v_{1} = z + (1, 1)$ 
and $z_{x} - z_{y} = \delta = 2$.
Therefore holes $v_{0}$, $v_{1}$ are critical and the pair 
$v_{0}$, $v_{1}$ is a critical pair.
Similarly, for the case $S = S_{5}$, $p = (1, 0)$, $\delta = -2$ too
there are two holes at $v_{0} = z$, $v_{1} = z + (1, 1)$ and 
$z_{x} - z_{y} = -2$ and the pair $v_{0}$, $v_{1}$ is a critical 
pair.
Hence the statement (2) is true for both cases.

Next, suppose that the statement (2) is true.
Let a pair $v_{0} = (x_{0}, y_{0})$, $v_{1} = v_{0} + (1, 1)$ 
be a critical pair.
We consider only the case $x_{0} - y_{0} = 2$.
Let $v$ be the position $(x_{0}, y_{0} + 1)$.
Then $v$ is in the maximal barrier $R$ consisting of 
the four positions 
$(x_{0}, y_{0})$ (the hole $v_{0}$), 
$(x_{0}, y_{0} + 1)$ (the node $v$), 
$(x_{0} + 1, y_{0})$ (a node), 
$(x_{0} + 1, y_{0} + 1)$ (the hole $v_{1}$).
Then we have $S = S_{5}$, $p = v - v_{0} = (0, 1)$, $\delta = x_{0} - y_{0} = 2$.
Therefore, 
by Table \ref{table:tab004} we have 
$T(v, C) = 2w + E(S, p, \delta) 
= 2w + E(S_{5}, (0, 1), 2) 
= 2w + \min \{ \delta - H - 1 + d_{0}, - \delta - W - 1 + d_{1} \}
= 2w + \min \{ 2 - 2 - 1 + 2, - 2 - 2 - 1 + 6 \} = 2w + 1$ 
and the statement (1) is true.
\hfill
$\Box$

\medskip

When we prove $2w + 1 \leq \abb{mft}(C)$ in the proof of 
the main theorem (Theorem \ref{theorem:thm012}) we prove this 
by showing existence of a sequence $C_{0}$, \ldots, $C_{n}$ 
of configurations of size $w$ 
($n \geq 0$) such that $C_{0} = C$, $C_{n}$ has a critical pair of holes, 
and $\pi(C_{i}, H) = \pi(C_{i+1}, H)$ ($H$ is one of $H_{0}$, $H_{1}$, $H_{2}$) 
for any $0 \leq i \leq n - 1$.

\subsection{Subsets $U$, $V$, $W$, $X$ of squares $S_{w}$}
\label{subsection:four_subsets}

We define four nonoverlapping subsets 
$U$, $V$, $W$, $X$ of $S_{w}$ as follows.
First we define $U \cup V$ and $U \cup V \cup W$ by 
\begin{align}
U \cup V = \{v \in S_{w} \mid 
\text{for any $v' \in S_{w}$, 
$\abb{d}_{\abb{MH}}(v_{\abb{gen}}, v) + 
\abb{d}_{\abb{MH}}(v, v') \leq 2w$}\},
\label{equation:eq017}
\end{align}
\begin{align}
U \cup V \cup W & = \{v \in S_{w} \mid 
\text{for any $v' \in S_{w}$ there is $v'' \in S_{w}$ such that} 
\nonumber \\
& \text{$\abb{d}_{\abb{MH}}(v, v'') \leq 1$ and 
$\abb{d}_{\abb{MH}}(v_{\abb{gen}}, v'') + 
\abb{d}_{\abb{MH}}(v'', v') \leq 2w$} \}.
\label{equation:eq018}
\end{align}

\begin{figure}[htbp]
\centering
\includegraphics[scale=1.0]{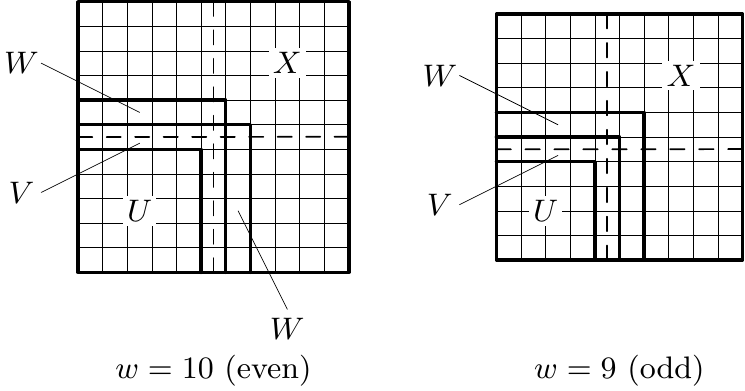}
\caption{The four sets $U$, $V$, $W$, $X$.}
\label{figure:fig066}
\end{figure}

We can define these two sets more explicitly as follows 
(see Fig. \ref{figure:fig066}).
For $U \cup V$ we define 
\begin{align}
U \cup V
= & 
\{ (x, y) \in S_{w} \mid 0 \leq x \leq \lfloor w / 2 \rfloor, 
                               0 \leq x \leq \lfloor w / 2 \rfloor\}.
\label{equation:eq019}
\end{align}
For $U \cup V \cup W$ we have different definitions 
for an even $w$ and for an odd $w$.
If $w$ is even then 
\begin{align}
U \cup V \cup W 
= & 
\{ (x, y) \in S_{w} \mid \text{either $0 \leq x \leq \lfloor w / 2 \rfloor$ and 
                                      $0 \leq y \leq \lfloor w / 2 \rfloor$, or} 
                                      \nonumber \\ 
                & \quad\quad \text{$0 \leq x \leq \lfloor w / 2 \rfloor$ and 
                                      $y = \lfloor w / 2 \rfloor + 1$, or}
                                      \nonumber \\ 
                & \quad\quad \text{$x = \lfloor w / 2 \rfloor + 1$ and 
                                      $0 \leq y \leq \lfloor w / 2 \rfloor$} \}
\label{equation:eq020}
\end{align}
and if $w$ is odd then 
\begin{align}
U \cup V \cup W 
= & 
\{ (x, y) \in S_{w} \mid 0 \leq x \leq \lfloor w / 2 \rfloor + 1, 
                               0 \leq y \leq \lfloor w / 2 \rfloor + 1 \}.
\label{equation:eq021}
\end{align}
We define $U$ explicitly by 
\begin{align}
U = \{ (x, y) \in S_{w} \mid 0 \leq x \leq \lfloor w / 2 \rfloor - 1, 
                             0 \leq y \leq \lfloor w / 2 \rfloor - 1 \}
\label{equation:eq022}
\end{align}
and define $X$ by 
\begin{equation}
X = S_{w} - U \cup V \cup W.
\label{equation:eq023}
\end{equation}

The equivalence of the two definitions \eqref{equation:eq017}, 
\eqref{equation:eq019} of $U \cup V$ is easy to show.
We show an outline of the proof of the equivalence of 
the first definition \eqref{equation:eq018} and 
the second definition \eqref{equation:eq020}, \eqref{equation:eq021} of 
$U \cup V \cup W$.

Suppose that $v = (x, y)$ is in $U \cup V \cup W$ of 
the first definition.
We show that $v$ is in $U \cup V \cup W$ of the second definition.
In this case $y \geq \lfloor w / 2 \rfloor + 2$ is not possible 
because for $v' = (w, 0)$ there does not exist $v''$ 
such that the statement of the first definition is true.
Similarly $x \geq \lfloor w / 2 \rfloor + 2$ is not possible.
Moreover, when $w$ is even $x = y = \lfloor w / 2 \rfloor + 1$ 
is not possible because for $v' = (0, 0)$ there does not exist $v''$ 
such that the statement of the first definition is true.
Therefore, $v$ must be in $U \cup V \cup W$ of 
the second definition.

Next suppose that $v = (x, y)$ is in $U \cup V \cup W$ of 
the second definition. 
We show that $v$ is in $U \cup V \cup W$ of the first definition.
We use the equivalence of the two definitions \eqref{equation:eq017}, 
\eqref{equation:eq019} of $U \cup V$ to show it.
If $v$ is in $U \cup V$ then the statement of the first definition 
is true because for any $v$ we can use $v$ itself as $v''$.
If $v$ is in $W$ and is adjacent to a position in $U \cup V$ 
then the statement of the first definition is true because 
for any $v'$ we can use the position in $U \cup V$ that is 
adjacent to $v$ as $v''$.
Finally, if $w$ is odd and $v = (\lfloor w / 2 \rfloor + 1, 
\lfloor w / 2 \rfloor + 1)$ 
then the statement of the first definition is true because 
for any $v' = (x', y')$ we can use $v - (1, 0)$, $v - (0, 1)$, or $v$ respectively 
as $v''$ according as $x' \leq \lfloor w / 2 \rfloor$, 
$y' \leq \lfloor w / 2 \rfloor$, or otherwise respectively.
Therefore, in any case $v$ is in $U \cup V \cup W$ of the first definition.

By $v_\abb{cnt}$ we denote the position 
$(\lfloor \tilde{w} / 2 \rfloor, \lfloor 
\tilde{w} / 2 \rfloor)$ 
(``\textrm{cnt}'' is for \textit{center}).
It is at the corner of $V$.
We have the following relations among 
$U \cup V \cup W$, $H_{0}$, $H_{1}$, $H_{2}$:
\begin{enumerate}
\item[$\bullet$] $U \cup V \cup W \subseteq H_{0}$ for both of 
even $w$ and odd $w$.
\item[$\bullet$] $U \cup V \cup W = H_{1} \cap H_{2} - 
\{v_{\abb{cnt}} + (1, 1)\}$ for even $w$ and 
$U \cup V \cup W = H_{1} \cap H_{2}$ for odd $w$.
\end{enumerate}

$U \cup V$ is the set of positions $v$ such that 
$\abb{d}_\abb{MH}(v_\abb{gen}, v) + \abb{d}_\abb{MH}(v, v') \leq 2w$ 
for any position $v'$.
Therefore, if $v$, $v'$ are nodes in a configuration $C$ such that 
$v \in U \cup V$ then we expect that 
$\abb{d}_\abb{MH}(v_\abb{gen}, v) + \abb{d}_{C}(v, v') \leq 2w$ 
is true except the case where $v'$ is near the four corners 
$(0, 0)$, $(0, w)$, $(w, 0)$, $(w, w)$.
The following theorem is a precise statement of this intuitive statement.

\begin{thm}
\label{theorem:thm011}
Suppose that $w \geq 5$, $C$ is a configuration of 
size $w$ of $\abb{SH}[2]$, 
and $v = (x, y)$, $v' = (x', y')$ are nodes in $C$ such that 
$v \in U \cup V$.
Then $\abb{d}_\abb{MH}(v_\abb{gen}, v) + 
\abb{d}_{C}(v, v') \leq 2w$ except the following cases.
\begin{enumerate}
\item[{\rm (1)}] Two holes are at $v + (0, 1)$, $v + (1, 0)$ 
and $v'$ is one of $(w - 1, w)$, $(w, w - 1)$, $(w, w)$.
\item[{\rm (2)}] $x = \lfloor w / 2 \rfloor$, 
two holes are at $v - (1, 0)$, $v + (0, 1)$, and 
$v'$ is one of $(0, w - 1)$, $(1, w)$, $(0, w)$ if 
$w$ is even and is $(0, w)$ if $w$ is odd.
\item[{\rm (3)}] $y = \lfloor w / 2 \rfloor$, 
two holes are at $v - (0, 1)$, $v + (1, 0)$, and 
$v'$ is one of $(w - 1, 0)$, $(w, 1)$, $(w, 0)$ if 
$w$ is even and is $(w, 0)$ if $w$ is odd.
\item[{\rm (4)}] $w$ is even, 
$x = y = \lfloor w / 2 \rfloor$, 
two holes are at $v - (0, 1)$, $v - (1, 0)$, and 
$v'$ is one of $(1, 0)$, $(0, 1)$, $(0, 0)$.
\end{enumerate}
\end{thm}

\begin{figure}[htbp]
\centering
\includegraphics[scale=1.0]{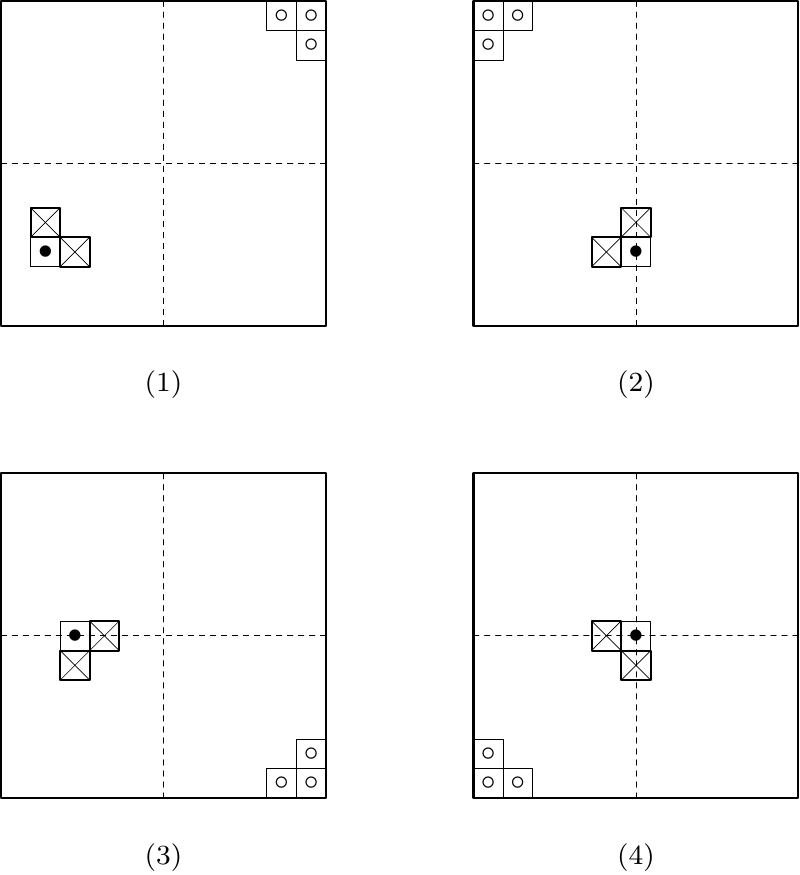}
\caption{The four exceptions in Theorem 
\ref{theorem:thm011} for an even $w$.}
\label{figure:fig022}
\end{figure}

In Fig. \ref{figure:fig022} we show examples of these four 
exceptional cases for an even $w$. 
A bullet denotes a position of $v$ and 
a small circle denotes a position of $v'$ of the four exceptions.

The proof of this theorem is not essentially difficult 
but is tedious. 
Therefore we show it in 
Appendix \ref{section:tedious_proof}.

\subsection{An idea to construct partial solutions 
for proving upper bounds of $\abb{mft}_{\abb{SH}[k]}(C)$}
\label{subsection:partial_solution_strategy}

Suppose that $\tilde{C}$ is a fixed configuration of $\abb{SH}[k]$ 
of size $\tilde{w}$, $s$ ($\geq 0$) is a fixed constant, 
and we want to prove an upper bound 
$\abb{mft}_{\abb{SH}[k]}(\tilde{C}) \leq 2 \tilde{w} + s$ 
of $\abb{mft}_{\abb{SH}[k]}(\tilde{C})$.
We prove it if we can construct a partial solution $A$ of 
$\abb{SH}[k]$ that fires $\tilde{C}$ at time $2 \tilde{w} + s$.
Below we show one idea to construct such a partial solution $A$.
Assume that $C$ is a configuration of $\abb{SH}[k]$ 
of size $w$ and nodes of $C$ are copies of $A$.

$A$ uses the size check messages $\abb{W}_{0}$, $\abb{W}_{1}$ 
that were used in the proofs of Theorems \ref{theorem:thm001}, 
\ref{theorem:thm003} (see also the comment after 
Theorem \ref{theorem:thm001}).
These messages are generated if and only if $w = \tilde{w}$.
If $\abb{W}_{0}$ or $\abb{W}_{1}$ is generated then 
it is generated at $(0, w)$ or $(w, 0)$ respectively and 
at the time $w$.

We define a pattern $\tilde{\pi}$ such that $\tilde{C}$ has $\tilde{\pi}$.
Moreover we define patterns $\pi_{i, j}$, 
positions $v_{i, j}$, values $r_{i, j}$ ($\geq 0$) 
for $0 \leq i \leq n - 1$, $0 \leq j \leq m_{i} - 1$ 
($n \geq 0$, $m_{i} \geq 1$) 
and design $A$ so that all of 
the following five statements are true under the assumption 
that $w = \tilde{w}$.

\begin{enumerate}
\item[(C1)] If $C$ has $\tilde{\pi}$ then 
$\max_{v \in C} T(v, C) \leq 2\tilde{w} + s$.
\item[(C2)] For any $i$, 
$C$ has $\tilde{\pi}$ 
if and only if $C$ has all of 
$\pi_{i,0}, \ldots, \pi_{i, m_{i} - 1}$ ($0 \leq i \leq n - 1$).
\item[(C3)] For any $i, j$, 
$C$ has $\pi_{i,j}$ 
if and only if the message $\abb{M}_{i,j}$ is generated 
($0 \leq i \leq n - 1$, $0 \leq j \leq m_{i} - 1$).
\item[(C4)] For any $i, j$, if $\abb{M}_{i,j}$ is 
generated then it is generated at the node $v_{i,j}$ 
at time $\abb{d}_\abb{MH}(v_\abb{gen}, v_{i,j}) + r_{i, j}$ 
($0 \leq i \leq n - 1$, $0 \leq j \leq m_{i} - 1$).
\item[(C5)] If $C$ has $\tilde{\pi}$ then 
for any node $v$ in $C$ there exists $i$ ($0 \leq i \leq n - 1$) 
such that 
$\abb{d}_{\abb{MH}}(v_{\abb{gen}}, v_{i, j}) + r_{i, j} + 
\abb{d}_{C}(v_{i, j}, v) \leq 2 \tilde{w} + s$ for any $j$ 
($0 \leq j \leq m_{i} - 1$).
\end{enumerate}

A node $v$ of $C$ fires if and only if 
the current time is $2\tilde{w} + s$ and the following statement 
is true: 
the node has received at least one of $\abb{W}_{0}$, $\abb{W}_{1}$ 
before or at time $2\tilde{w} + s$ and there is $i$ ($0 \leq i \leq n - 1$) 
such that the node has received all of 
$\abb{M}_{i,0}, \ldots, \abb{M}_{i, m_{i} - 1}$ 
before or at $2\tilde{w} + s$.
From now on, we denote the above statement by 
a logical formula-like expression 
\begin{equation}
(\abb{W}_{0} \vee \abb{W}_{1}) \wedge 
((\abb{M}_{0,0} \wedge \ldots \wedge 
\abb{M}_{0, m_{0} - 1}) 
\vee \ldots \vee (\abb{M}_{n-1, 0} \wedge \ldots 
\wedge \abb{M}_{n-1, m_{n-1} - 1})).
\label{equation:eq004}
\end{equation}

This completes the explanation of our idea for constructing $A$.
In the following theorem we show that 
if we can successfully construct a finite automaton $A$ using the above idea 
then $A$ is a desired partial solution.

\begin{thm}
\label{theorem:thm016}
If $A$ is a finite automaton constructed by the above idea for 
a configuration $\tilde{C}$ of $\abb{SH}[k]$ of size $\tilde{w}$ 
then $A$ is a partial solution of $\abb{SH}[k]$ that has the set 
\[
X = \{ C \mid \text{$C$ is of size $\tilde{w}$ and has $\tilde{\pi}$} \}
\]
as its domain and that fires any configuration $C$ in the domain $X$ 
{\rm (}including $\tilde{C}${\rm )} 
at time $2 \tilde{w} + s$.
\end{thm}

\noindent
{\it Proof}. 
(1) We show that if $C$ is in $X$ then any node in it fires at time $2 \tilde{w} + s$.

$C$ is of size $\tilde{w}$.
Hence all of (C1) -- (C5) are true.
Both of the two messages $\abb{M}_{0}$, $\abb{M}_{1}$ 
are generated because $w = \tilde{w}$.
Moreover, by (C1) $T(v, C) \leq 2 \tilde{w} + s$ for any node $v$ in $C$.
Hence any node $v$ in $C$ receives at least one of $\abb{M}_{0}$, $\abb{M}_{1}$ 
before or at $2 \tilde{w} + s$.

$C$ has $\tilde{\pi}$.
Therefore by (C2) $C$ has $\pi_{i, j}$ for all $i$, $j$ and hence 
by (C3), (C4) the message $\abb{M}_{i, j}$ is generated 
at $v_{i, j}$ at time $\abb{d}_{\abb{MH}}(v_{\abb{gen}}, v_{i, j}) + r_{i, j}$ 
for all $i$, $j$.
Then, by (C5), for any node in $C$ there is $i$ such that 
the node receives all of the messages $\abb{M}_{i, 0}$, \ldots, $\abb{M}_{i, m_{i} - 1}$ 
before or at time $2 \tilde{w} + s$.
Therefore, any node in $C$ fires at time $2 \tilde{w} + s$.

\medskip

\noindent
(2) We show that if a node $v$ in $C$ fires at some time then $C$ is in $X$.
(This means that if $C$ is not in $X$ then any node in $C$ never fires.)

The node $v$ received at least one of $\abb{M}_{0}$, $\abb{M}_{1}$.
Hence $w = \tilde{w}$ is true.
Hence all of (C1) -- (C5) are true.
Moreover, for some $i$, $v$ received all of $\abb{M}_{i, 0}$, \ldots, 
$\abb{M}_{i, m_{i} - 1}$, $C$ has all of $\pi_{i, 0}$, \ldots, $\pi_{i, m_{i} - 1}$ 
by (C3), and $C$ has $\tilde{\pi}$ by (C2).
Therefore $C$ is in $X$.
\hfill
$\Box$

\medskip

In the following subsection we prove all the upper bounds 
$\abb{mft}_{\abb{SH}[2]}(\tilde{C}) \leq 2 \tilde{w}$ 
in the proof of the main result 
(Theorem \ref{theorem:thm012}) using the above 
idea with $k = 2$, $s = 0$.

\subsection{The statement of the result and its proof}
\label{subsection:statement_and_proof}

Now we are ready to state our main result.

\begin{figure}[htbp]
\centering
\includegraphics[scale=1.0]{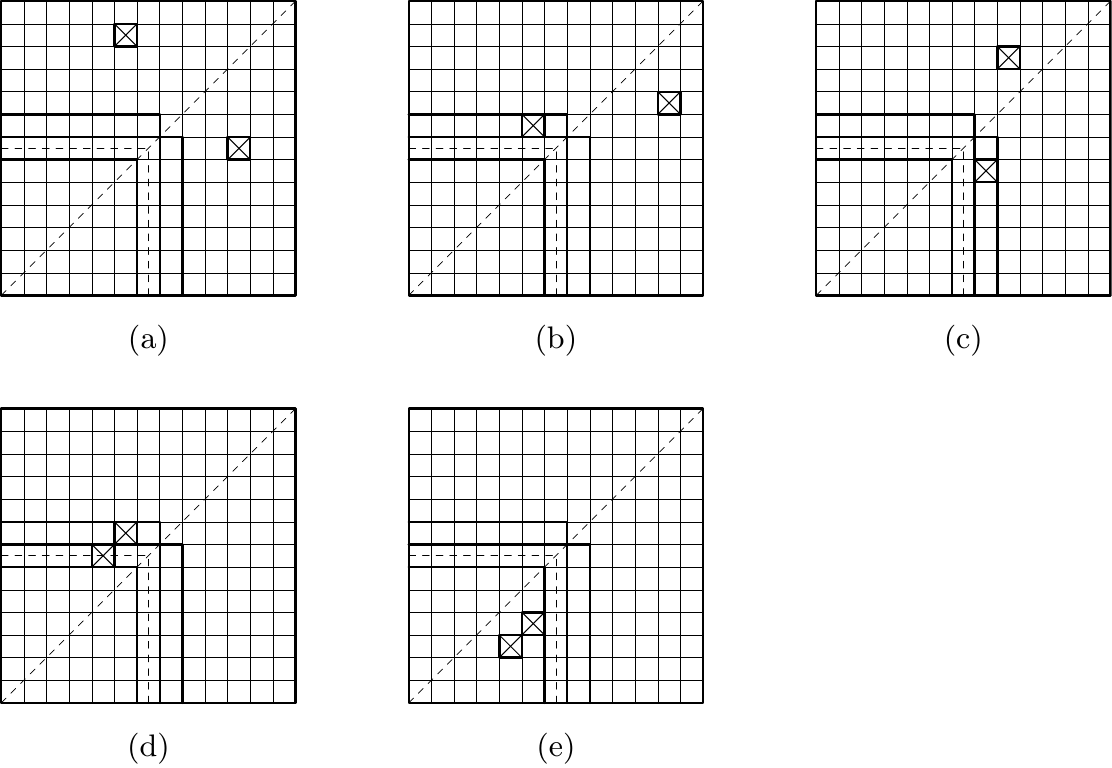}
\caption{Five examples of configurations mentioned in 
the statement (1) of Theorem \ref{theorem:thm012} 
for an even value of $\tilde{w}$ ($\tilde{w} = 12$).
(a) $\tilde{C}$ has no holes in $U \cup V \cup W$.
(b), (c) $\tilde{C}$ has no holes in $U \cup V$, has one hole in $W$ 
and the hole ($(5, 7)$ for (b) and $(7, 5)$ for (c))
is critical.  
(d) , (e) $\tilde{C}$ has a critical pair of holes 
($(4, 6)$, $(5, 7)$ for (d) and $(4, 2)$, $(5, 3)$ for (e))
in $U \cup V \cup W$.
}
\label{figure:fig047}
\end{figure}

\begin{thm}
\label{theorem:thm012}
Let $\tilde{C}$ be a configuration of size $\tilde{w}$ of 
$\abb{SH}[2]$ 
and assume that $\tilde{w} \geq 11$.
\begin{enumerate}
\item[{\rm (1)}] If one of the following is true 
then $\abb{mft}_{\abb{SH}[2]}(\tilde{C}) = 2\tilde{w} + 1$. 

\begin{enumerate}
\item[$\bullet$] $\tilde{C}$ has no holes in $U \cup V \cup W$.
\item[$\bullet$] $\tilde{C}$ has no holes in $U \cup V$, 
has one hole in $W$, and the hole is critical.
\item[$\bullet$] $\tilde{C}$ has a critical pair of holes in $U \cup V \cup W$.
\end{enumerate}
{\rm (}See the five examples of such configurations 
for an even value of $\tilde{w}$ shown in Fig. $\ref{figure:fig047}$.{\rm )}
\item[{\rm (2)}] Otherwise, 
$\abb{mft}_{\abb{SH}[2]}(\tilde{C}) = 2 \tilde{w}$.
\end{enumerate}
\end{thm}

\noindent
{\it Proof}. 

\medskip

The statement of this theorem is very simple. 
However our proof for it is by a very detailed 
and tedious case analysis.

\medskip

\noindent
Part I: Proof of the statement (1).

\medskip

By Corollary \ref{corollary:cor001} and $c_{2} = 1$ 
(Table \ref{table:tab002}), 
to prove 
$\abb{mft}_{\abb{SH}[2]}(\tilde{C}) = 2\tilde{w} + 1$ 
it is sufficient to prove 
$\abb{mft}_{\abb{SH}[2]}(\tilde{C}) \geq 2\tilde{w} + 1$.

\medskip

\noindent
(Case 1) $\tilde{C}$ has no holes in $U \cup V \cup W$.

\medskip

\noindent
(Case 1.1) There is no hole at $v_\abb{cnt} + (1, 1)$ 
(this is always true for odd $\tilde{w}$).

\medskip

\begin{figure}[htbp]
\centering
\includegraphics[scale=1.0]{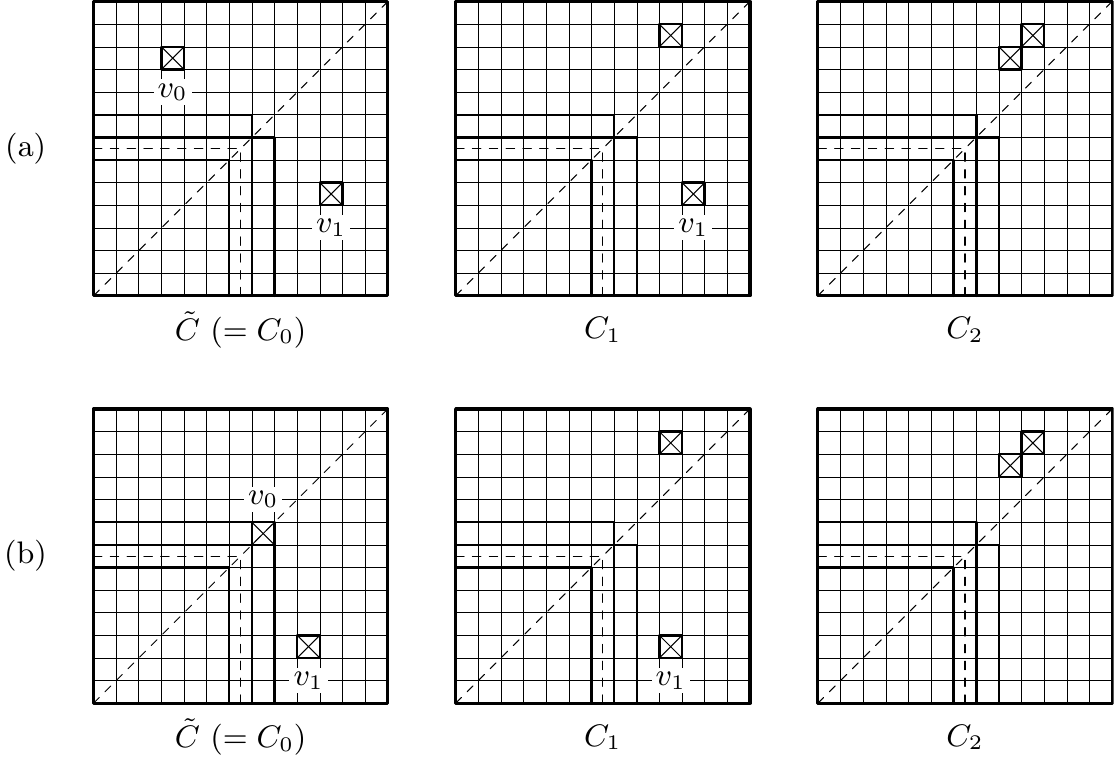}
\caption{Examples of configurations used in Case 1.1 and 
Case 1.2.}
\label{figure:fig024}
\end{figure}

In this case the two holes are in 
$\overline{H_{1}} \cup \overline{H_{2}}$.
We consider the case where a hole $v_{0}$ is in 
$H_{1} \cap \overline{H_{2}}$ 
and another hole $v_{1}$ is in 
$\overline{H_{1}} \cap H_{2}$. 
(The proofs for other cases are simpler.)
In Fig. \ref{figure:fig024} (a) we show an example of 
such $\tilde{C}$.
For this $\tilde{C}$ we define three configurations 
$C_{0}$, $C_{1}$, $C_{2}$ shown in the figure (a).

$C_{0}$ is $\tilde{C}$ itself.
$C_{1}$ is obtained from $C_{0}$ by moving $v_{0}$ to 
$(\tilde{w} - 3, \tilde{w} - 1)$.
Both of the old and the new positions of $v_{0}$ are 
in $\overline{H_{2}}$ and hence 
we have $\pi(C_{0}, H_{2}) = \pi(C_{1}, H_{2})$ 
and $C_{0} \equiv_{2 \tilde{w}}' C_{1}$ by Theorem \ref{theorem:thm007}.
$C_{2}$ is obtained from $C_{1}$ by moving $v_{1}$ to 
$(\tilde{w} - 4, \tilde{w} - 2)$.
Both of the old and the new positions of 
$v_{1}$ are in $\overline{H_{1}}$ 
because we assume $\tilde{w} \geq 11$, 
and hence 
we have $\pi(C_{1}, H_{1}) = \pi(C_{2}, H_{1})$ 
and $C_{1} \equiv_{2 \tilde{w}}' C_{2}$ by Theorem \ref{theorem:thm007}.
For the last configuration $C_{2}$, 
it has a critical pair of holes and hence 
$2 \tilde{w} + 1 \leq \max_{v \in C_{2}} T(v, C_{2})$ 
by Theorem \ref{theorem:thm013}.
Therefore we have $2 \tilde{w} + 1 \leq \abb{mft}(C_{0}) 
= \abb{mft}(\tilde{C})$ by Corollary \ref{corollary:cor000}.

For other cases too we use the same reasoning.
We define a sequence of configurations $C_{0}$, 
\ldots, $C_{n-1}$ such that $\tilde{C} = C_{0}$, 
$\pi(C_{i}, H) = \pi(C_{i+1}, H)$ 
for each $i$ ($0 \leq i \leq n-2$) and for some 
$H = H_{0}, H_{1}, H_{2}$, 
and $C_{n - 1}$ has a critical pair of holes.
Then using Theorem \ref{theorem:thm007}, Theorem \ref{theorem:thm013} and 
Corollary \ref{corollary:cor000} 
we can derive $2 \tilde{w} + 1 \leq \abb{mft}(\tilde{C})$.
Therefore, we will show only the sequence 
$C_{0}$, \ldots, $C_{n-1}$.

\medskip

\noindent
(Case 1.2) There is a hole at $v_{0} = v_\abb{cnt} + (1, 1)$ 
(this is possible only for even $\tilde{w}$).

\medskip

We consider the case where the other hole $v_{1}$ is 
in $\overline{H_{1}} \cap H_{2}$.
(The proofs for other cases are similar.)

In Fig. \ref{figure:fig024} (b) we show an example of 
such $\tilde{C}$ ($= C_{0}$) and the corresponding 
$C_{1}$, $C_{2}$.
In this case we move $v_{0}$ to 
$(\tilde{w} - 3, \tilde{w} - 1)$ and 
move $v_{1}$ to $(\tilde{w} - 4, \tilde{w} - 2)$ 
to obtain $C_{1}$ and $C_{2}$.
We have $\pi(C_{0}, H_{0}) = \pi(C_{1}, H_{0})$, 
$\pi(C_{1}, H_{1}) = \pi(C_{2}, H_{1})$ 
and $C_{2}$ has a critical pair of holes.

\noindent
(Case 2) $\tilde{C}$ has no holes in $U \cup V$ but 
has one hole in $W$ and the hole is critical.

\medskip

\noindent
(Case 2.1) There is no hole at $v_\abb{cnt} + (1, 1)$ 
(this is always true for odd $\tilde{w}$).

\medskip

\begin{figure}[htbp]
\centering
\includegraphics[scale=1.0]{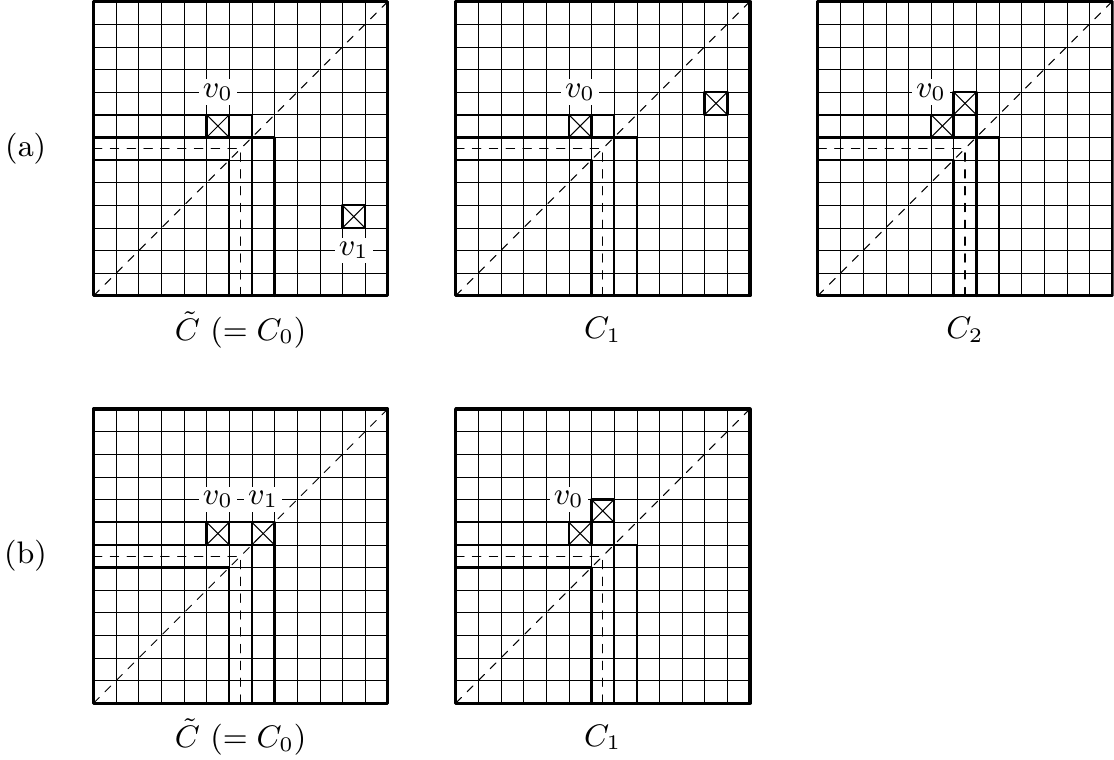}
\caption{Examples of configurations used in Case 2.1 and 
Case 2.2.}
\label{figure:fig025}
\end{figure}

In Fig. \ref{figure:fig025} (a) we show 
an example of such $\tilde{C}$ ($= C_{0}$) and 
the corresponding $C_{1}$, $C_{2}$.
We use 
$\pi(C_{0}, H_{1}) = \pi(C_{1}, H_{1})$, 
$\pi(C_{1}, H_{2}) = \pi(C_{2}, H_{2})$.

\medskip

\noindent
(Case 2.2) There is a hole at $v_\abb{cnt} + (1, 1)$ 
(this is possible only for even $\tilde{w}$).

\medskip

In Fig. \ref{figure:fig025} (b) we show 
an example of such $\tilde{C}$ ($= C_{0}$) 
and the corresponding $C_{1}$.
We use 
$\pi(C_{0}, H_{0}) = \pi(C_{1}, H_{0})$.

\medskip

\noindent
(Case 3) $\tilde{C}$ has a critical pair of holes 
in $U \cup V \cup W$.

\medskip

In this case we have 
$2 \tilde{w} + 1 \leq \max_{v \in \tilde{C}} T(v, \tilde{C}) 
\leq \abb{mft}(\tilde{C})$ by Theorem \ref{theorem:thm013}.

\medskip

\noindent
Part II: Proof of the statement (2).

\medskip

In the statement of the theorem we include the assumption 
``$\tilde{w} \geq 11$.''
In Part I this assumption was essentially used.
However in the proofs in Part II we need only a weaker assumption 
``$\tilde{w} \geq 5$.''  
(We need it when we use Theorem \ref{theorem:thm011}.)
Therefore, as example configurations $\tilde{C}$ we may use 
configurations of sizes $\tilde{w}$ smaller than $11$ to save space.

By Theorem \ref{theorem:thm001}, to prove 
$\abb{mft}_{\abb{SH}[2]}(\tilde{C}) = 2\tilde{w}$ 
it is sufficient to prove 
$\abb{mft}_{\abb{SH}[2]}(\tilde{C})$ $\leq 2\tilde{w}$.
We prove this by constructing a partial solution $A$ 
that fires $\tilde{C}$ at time $2\tilde{w}$ 
using the idea explained previously with $k = 2$, $s = 0$.

Let $\#U$ denote the number of holes of $\tilde{C}$ 
in $U$, and similarly for $\#V$, $\#W$, $\#X$.
For a configuration $\tilde{C}$, 
we call the 4-tuple $(\#U, \#V, \#W, \#X)$ 
the \textit{type} of $\tilde{C}$.
A type is a 4-tuple $(a, b, c, d)$ of nonnegative integers 
such that $a + b + c + d = 2$.
Hence there are ten types $(2, 0, 0, 0)$, $(1, 1, 0, 0)$, 
\ldots, $(0, 0, 0, 2)$.

Using this notation we can represent the three conditions in 
the statement (1) of the theorem as follows:
\begin{enumerate}
\item[$\bullet$] The type of $\tilde{C}$ is $(0, 0, 0, 2)$.
\item[$\bullet$] The type of $\tilde{C}$ is $(0, 0, 1, 1)$ 
and the unique hole in $W$ is critical.
\item[$\bullet$] The type of $\tilde{C}$ is of the form 
$(a, b, c, 0)$ and $\tilde{C}$ has a critical pair in 
$U \cup V \cup W$.
\end{enumerate}
Therefore, the conditions in the statement (2) 
(that is, ``otherwise'') are as follow:
\begin{enumerate}
\item[$\bullet$] The type of $\tilde{C}$ is $(0, 0, 1, 1)$ 
and the unique hole in $W$ is not critical.
\item[$\bullet$] The type of $\tilde{C}$ is $(0, 1, 1, 0)$ 
and $\tilde{C}$ has no critical pairs in $V \cup W$.
\item[$\bullet$] The type of $\tilde{C}$ is $(1, 1, 0, 0)$ 
and $\tilde{C}$ has no critical pairs in $U \cup V$.
\item[$\bullet$] The type of $\tilde{C}$ is $(2, 0, 0, 0)$ 
and $\tilde{C}$ has no critical pairs in $U$.
\item[$\bullet$] The type of $\tilde{C}$ is one of 
$(0, 1, 0, 1)$, $(1, 0, 0, 1)$, $(0, 0, 2, 0)$, 
$(1, 0, 1, 0)$, $(0, 2, 0,$ $0)$.
(Note that in these cases $\tilde{C}$ has no critical pairs.)
\end{enumerate}
We merge these conditions to the following conditions:
\begin{enumerate}
\item[$\bullet$] The type of $\tilde{C}$ is of the form 
$(a, 0, c, d)$ ($a \geq 1$) and $\tilde{C}$ has no critical pairs 
in $U$.
\item[$\bullet$] The type of $\tilde{C}$ is of the form 
$(0, 0, c, d)$ ($c \geq 1$) and 
``$c = 1$ and the unique hole in $W$ is critical'' is false.
\item[$\bullet$] The type of $\tilde{C}$ is of the form 
$(0, b, c, d)$ ($b \geq 1$) and $\tilde{C}$ has no critical pair in 
$V \cup W$.
\item[$\bullet$] The type of $\tilde{C}$ is $(1, 1, 0, 0)$ 
and $\tilde{C}$ has no critical pairs in $U \cup V$.
\end{enumerate}

\medskip

\noindent
(Case 1) The type of $\tilde{C}$ is of the form $(a, 0, c, d)$ ($a \geq 1$) 
and $\tilde{C}$ has no critical pairs in $U$.

\medskip

We construct a partial solution $A$ 
that fires $\tilde{C}$ at time $2\tilde{w}$ using 
$\tilde{\pi} = \pi(\tilde{C}, U \cup V)$, 
$\pi_{0,0} = \tilde{\pi}$, $v_{0,0} = v_\abb{cnt}$, $r_{0, 0} = 0$.
The firing rule is $(\abb{W}_{0} \vee \abb{W}_{1}) 
\wedge \abb{M}_{0,0}$.
More precisely, a node in a configuration $C$ fires at a time 
if and only if the current time is $2\tilde{w}$ and 
it has received at least one of the two size check messages 
$\abb{W}_{0}$, $\abb{W}_{1}$ and also the message 
$\abb{M}_{0, 0}$ before or at that time.

It is obvious that $\tilde{C}$ has $\tilde{\pi}$ 
because $\tilde{\pi} = \pi(\tilde{C}, U \cup V)$.
We prove that all of the five statements (C1), ..., (C5) mentioned 
in the explanation of our idea for constructing partial solutions 
are true. 
The condition (C2) is true by our definition of 
$\tilde{\pi}$, $\pi_{0, 0}$.
To prove (C1), (C3), (C4), (C5) we assume that $C$ is a configuration 
of size $\tilde{w}$.

The proof of (C1) is as follows.
Let $(a, 0, c, d)$ ($a \geq 1$) be the type of $\tilde{C}$.
Suppose that $C$ has $\tilde{\pi}$.
Then, by the definition $\tilde{\pi} = \pi(\tilde{C}, U \cup V)$, 
for any position $v$ in $U \cup V$ whether $v$ is a node or a hole 
is the same in $\tilde{C}$ and $C$.
Therefore the type of $C$ is of the form $(a, 0, c', d')$.
If $a = 1$ then $C$ has one hole in $U$ and no holes in $V$.
Therefore $C$ has an isolated hole and consequently $C$ has 
no critical pairs.
If $a = 2$ then $C$ has two holes in $U$ but the pair of these two holes 
is not a critical pair because $\tilde{C}$ has 
no critical pairs in $U$.
Therefore $C$ has no critical pairs.
In both cases $C$ has no critical pairs and hence 
$\max_{v \in C} T(v, C) \leq 2 \tilde{w}$ by Theorem 
\ref{theorem:thm013}.
This shows (C1).  
(By an \textit{isolated hole} we mean a hole $v$ such that 
any position that is adjacent to $v$ or touches $v$ with corners 
is a node.  
An isolated hole cannot be a hole of a critical pair.)

Next we explain how to generate the message $\abb{M}_{0, 0}$ 
so that (C3), (C4) are true.

Suppose that $C$ has $\pi_{0, 0}$ and copies of $A$ are placed 
in $C$ as nodes.  
Then $C$ has no holes in $V$.
Therefore the part of $C$ in $U \cup V$ satisfies the definition of 
configurations of $\abb{SH}[k]$ ($k \leq 2$).  
We can apply Theorem \ref{theorem:thm004} to this 
configuration-like region and show the following for any position $v$ 
in $U \cup V$: 
(1) if $v$ is a node that is not in maximal barriers then it is 
on a path from $v_\abb{gen}$ to $v_{0,0}$ 
of the MH distance length, and 
(2) if $v$ is a node or a hole in a maximal barrier then it is 
adjacent to a node $v'$ that is on such a path.
(To prove (2) we essentially use the fact that 
maximal barriers of $\abb{SH}[2]$ are of the forms 
$S_{1}, \ldots, S_{5}$ shown in Fig. \ref{figure:fig010}.  
We cannot prove (2) for general $\abb{SH}[k]$.)

Using the above fact we can design the finite automaton $A$ so that $A$ generates 
a finite number of signals such that 
\begin{enumerate}
\item[$\bullet$] each signal starts at $v_{\abb{gen}}$ at time $0$ 
and proceeds along one specific path from $v_{\abb{gen}}$ to $v_{0, 0}$ 
of the MH distance length with speed $1$, 
\item[$\bullet$] if $C$ has $\pi_{0, 0}$ then all of the signals arrive at 
$v_{0, 0}$, and 
\item[$\bullet$] if $C$ has not $\pi_{0, 0}$ then at least one of the signals 
vanishes and fails to arrive at $v_{0, 0}$.
\end{enumerate}
We define the rule to generate the message $\abb{M}_{0, 0}$ as follows: 
$\abb{M}_{0, 0}$ is generated at a node at a time if and only if the time is $2 \tilde{w}$ and 
all of the above mentioned signals arrive at the node at the time.
Then we can easily prove (C3), (C4) using 
$\abb{d}_{\abb{MH}}(v_{\abb{gen}}, v_{0, 0}) = 2 \tilde{w}$.

\begin{figure}[htbp]
\centering
\includegraphics[scale=1.0]{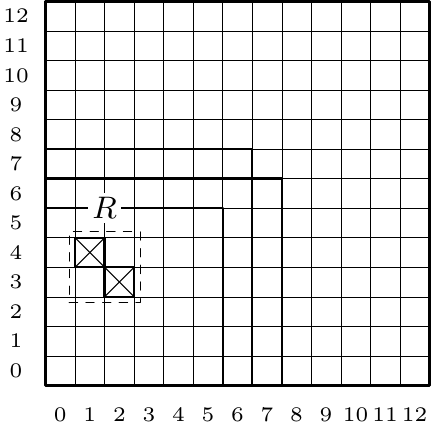}
\caption{An example configuration $\tilde{C}$.}
\label{figure:fig026}
\end{figure}

We explain what signals to use for generating $\abb{M}_{0, 0}$ using 
the configuration $\tilde{C}$ shown in Fig. \ref{figure:fig026} 
as an example.
This is a configuration of size $12$ of type $(2, 0, 0, 0)$ 
and the region $U \cup V$ has one maximal barrier $R$ consisting of 
the four positions 
$(1, 3)$ (a node), 
$(1, 4)$ (a hole), 
$(2, 3)$ (a hole), 
$(2, 4)$ (a node).
The barrier is of the form $S_{4}$ in Fig. \ref{figure:fig010}.

\begin{figure}[htbp]
\centering
\includegraphics[scale=1.0]{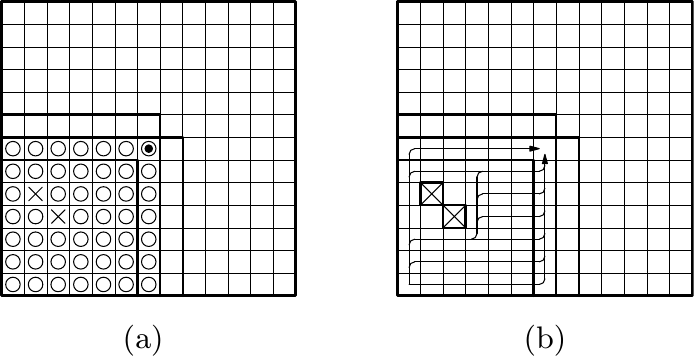}
\caption{(a) The pattern 
$\pi_{0, 0}$ ($ = \tilde{\pi} = 
\pi(\tilde{C}, U \cup V)$) 
and the position $v_{0,0}$ ($= v_\abb{cnt}$).
(b) The eight paths along which the eight signals travel.
}
\label{figure:fig067}
\end{figure}

Fig. \ref{figure:fig067} (a) shows the pattern 
$\pi_{0, 0}$ ($= \tilde{\pi} = \pi(\tilde{C}, U \cup V)$) 
and the position $v_{0, 0}$ ($= v_\abb{cnt}$).
(From now on, to represent a pattern $\pi_{i, j}$ by a figure, 
we write a circle or a cross at 
a position if the value of the pattern at the position is 
``$\abb{N}$'' (a node) or ``$\abb{H}$'' (a hole) respectively.
We also show the position $v_{i, j}$ by a bullet.)

In Fig. \ref{figure:fig067} (b) we show eight paths 
from $v_{\abb{gen}}$ to $v_{0, 0}$ by arrow lines.
We selected these paths so that 
(1) for any position in $U \cup V - R$ 
there is at least one path that passes it, 
(2) no paths enter $R$, and 
(3) each path goes only to the north and to the east 
so that it is of the MH distance length.

For each of these path a signal proceeds from $v_{\abb{gen}}$ 
to $v_{0, 0}$ along it.
If a hole is on the path the signal vanished at the position.
There are $8$ positions that are not in $R$ but are adjacent 
to positions in $R$.
They are $(0, 3)$, $(0, 4)$, $(1, 2)$ $(2, 2)$, $(1, 5)$, $(2, 5)$, 
$(3, 3)$, $(3, 4)$.
If a signal is at one of these positions it checks that 
the position in $R$ adjacent to itself is a node or a hole 
according as the position in the pattern $\pi_{0, 0}$ 
is a node or a hole.
If the check fails the signal vanishes.

It is evident that if $C$ has $\pi_{0, 0}$ then no signals vanish.
If $C$ has not $\pi_{0, 0}$ then either (1) there is at least one hole in 
$U \cup V - \{(x, y) \mid \text{$x = 0$ or $y = 0$} \} - R$ or 
(2) the part $R$ is not a barrier of the form $S_{4}$.  
In both of the cases at least one signal vanishes.

Finally we prove the statement (C5).  
Suppose that $C$ has $\tilde{\pi}$.  
Let $\tilde{v}_{0} = v_{0, 0} + (1, 0)$, 
$\tilde{v}_{1} = v_{0, 0} + (0, 1)$, 
$\tilde{v}_{2} = v_{0, 0} + (-1, 0)$, 
$\tilde{v}_{3} = v_{0, 0} + (0, -1)$ 
be the four positions that are adjacent to $v_{0, 0}$.
Both of $v_{2}$, $v_{3}$ are nodes in $C$ because 
they are in $V$.
At least one of $v_{0}$, $v_{1}$ is a node in $C$ 
because there is at least 
one holes in $U$ and there are exactly two holes.
Therefore, we have none of the four exceptions when we 
use Theorem \ref{theorem:thm011} for $C$ with 
$v = v_{0, 0}$.
Hence by this theorem we have 
$\abb{d}_{\abb{MH}}(v_{\abb{gen}}, v_{0, 0}) + 
\abb{d}_{C}(v_{0, 0}, v') \leq 2 \tilde{w}$ 
for any node $v'$ in $C$.
Therefore (C5) is true.

This completes the proof of Case 1 of Part II.
In the remainder of the proof of Part II we have many cases.  
However the proofs for these cases are similar to that of Case 1.
Hence, from now on we only show 
\begin{enumerate}
\item[$\bullet$] an example configuration $\tilde{C}$, 
\item[$\bullet$] the pattern $\tilde{\pi}$ for that $\tilde{C}$, 
\item[$\bullet$] the patterns $\pi_{i,j}$ and the positions $v_{i,j}$ 
for that $\tilde{C}$, and 
\item[$\bullet$]  the firing rule
\end{enumerate}
to explain an idea to construct a partial solution.
The value $r_{i, j}$ is $0$ for all cases.

In all cases we define $\tilde{\pi}$ to be $\pi(\tilde{C}, Z)$ 
for some $Z \subseteq S_{\tilde{w}}$ 
and hence $\tilde{C}$ has $\tilde{\pi}$.
We give proofs for the statements (C1), \ldots, (C5) only when it is necessary.
Usually we can prove them as follows.
For (C1), we can easily prove that 
if $C$ has $\tilde{\pi}$ then $C$ has no critical pairs of holes.
This proves (C1) by Theorem \ref{theorem:thm013}.
The ``only if'' part of (C2) is obvious.
We can prove the ``if'' part of (C2) 
using our assumption that each configuration has exactly two holes.
The design of signals to generate messages $\abb{M}_{i, j}$ 
is obvious and the proofs of (C3), (C4) follow from the design of 
the signals.

In the proof of (C5) we use a value $D_{i, j}(C, v)$ defined by 
\begin{equation}
D_{i, j}(C, v) = \abb{d}_{\abb{MH}}(v_{\abb{gen}}, v_{i, j}) + 
                 \abb{d}_{C}(v_{i, j}, v).
\label{equation:eq024}
\end{equation}
Suppose that the firing rule is 
$(\abb{W}_{0} \vee \abb{W}_{1}) \wedge 
((\abb{M}_{0} \wedge \ldots \wedge \abb{M}_{0, m_{0} - 1}) \vee \ldots \vee 
(\abb{M}_{n - 1, 0} \wedge \ldots \wedge \abb{M}_{n - 1, m_{n - 1} - 1}))$.
Then, for (C5) we must prove that if a configuration $C$ of size $\tilde{w}$ has 
$\tilde{\pi}$ then for any node $v$ in $C$ there is $i$ such that 
$\max_{0 \leq j \leq m_{i} - 1} D_{i, j}(C, v) \leq 2 \tilde{w}$.
Usually we can prove this by Theorem \ref{theorem:thm011}.
However, in some cases (for example, in cases where $v_{i, j}$ is not 
in $U \cup V$) we cannot use that theorem and we need 
ad hoc analyses of the value $D_{i, j}(C, v)$.

\medskip

\noindent
(Case 2) The type of $\tilde{C}$ is of the form 
$(0, 0, c, d)$ ($c \geq 1$) and 
``$c = 1$ and the unique hole in $W$ is critical'' is false.

\medskip

\noindent
(Case 2.1) In $\tilde{C}$, at least one of 
$v_{\abb{cnt}} + (0, 1)$, 
$v_{\abb{cnt}} + (1, 0)$ is a node.

\medskip

\noindent
(Case 2.1.1) $\tilde{w}$ is even.

\medskip

As the type of a configuration $C$ 
we use $(a, b, (c_{0}, c_{1}), d)$ instead of $(a, b, c, d)$.
Here $c_{0}$ and $c_{1}$ are the number of critical holes of $C$ in $W$ 
and the number of noncritical holes of $C$ in $W$, respectively.
The possible pairs $(c_{0}, c_{1})$ are 
$(0, 1)$, $(1, 0)$, $(0, 2)$, $(1, 1)$, $(2, 0)$.
However at present we assume that 
``$c = 1$ and the unique hole in $W$ is critical'' is false.
Hence $(1, 0)$ is excluded.
Therefore the four pairs $(0, 1)$, $(0, 2)$, $(1, 1)$, $(2, 0)$ 
are possible.

\begin{figure}[htbp]
\centering
\includegraphics[scale=1.0]{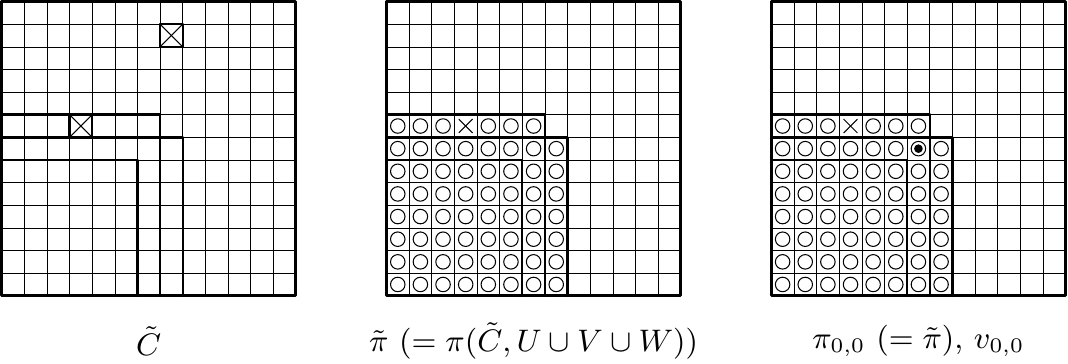}
\caption{An example configuration $\tilde{C}$ and 
$\tilde{\pi}$, $\pi_{0, 0}$, $v_{0, 0}$ for 
Case 2.1.1.}
\label{figure:fig069}
\end{figure}

Suppose that the type of $\tilde{C}$ is $(0, 0, (c_{0}, c_{1}), d)$ 
and $(c_{0}, c_{1})$ is one of $(0, 1)$, $(0, 2)$, $(1, 1)$, $(2, 0)$.
Then we can construct a partial solution $A$ that fires $\tilde{C}$ 
at $2 \tilde{w}$ by 
$\tilde{\pi} = \pi(\tilde{C}, U \cup V \cup W)$, 
$\pi_{0, 0} = \tilde{\pi}$, $v_{0, 0} = v_{\abb{cnt}}$ 
and the firing rule $(\abb{W}_{0} \vee \abb{W}_{1}) 
\wedge \abb{M}_{0, 0}$.
In Fig. \ref{figure:fig069} we show an example configuration 
$\tilde{C}$ of size $12$ and $\tilde{\pi}$, $\pi_{0, 0}$, $v_{0, 0}$ 
for this $\tilde{C}$.
$\tilde{C}$ has one hole $(3, 7)$ in $W$ and it is a noncritical hole.
Therefore the type of $\tilde{C}$ is $(0, 0, (0, 1), 1)$.

(C1) is true as follows.
If $(c_{0}, c_{1})$ is one of $(0, 1)$, $(0, 2)$, $(1, 1)$ then 
$C$ has at least one noncritical hole in $W$.
If $(c_{0}, c_{1})$ is $(2, 0)$ then $C$ has two critical holes in $W$ 
(that is, $v_{\abb{cnt}} + (-1, 1)$, $v_{\abb{cnt}} + (1, -1)$)
but they cannot constitute a critical pair.
Therefore, in both cases $C$ has no critical pairs and hence 
$\max_{v \in C} T(v, C) \leq 2 \tilde{w}$ by Theorem \ref{theorem:thm013}.

It is easy to generate a message $\abb{M}_{0, 0}$ so that 
(C3), (C4) are true.
A finite number of signals like those shown in Fig. \ref{figure:fig067} (b) 
can check that $C$ has no holes in $U \cup V$.
Let $P_{0}$, $P_{1}$ be the paths from $v_{\abb{gen}}$ to $v_{0, 0}$ 
via $(0, \lfloor \tilde{w} / 2 \rfloor)$ or 
via $(\lfloor \tilde{w} / 2 \rfloor, 0)$ respectively 
of the MH distance length.
Then each position in $W$ is adjacent to a position in $P_{0}$ or $P_{1}$.
Therefore two signals that travel on these two paths can check 
that the distribution of holes in $W$ in $C$ is consistent 
with that specified by $\tilde{\pi}$. 
Here we essentially use our assumption of Case 2.1.1 
that $\tilde{w}$ is even and 
hence the position $v_{\abb{cnt}} + (1, 1)$ is not in $W$.

The proof of (C5) is as follows.
$C$ has no holes at $v_{0, 0} - (0, 1)$, $v_{0, 0} - (1, 0)$ 
because they are in $V$.
$C$ has a node at at least one of $v_{0, 0} + (0, 1)$, 
$v_{0, 0} + (1, 0)$ by our assumption of Case 2.1.
Therefore, we have none of the four exceptions when we apply 
Theorem \ref{theorem:thm011} to $C$ with $v = v_{0, 0}$ and 
this shows (C5).

\medskip

\noindent
(Case 2.1.2) $\tilde{w}$ is odd.

\medskip 

When $\tilde{w}$ is odd, $W$ contains the position 
$v_{\abb{cnt}} + (1, 1)$ 
(see Fig. \ref{figure:fig066}) and this position is not adjacent 
to any position in $U \cup V$.
This makes the design of the partial solution a little complicated.

By $W'$ we denote the set $W - \{v_{\abb{cnt}} + (1, 1)\}$.
As the type of a configuration $C$ we use 
$(a, b, (c_{0}, c_{1}, c_{2}), d)$ instead of 
$(a, b, c, d)$.
Here, $c_{0}$ is the number of critical holes in $W'$, 
$c_{1}$ is the number of noncritical holes in $W'$, 
and $c_{2}$ is the number of noncritical holes in 
the set $\{v_{\abb{cnt}} + (1, 1)\}$ 
(that is, $c_{2}$ is $0$ or $1$ according as the position 
$v_{\abb{cnt}} + (1, 1)$ is a node or a hole).

Suppose that the type of $\tilde{C}$ is $(0, 0, (c_{0}, c_{1}, c_{2}), d)$.
We have $0 \leq c_{0} \leq 2$, $0 \leq c_{1} \leq 2$, $0 \leq c_{2} \leq 1$, 
$1 \leq c_{0} + c_{1} + c_{2} \leq 2$.
The possible triples $(c_{0}, c_{1}, c_{2})$ are 
$(0, 0, 1)$, 
$(0, 1, 0)$, 
$(1, 0, 0)$, 
$(0, 1, 1)$, 
$(1, 0, 1)$, 
$(0, 2, 0)$, 
$(1, 1, 0)$, 
$(2, 0, 0)$.
However, $(1, 0, 0)$ is excluded 
by the same reason as in Case 2.1.1.
Therefore, the seven triples 
$(0, 0, 1)$, 
$(0, 1, 0)$, 
$(0, 1, 1)$, 
$(1, 0, 1)$, 
$(0, 2, 0)$, 
$(1, 1, 0)$, 
$(2, 0, 0)$ 
are possible.

\medskip

\noindent
(Case 2.1.2.1) $(c_{0}, c_{1}, c_{2})$ is one of 
$(0, 1, 0)$, 
$(0, 1, 1)$, 
$(0, 2, 0)$, 
$(1, 1, 0)$, 
$(2, 0, 0)$. 

\medskip

\begin{figure}[htbp]
\centering
\includegraphics[scale=1.0]{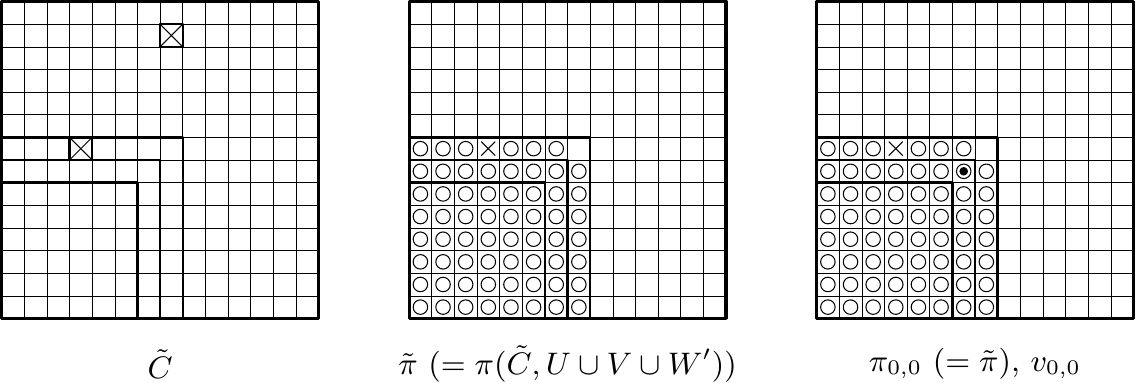}
\caption{An example configuration $\tilde{C}$ and 
$\tilde{\pi}$, $\pi_{0, 0}$, $v_{0, 0}$ for 
Case 2.1.2.1.}
\label{figure:fig070}
\end{figure}

We construct a partial solution $A$ that fires $\tilde{C}$ 
at $2 \tilde{w}$ by 
$\tilde{\pi} = \pi(\tilde{C}, U \cup V \cup W')$, 
$\pi_{0, 0} = \tilde{\pi}$, 
$v_{0, 0} = v_{\abb{cnt}}$ and the firing rule 
$(\abb{W}_{0} \vee \abb{W}_{1}) \wedge \abb{M}_{0, 0}$.
In Fig. \ref{figure:fig070} we show an example configuration $\tilde{C}$ 
of size $13$ and $\tilde{\pi}$, $\pi_{0, 0}$, $v_{0, 0}$ for this 
$\tilde{C}$.
The type of $\tilde{C}$ is $(0, 0, (0, 1, 0), 1)$.  
The hole at $(3, 7)$ is the unique hole in $W'$ and it is not a critical hole.

(C1) is shown as follows.
Let $(0, 0, (c_{0}, c_{1}, c_{2}), d)$ be the type of 
$\tilde{C}$.  
Suppose that $C$ of size $\tilde{w}$ has $\tilde{\pi}$.
If $(c_{0}, c_{1}, c_{2})$ is one of 
$(0, 1, 0)$, $(0, 1, 1)$, $(0, 2, 0)$, $(1, 1, 0)$ 
then $C$ has at least one noncritical hole in $W'$ and hence 
$C$ has no critical pairs.
If $(c_{0}, c_{1}, c_{2})$ is $(2, 0, 0)$ then 
$C$ has two critical holes in $W'$ and 
hence $C$ has no critical pairs.
Therefore, in both cases we have (C1) by Theorem \ref{theorem:thm013}.

The generation of the message $\abb{M}_{0, 0}$ 
and the proof of (C5) are completely the same as for Case 2.1.1.

Now there remain two cases $(c_{0}, c_{1}, c_{2}) = (1, 0, 1), (0, 0, 1)$.
We already know that if $(c_{0}, c_{1}, c_{2}) = (1, 0, 0), (0, 0, 0)$ 
then $\abb{mft}(\tilde{C}) \geq 2 \tilde{w} + 1$ (Part I of the proof).
Therefore, in any partial solution for these two cases some signal must 
check that $v_{\abb{cnt}} + (1, 1)$ is really a hole and hence $v_{i, j}$ 
must be out of $U \cup V$ for some $i, j$.

\medskip

\noindent
(Case 2.1.2.2) $(c_{0}, c_{1}, c_{2})$ is $(1, 0, 1)$.

\medskip

\begin{figure}[htbp]
\centering
\includegraphics[scale=1.0]{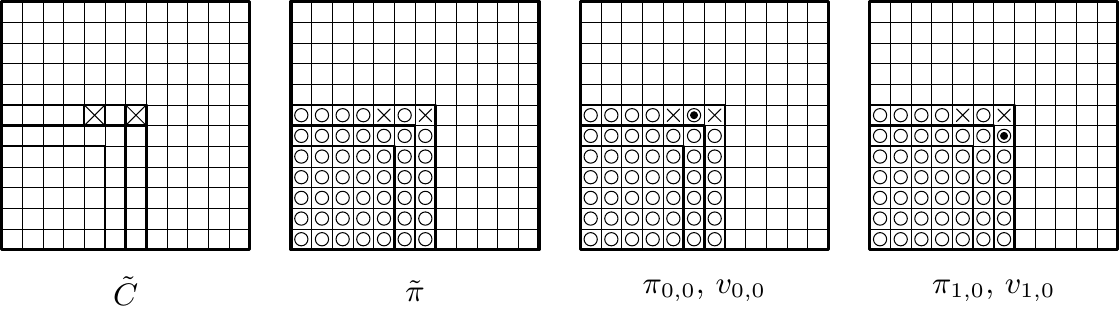}
\caption{An example configuration $\tilde{C}$ and 
$\tilde{\pi}$, $\pi_{0, 0}$, $v_{0, 0}$, $\pi_{1, 0}$, 
$v_{1, 0}$ for $\tilde{C}$ 
for Case 2.1.2.2.}
\label{figure:fig071}
\end{figure}

We construct a partial solution $A$ by 
$\tilde{\pi} = \pi(\tilde{C}, U \cup V \cup W)$, 
$\pi_{0, 0} = \pi_{1, 0} = \tilde{\pi}$, 
$v_{0, 0} = v_{\abb{cnt}} + (0, 1)$, 
$v_{1, 0} = v_{\abb{cnt}} + (1, 0)$ 
and the firing rule 
$(\abb{W}_{0} \vee \abb{W}_{1}) \wedge 
(\abb{M}_{0, 0} \vee \abb{M}_{1, 0})$.
In Fig. \ref{figure:fig071} 
we show $\tilde{\pi}$, $\pi_{i, j}$, $v_{i, j}$ for 
an example configuration $\tilde{C}$ of size $11$.

(C1) is obvious because $\tilde{\pi}$ has 
one critical hole ($v_{\abb{cnt}} + (-1, 1)$ or $v_{\abb{cnt}} + (1, -1)$) 
and one noncritical hole ($v_{\abb{cnt}} + (1, 1)$) and hence 
if $C$ has $\tilde{\pi}$ then $C$ has no critical pairs of holes.

To generate the message $\abb{M}_{0, 0}$ satisfying (C3), (C4) 
we use the fact that each position in $W$ (including $v_{\abb{cnt}} + (1, 1)$) 
has an adjacent position in one of the following two paths: 
(1) $v_{\abb{gen}}$ $\rightarrow$ $(0, \lfloor \tilde{w} / 2 \rfloor)$ 
$\rightarrow$ $v_{\abb{cnt}}$ $\rightarrow$ $v_{0, 0}$, 
(2) $v_{\abb{gen}}$ $\rightarrow$ $(\lfloor \tilde{w} / 2 \rfloor, 0)$ 
$\rightarrow$ $v_{\abb{cnt}}$ $\rightarrow$ $v_{0, 0}$.  
The idea for generating $\abb{M}_{1, 0}$ is similar.  
(The destinations of the two paths are $v_{1, 0}$ instead of $v_{0, 0}$.)

(C5) is the only step we need a proof.
For the proof we cannot use Theorem \ref{theorem:thm011} because 
$v_{0, 0}$, $v_{1, 0}$ are not in $U \cup V$.
The firing rule is $(\abb{W}_{0} \vee \abb{W}_{1}) \wedge 
(\abb{M}_{0, 0} \vee \abb{M}_{1, 0})$.
Therefore we must prove that if $C$ 
has $\tilde{\pi}$ then for any $v \in C$ 
either $D_{0, 0}(C, v) \leq 2 \tilde{w}$ or 
$D_{1, 0}(C, v) \leq 2 \tilde{w}$.  
(See \eqref{equation:eq024} for the definition of $D_{i, j}(C, v)$.)
However, if $C$ has $\tilde{\pi}$ then $C = \tilde{C}$ 
because $\tilde{\pi}$ has two holes.
Therefore the value of the two argument function $D_{i, j}(C, v)$ depends only on $v$. 
Using this, for example we can determine the value $D_{0, 0}(C, (\tilde{w}, 0))$ by 
$D_{0, 0}(C, (\tilde{w}, 0)) = 
\lfloor \tilde{w} / 2 \rfloor + 
(\lfloor \tilde{w} / 2 \rfloor + 1) + 
(\lfloor \tilde{w} / 2 \rfloor + 1) + 
(\tilde{w} - \lfloor \tilde{w} / 2 \rfloor) 
= 2 \tilde{w} + 1$.
By similar elementary calculation we have the following conclusion.  
\begin{enumerate}
\item[(a1)] $D_{0, 0}(v) \leq 2 \tilde{w}$ except for $v = (\tilde{w}, 0)$ and 
\item[(a2)] $D_{1, 0}(v) \leq 2 \tilde{w}$ except for $v = (0, \tilde{w})$.
\end{enumerate}
This shows (C5).

\medskip

\noindent
(Case 2.1.2.3) $(c_{0}, c_{1}, c_{2})$ is $(0, 0, 1)$.

\medskip

\begin{figure}[htbp]
\centering
\includegraphics[scale=1.0]{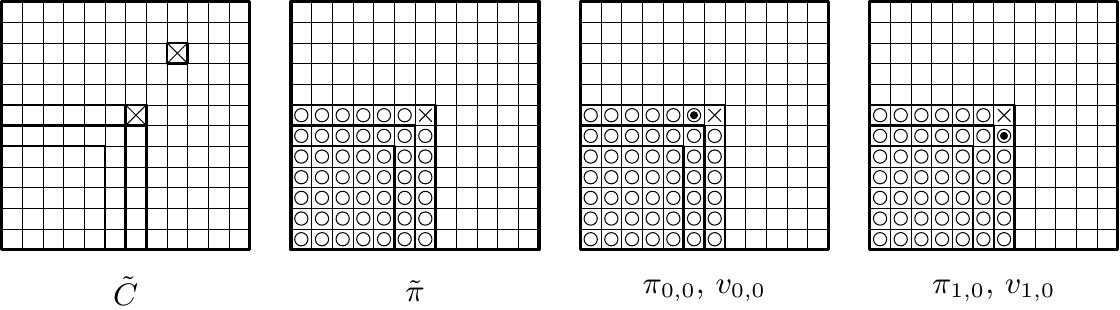}
\caption{An example configuration $\tilde{C}$ and 
$\tilde{\pi}$, $\pi_{0, 0}$, $v_{0, 0}$, $\pi_{1, 0}$, 
$v_{1, 0}$ for $\tilde{C}$ 
for Case 2.1.2.3.}
\label{figure:fig072}
\end{figure}

We construct a partial solution $A$ by 
$\tilde{\pi} = \pi(\tilde{C}, U \cup V \cup W)$, 
$\pi_{0, 0} = \pi_{1, 0} = \tilde{\pi}$, 
$v_{0, 0} = v_{\abb{cnt}} + (0, 1)$, 
$v_{1, 0} = v_{\abb{cnt}} + (1, 0)$ and the firing rule 
$(\abb{W}_{0} \vee \abb{W}_{1}) 
\wedge (\abb{M}_{0, 0} \vee \abb{M}_{1, 0})$.
We show an example in Fig. \ref{figure:fig072}.

\begin{figure}[htbp]
\centering
\includegraphics[scale=1.0]{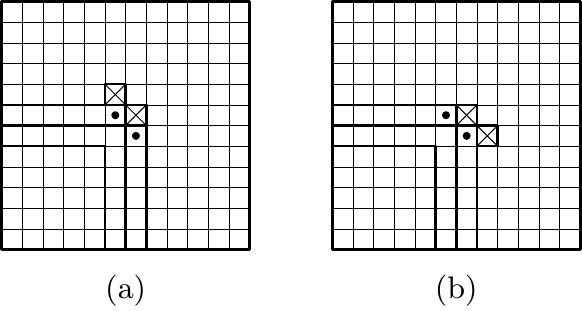}
\caption{Two exceptional cases in the proof of (C6) of Case 2.1.2.3.}
\label{figure:fig073}
\end{figure}

For (C5) we must prove that 
if a configuration $C$ has $\tilde{\pi}$ 
then for any $v \in C$ either 
$D_{0, 0}(C, v) \leq 2 \tilde{w}$ or 
$D_{1, 0}(C, v) \leq 2 \tilde{w}$.
In this case $D_{i, 0}(C, v)$ depends on $C$.
If $C$ has $\tilde{\pi}$ then $C$ has one hole 
at $v_{\abb{cnt}} + (1, 1)$ and another hole at another position 
in $S_{w} - U \cup V \cup W$ which we will denote by $v_{C}$.  
We can determine the value $D_{i, 0}(C, v)$ for each of the five possible relative 
relations between $v_{C}$ and $v_{i, 0}$: 
(1) $v_{C}$ is northwest of $v_{i, 0}$, 
(2) $v_{C}$ is north of $v_{i, 0}$, 
(3) $v_{C}$ is northeast of $v_{i, 0}$, 
(4) $v_{C}$ is east of $v_{i, 0}$, 
(3) $v_{C}$ is southeast of $v_{i, 0}$.
Such analysis gives the same conclusion (a1), (a2) of Case 2.1.2.2 
except two cases.
The first exceptional case is when $v_{C} = v_{\abb{cnt}} + (0, 2)$ 
(Fig. \ref{figure:fig073} (a)) and we have the following conclusion.
\begin{enumerate}
\item[(b1)] $D_{0, 0}(C, v) \leq 2 \tilde{w}$ except for $v = (\tilde{w}, 0), 
(\tilde{w} - 1, \tilde{w}), (\tilde{w}, \tilde{w} - 1), 
(\tilde{w}, \tilde{w})$ and 
\item[(b2)] $D_{1, 0}(C, v) \leq 2 \tilde{w}$ except for $v = (0, \tilde{w})$.
\end{enumerate}
The second exceptional case is when $v_{C} = v_{\abb{cnt}} + (2, 0)$ 
(Fig. \ref{figure:fig073} (b)) and we have the following conclusion.
\begin{enumerate}
\item[(c1)] $D_{0, 0}(C, v) \leq 2 \tilde{w}$ except for $v = (\tilde{w}, 0)$ and 
\item[(c2)] $D_{1, 0}(C, v) \leq 2 \tilde{w}$ except for $v = (0, \tilde{w}), 
(\tilde{w} - 1, \tilde{w}), (\tilde{w}, \tilde{w} - 1), 
(\tilde{w}, \tilde{w})$.
\end{enumerate}
All of the conclusions (a1), (a2), (b1), (b2), (c1), (c2) show (C5).  

\medskip

\noindent
(Case 2.2) $\tilde{C}$ has the two holes at $v_{\abb{cnt}} + (0, 1)$, 
$v_{\abb{cnt}} + (1, 0)$.

\medskip

\begin{figure}[htbp]
\centering
\includegraphics[scale=1.0]{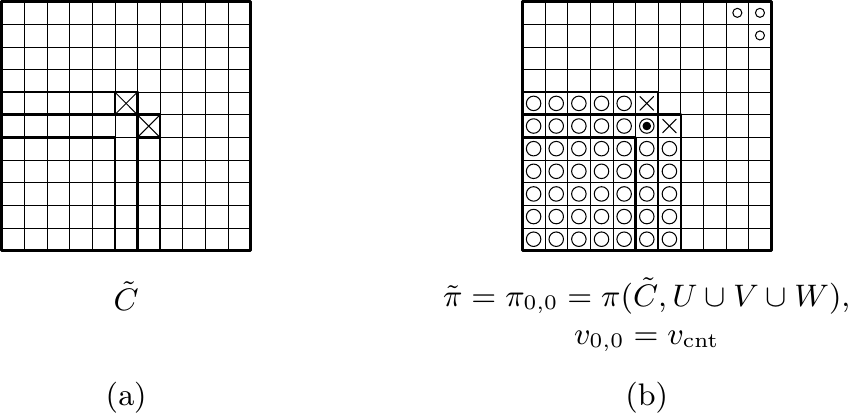}
\caption{(a) An example configuration $\tilde{C}$ of Case 2.2.
(b) One idea for constructing a partial solution $A$ for $\tilde{C}$ uses these 
patterns and so on but it fails.
}
\label{figure:fig065}
\end{figure}

First we consider the case where $\tilde{w}$ is even and later 
we explain how to modify the proof for odd $\tilde{w}$.
For each $\tilde{w}$ there is only one configuration $\tilde{C}$ of this form. 
Its type is $(0, 0, 2, 0)$.
In Fig. \ref{figure:fig065} (a) we show the configuration for $\tilde{w} = 10$. 

One idea to construct a partial solution $A$ for $\tilde{C}$ is to use 
$\tilde{\pi} = \pi(\tilde{C}, U \cup V \cup W)$, 
$\pi_{0, 0} = \tilde{\pi}$, 
$v_{0, 0} = v_{\abb{cnt}}$ 
and the firing rule $(\abb{W}_{0} \vee \abb{W}_{1}) \wedge 
\abb{M}_{0, 0}$.
We show these patterns and so on for the example configuration $\tilde{C}$ 
in Fig. \ref{figure:fig065} (b).
However, for this $A$ we cannot prove (C5) because 
for a configuration $C$ of size $\tilde{w}$ that has $\tilde{\pi}$, 
we have 
$D_{0, 0}(C, v) > 2 \tilde{w}$ 
for the three nodes 
$v = (\tilde{w} - 1, \tilde{w})$, 
$(\tilde{w}, \tilde{w} - 1)$, 
$(\tilde{w}, \tilde{w})$ 
(the small circles in Fig. \ref{figure:fig065} (b)).  
(We have the exception (1) in Theorem \ref{theorem:thm011}.)

\begin{figure}[htbp]
\centering
\includegraphics[scale=1.0]{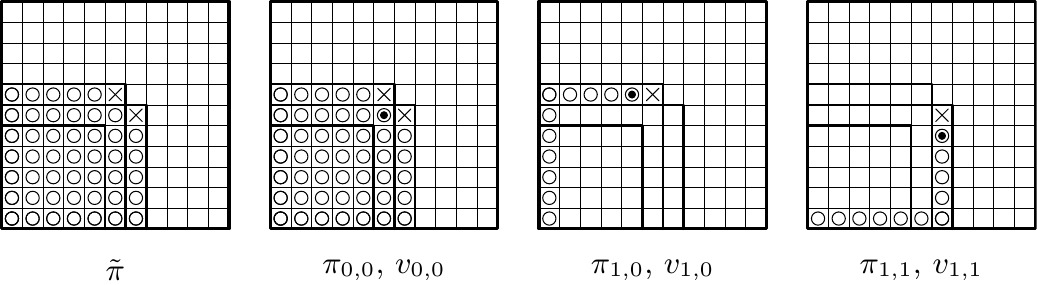}
\caption{The patterns and so on we use for the example configuration 
$\tilde{C}$ shown in Fig. \ref{figure:fig065} (a).}
\label{figure:fig074}
\end{figure}

Instead of the above idea we use another idea for constructing the desired 
partial solution $A$.
In Fig. \ref{figure:fig074} we show patterns and nodes 
$\tilde{\pi}$ ($= \pi(\tilde{C}, U \cup V \cup W)$), 
$\pi_{0, 0}$ ($= \tilde{\pi}$), 
$\pi_{1, 0}$, $\pi_{1, 1}$, 
$v_{0, 0}$ ($= v_{\abb{cnt}}$), 
$v_{1, 0}$ ($= v_{\abb{cnt}} + (-1, 1)$), 
$v_{1, 1}$ ($= v_{\abb{cnt}} + (1, -1)$) 
used for $A$.  
The firing rule is 
$(\abb{W}_{0} \vee \abb{W}_{1}) \wedge 
(\abb{M}_{0, 0} \vee (\abb{M}_{1, 0} \wedge \abb{M}_{1, 1}))$.

The message $\abb{M}_{0, 0}$ implies that 
$C$ has $\tilde{\pi}$.
Each of the two messages $\abb{M}_{1, 0}$, $\abb{M}_{1, 1}$ 
has only a partial information on $C$.  
However the two messages as a whole imply that 
$C$ has $\tilde{\pi}$.
Therefore a node knows that $C$ has $\tilde{\pi}$ 
if either (i) it receives $\abb{M}_{0, 0}$ or 
(ii) it receives both of $\abb{M}_{1, 0}$, $\abb{M}_{1, 1}$.
Hence the main part of the firing rule is 
$\abb{M}_{0, 0} \vee (\abb{M}_{1, 0} \wedge \abb{M}_{1, 1})$.

We can prove the ``if'' part of (C2) for $\pi_{1, 0}$, $\pi_{1, 1}$ 
using the fact that any $C$ has exactly two holes.
The message $\abb{M}_{1, 0}$ is generated by the signal that goes 
from $v_{\abb{gen}}$ to $v_{1, 0}$ via 
$(0, \lfloor \tilde{w} / 2 \rfloor + 1)$ and similarly for $M_{1, 1}$.

The proof of (C5) is as follows.
We must prove that if $C$ of size $\tilde{w}$ has $\tilde{\pi}$ 
then for any $v \in C$ either (1) $D_{0, 0}(C, v) \leq 2 \tilde{w}$ 
or (2) $D_{1, 0}(C, v) \leq 2 \tilde{w}$ and 
$D_{1, 1}(C, v) \leq 2 \tilde{w}$.
However, if $C$ has $\tilde{\pi}$ then $C = \tilde{C}$.
Using this we can show the following:
\begin{enumerate}
\item[$\bullet$] $D_{0, 0}(C, v) \leq 2 \tilde{w}$ except for 
$v = (\tilde{w} - 1, \tilde{w})$, 
$(\tilde{w}, \tilde{w} - 1)$, 
$(\tilde{w}, \tilde{w})$ 
(Theorem \ref{theorem:thm011}, exception (1)),
\item[$\bullet$] $D_{1, 0}(C, v) \leq 2 \tilde{w}$ and 
$D_{1, 1}(C, v) \leq 2 \tilde{w}$ for 
$v = (\tilde{w} - 1, \tilde{w})$, 
$(\tilde{w}, \tilde{w} - 1)$, 
$(\tilde{w}, \tilde{w})$ (among others) 
(by elementary calculation and our assumption $\tilde{w} \geq 11$).
\end{enumerate}
This shows (C5).

\medskip

The proof for odd $\tilde{w}$ is obtained from the above proof 
by changing the definition of $\tilde{\pi}$ 
to $\tilde{\pi} = \pi(C, U \cup V \cup (W - \{v_{\abb{cnt}} + (1, 1)\}))$ 
(see $\tilde{\pi}$ shown in Fig. \ref{figure:fig070}).

\medskip

\noindent
(Case 3) The type of $\tilde{C}$ is of the form 
$(0, b, c, d)$ ($b \geq 1$) and $\tilde{C}$ has no critical pairs 
in $V \cup W$.

\medskip

Case 3 is a ``shrunk'' version of Case 2 for odd $\tilde{w}$.
In Case 3, $U$, $V$, $v_{\abb{cnt}}$ (the corner of $V$) 
play the roles of $U \cup V$, $W$, $v_{\abb{cnt}} + (1, 1)$ 
(the corner of $W$) respectively of Case 2 for odd $\tilde{w}$.
However the proofs for Case 3 are easier than those for Case 2 
for odd $\tilde{w}$ because $v_{\abb{cnt}} + (1, 1)$ is 
out of $U \cup V$ but $v_{\abb{cnt}}$ is in $U \cup V$.

The proofs in Case 3 are independent of whether $\tilde{w}$ is 
even or odd.
As example configurations we show configurations with even $\tilde{w}$.

\medskip

\noindent
(Case 3.1) In $\tilde{C}$, at least one of 
$v_{\abb{cnt}} - (1, 0)$, 
$v_{\abb{cnt}} - (0, 1)$ is a node.

\medskip

By $V'$ we denote the set $V - \{v_{\abb{cnt}}\}$.
As the type of a configuration $C$ we use 
$(a, (b_{0}, b_{1}, b_{2}), c, d)$ instead of 
$(a, b, c, d)$.
Here, $b_{0}$ is the number of critical holes in $V'$, 
$b_{1}$ is the number of noncritical holes in $V'$, 
and $b_{2}$ is the number of noncritical holes in 
the set $\{v_{\abb{cnt}}\}$ 
(that is, $b_{2}$ is $0$ or $1$ according as the position 
$v_{\abb{cnt}}$ is a node or a hole).
As was in Case 2.1.2, the eight triples 
$(0, 0, 1)$, 
$(0, 1, 0)$, 
$(1, 0, 0)$, 
$(0, 1, 1)$, 
$(1, 0, 1)$, 
$(0, 2, 0)$, 
$(1, 1, 0)$, 
$(2, 0, 0)$ 
are possible as $(b_{0}, b_{1}, b_{2})$. 
(We do not exclude $(1, 0, 0)$.) 

Suppose that the type of $\tilde{C}$ is 
$(0, (b_{0}, b_{1}, b_{2}), c, d)$ 
and $(b_{0}, b_{1}, b_{2})$ is one of 
$(0, 0, 1)$, 
$(0, 1, 0)$, 
$(0, 1, 1)$, 
$(1, 0, 1)$, 
$(0, 2, 0)$, 
$(1, 1, 0)$, 
$(2, 0, 0)$ 
(that is, we ignore $(1, 0, 0)$).
Then we can construct a desired partial solution 
using the idea we used in Case 2.1.2 for the type 
$(0, 0, (b_{0}, b_{1}, b_{2}), d)$.
The idea used in Fig. \ref{figure:fig070} 
can be used for 
$(0, 1, 0)$, 
$(0, 1, 1)$, 
$(0, 2, 0)$, 
$(1, 1, 0)$, 
$(2, 0, 0)$ 
(that is, cases where either $V'$ contains 
at least one noncritical hole or $V'$ contains two critical holes).
The idea used in Fig. \ref{figure:fig071} can be used for 
$(1, 0, 1)$ and 
that used in Fig. \ref{figure:fig072} can be used for 
$(0, 0, 1)$.

\begin{figure}[htbp]
\centering
\includegraphics[scale=1.0]{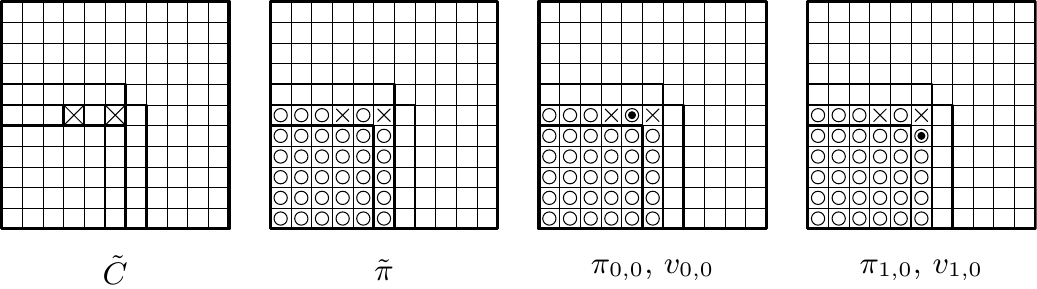}
\caption{An example for the type of the form 
$(0, (1, 0, 1), c, d)$.}
\label{figure:fig075}
\end{figure}

We show an example for the type $(0, (1, 0, 1), 0, 0)$ 
in Fig. \ref{figure:fig075}.
(Compare this figure with Fig. \ref{figure:fig071}, 
an example for the type $(0, 0, (1, 0, 1), 0)$.)
In this case we use $\tilde{\pi} = \pi(\tilde{C}, U \cup V)$, 
$\pi_{0, 0} = \pi_{1, 0} = \tilde{\pi}$, 
$v_{0, 0} = v_{\abb{cnt}} - (1, 0)$, 
$v_{1, 0} = v_{\abb{cnt}} - (0, 1)$.
The firing rule is $(\abb{W}_{0} \vee \abb{W}_{1}) 
\wedge (\abb{M}_{0, 0} \vee \abb{M}_{1, 0})$.
The proof of (C5) is simpler.  
(We can use Theorem \ref{theorem:thm011} because 
both of $v_{0, 0}$, $v_{1, 0}$ are in $U \cup V$.)

\begin{figure}[htbp]
\centering
\includegraphics[scale=1.0]{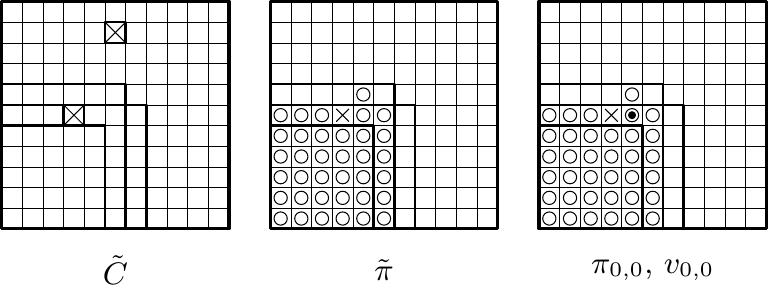}
\caption{An example for the type $(0, (1, 0, 0), c, d)$.}
\label{figure:fig076}
\end{figure}

Now there remains the case $(b_{0}, b_{1}, b_{2}) = (1, 0, 0)$.
The following two statements are equivalent:
\begin{enumerate}
\item[$\bullet$] The type of $\tilde{C}$ is 
$(0, (1, 0, 0), c, d)$ and 
$\tilde{C}$ has no critical pairs in $V \cup W$.
\item[$\bullet$] The type of $\tilde{C}$ is 
$(0, (1, 0, 0), c, d)$ and $v_{0} + (1, 1)$ is a node. 
Here $v_{0}$ is the unique critical hole in $V'$.
\end{enumerate}
Using the latter statement as the definition of 
the present case we can construct a partial solution.
In Fig. \ref{figure:fig076} we show an example.
In this case we use 
$\tilde{\pi} = \pi(\tilde{C}, U \cup V \cup 
\{v_{\abb{cnt}} + (-1, 1)\})$, 
$\pi_{0, 0} = \tilde{\pi}$, $v_{0, 0} = v_{\abb{cnt}} - (1, 0)$.
(We assume that $v_{\abb{cnt}} - (2, 0)$ is the unique 
critical hole in $V'$.)
The firing rule is $(\abb{W}_{0} \vee \abb{W}_{1}) \wedge 
\abb{M}_{0, 0}$.

\medskip

\noindent
(Case 3.2) $\tilde{C}$ has the two holes at $v_{\abb{cnt}} - (1, 0)$, 
$v_{\abb{cnt}} - (0, 1)$.

\medskip

The idea for constructing a partial solution is completely the same 
as for Case 2.2. (Consider to modify the idea shown in 
Fig. \ref{figure:fig074}.)

\medskip

\noindent
(Case 4) The type of $\tilde{C}$ is $(1, 1, 0, 0)$ and 
$\tilde{C}$ has no critical pairs in $U \cup V$.

\medskip

Let $v_{0} = (x_{0}, y_{0})$ be the unique hole in $U$ and 
$v_{1} = (x_{1}, y_{1})$ be the unique hole in $V$.
We assume that 
$v_{1}$ is in the horizontal part of $V$ 
(that is, $y_{1} = \lfloor \tilde{w} / 2 \rfloor$).

\medskip

\noindent
(Case 4.1) $v_{1} \not= v_{0} + (1, 1)$.

\medskip

\begin{figure}[htbp]
\centering
\includegraphics[scale=1.0]{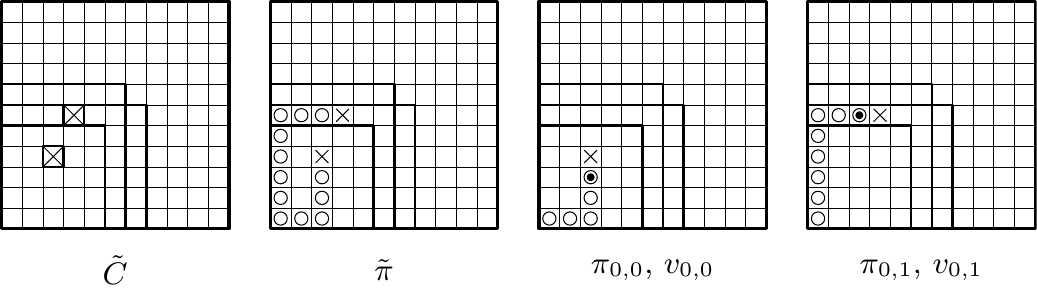}
\caption{An example configuration $\tilde{C}$ of Case 4.1 and 
$\tilde{\pi}$, 
$\pi_{0, 0}$, $v_{0, 0}$, $\pi_{0, 1}$, $v_{0, 1}$ for it.}
\label{figure:fig077}
\end{figure}

In Fig. \ref{figure:fig077} we show an example configuration $\tilde{C}$ of this case 
and $\tilde{\pi}$, $\pi_{0, 0}$, $v_{0, 0}$ ($= v_{0} - (0, 1)$), 
$\pi_{0, 1}$, $v_{0, 1}$ ($= v_{1} - (1, 0)$) 
used to construct a partial solution $A$ for $\tilde{C}$.
The firing rule is $(\abb{W}_{0} \vee \abb{W}_{1}) 
\wedge (\abb{M}_{0, 0} \wedge \abb{M}_{0, 1})$.
The message $\abb{M}_{0, 0}$ is generated by a signal that goes 
from $v_{\abb{gen}}$ to $v_{0, 0}$ via $(x_{0}, 0)$ 
and the message $\abb{M}_{0, 1}$ is generated by a signal that goes 
from $v_{\abb{gen}}$ to $v_{0, 1}$ via $(0, \lfloor \tilde{w} / 2 \rfloor)$.
We can prove all of the statements (C1), \ldots, (C5) for $A$.
(Note that if $v_{1} = v_{0} + (1, 1)$ we cannot prove (C5) 
because we have the exception (3) when we use Theorem \ref{theorem:thm011} 
with $v = v_{0, 1}$.)

\medskip

\noindent
(Case 4.2) $v_{1} = v_{0} + (1, 1)$.

\medskip

\begin{figure}[htbp]
\centering
\includegraphics[scale=1.0]{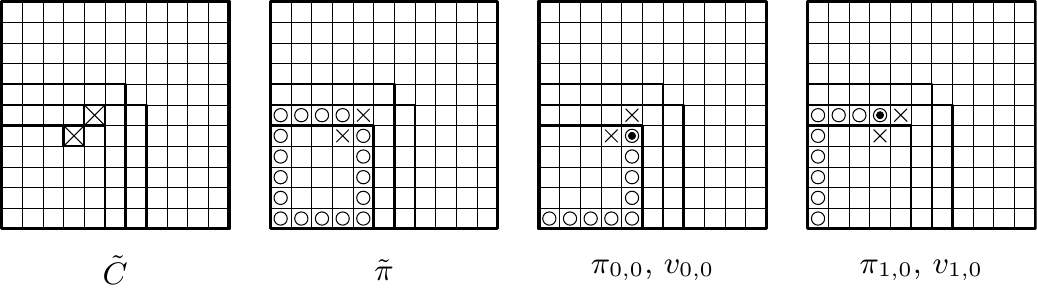}
\caption{An example configuration $\tilde{C}$ of Case 4.2 and 
$\tilde{\pi}$, 
$\pi_{0, 0}$, $v_{0, 0}$, $\pi_{1, 0}$, $v_{1, 0}$ for it.}
\label{figure:fig078}
\end{figure}

In Fig. \ref{figure:fig078} we show an example configuration 
$\tilde{C}$ of this case 
and 
$\tilde{\pi}$, 
$\pi_{0, 0}$, 
$v_{0, 0}$ ($= v_{0} + (1, 0)$), 
$\pi_{1, 0}$, 
$v_{1, 0}$ ($= v_{1} - (1, 0)$) 
used to construct a partial solution $A$ for $\tilde{C}$.
The firing rule is $(\abb{W}_{0} \vee \abb{W}_{1}) 
\wedge (\abb{M}_{0, 0} \vee \abb{M}_{1, 0})$.
We can easily prove (C1), \ldots, (C4).

The proof of (C5) is as follows.
We must show that either 
$D_{0, 0}(C, v) \leq 2 \tilde{w}$ or 
$D_{1, 0}(C, v) \leq 2 \tilde{w}$ for any $v \in C$.
Using Theorem \ref{theorem:thm011} we have the following 
estimations of $D_{i, 0}(C, v)$.
\begin{enumerate}
\item[$\bullet$] $D_{0, 0}(C, v) \leq 2 \tilde{w}$ except for 
$(0, \tilde{w} - 1)$, 
$(1, \tilde{w})$, 
$(0, \tilde{w})$ when $\tilde{w}$ is even and 
except for 
$(0, \tilde{w})$ when $\tilde{w}$ is odd.
\item[$\bullet$] $D_{1, 0}(C, v) \leq 2 \tilde{w}$ except for 
$(\tilde{w} - 1, 0)$, 
$(\tilde{w}, 1)$, 
$(\tilde{w}, 0)$ when $\tilde{w}$ is even and 
except for $(\tilde{w}, 0)$ when $\tilde{w}$ is odd.
\end{enumerate}
Therefore (C5) is true.

\hfill
$\Box$

\medskip

\section{Discussions and conclusion}
\label{section:discussions_and_conclusion}

We introduced a variation $\abb{SH}[k]$ of FSSP 
and showed some results on it.
We summarized the results in Section \ref{section:introduction}.
The most important of them are the following:
\begin{enumerate}
\item[$\bullet$] We showed a minimal-time solution of $\abb{SH}[1]$ 
(Theorem \ref{theorem:thm000}).
\item[$\bullet$] We determined the value $\abb{mft}_{\abb{SH}[2]}(C)$ 
(Theorem \ref{theorem:thm012}).
\end{enumerate}
In the remainder of this section we list open problems 
and give some comments on them.

First we list two rather technical open problems on $\abb{SH}[k]$ 
for the general values of $k$.
\begin{enumerate}
\item[$\bullet$] Prove $c_{k} = k - 2$ for $k \geq 10$.
To prove this is equivalent to prove that the barriers shown 
in Fig. \ref{figure:fig011} (a), (b) 
give the maximum value of $E_{\abb{max}}(S, p)$ 
for $S \in \mathcal{S}_{k}$, $p \in S$.
\item[$\bullet$] For any $k \geq 4$, any $s$ such that $0 \leq s \leq c_{k}$ 
and any sufficiently large $\tilde{w}$, 
prove that there exists a configuration $\tilde{C}$ 
of size $\tilde{w}$ of $\abb{SH}[k]$ such that $\abb{mft}_{\abb{SH}[k]}(\tilde{C}) = 
2 \tilde{w} + s$. 
We proved this only for $s = 0$ (Theorem \ref{theorem:thm001} (2)) 
and $s = c_{k}$ (Theorem \ref{theorem:thm003} (2), Theorem \ref{theorem:thm006} (2)).

\end{enumerate}

Our ultimate goal is either to construct 
a minimal-time solution of $\abb{SH}[k]$ or to prove its 
nonexistence for each $k$ ($\geq 2$).
The author has the conjecture that $\abb{SH}[k]$ has a minimal-time solution 
for all $k$.

We have two problems to start with to realize the above goal.
The first is to construct a minimal-time solution of $\abb{SH}[2]$.
We already know the value of $\abb{mft}_{\abb{SH}[2]}(C)$.
Hence our goal is reduced to construct a solution that fires 
$C$ before or at this time.

The second is to determine the value of $\abb{mft}_{\abb{SH}[k]}(C)$ 
for general values of $k \geq 3$ or for some specific values 
$k = 3$, $k = 4$, \ldots.
We determined this value for $k = 2$ in Theorem \ref{theorem:thm012}.
However its derivation essentially used the fact $k = 2$ 
and it cannot be readily modified for $k \geq 3$.

By analyzing the proof of Theorem \ref{theorem:thm012} we notice that 
one of the difficult parts in it is to estimate the range of 
the value $\abb{d}_{C}(v, v')$ for two positions $v, v'$ in 
a situation where we have only partial information on $C$ 
(such as ``$C$ has a pattern $\pi$'').
For the proof of Theorem \ref{theorem:thm012} we derived the range 
by an ad hoc analysis essentially using the fact $k = 2$ 
exemplified in Fig. \ref{figure:fig045}.
However this type of ad hoc analyses becomes very difficult 
even for $k = 3$ and nearly impossible for $k = 4$.
Hence, in order to derive the value of $\abb{mft}_{\abb{SH}[k]}(C)$ 
for $k \geq 3$ using our idea, it is mandatory 
to construct a theory for estimating $\abb{d}_{C}(v, v')$ with partial information on $C$.
This is itself an interesting problem in discrete mathematics.

Although the author has the conjecture that $\abb{SH}[k]$ has 
minimal-time solutions for all $k$, 
we should include the problem to prove nonexistence of 
minimal-time solutions of $\abb{SH}[k]$ for some $k$ 
in the list of open problems.

This completes our list of open problems and comments on $\abb{SH}[k]$.
Finally we give more general comments on variations of FSSP 
restricting ourselves to the problem to know 
existence/nonexistence of minimal-time solutions and to find one 
if they exist.

For the two variations $\abb{ORG}$, $\abb{SQ}$ we found minimal-time solutions 
long ago and by slightly modifying them we obtain very difficult 
variations $\abb{2PATH}$, $\abb{3PATH}$, $\abb{gSQ}$, $\abb{SH}[k]$. 
Hence generally speaking variations are very difficult 
and we know minimal-time solutions only for some rare simple cases.
This observation leads us to the following research themes.
\begin{enumerate}
\item[(1)] To study these and other difficult variations and 
develop basic tools to solve these variations.
\item[(2)] To construct a general theory of variations of FSSP.
\end{enumerate}
These two research themes are not independent and development in 
one will contribute to development of the other.

As for (2) we are not sure whether such a general theory exists.
However we have at least one encouraging result.  
We know that there is an algorithm for computing 
$\abb{mft}_{\Gamma}(C)$ for any natural variation $\Gamma$ 
and it is based on one simple idea ``if a node of a configuration $C$ 
fires as soon as it finds no reasons not to fire 
then the node fires at the time $\abb{mft}_{\Gamma}(C)$'' 
(\cite{Kobayashi_TCS_2014}).  
This means that the value $\abb{mft}_{\Gamma}(C)$ is determined by 
a simple mechanism that is common for all natural variations $\Gamma$.
This result is one that should be in the above mentioned general theory.

\bibliographystyle{plain}
\bibliography{ms}

\begin{thebibliography}{10}

\bibitem{Balzer_1967}
R.~Balzer.
\newblock An $8$-state minimal time solution to the firing squad
  synchronization problem.
\newblock {\em Information and Control}, 10:22--42, 1967.

\bibitem{Culik_1987}
K.~{Culik II}.
\newblock Variations of the firing squad problem and applications.
\newblock {\em Information Processing Letters}, 30:153--157, 1989.

\bibitem{Dimitriadis_Kutrib_Sirakoulis_2018}
A.~Dimitriadis, M.~Kutrib, and G.~C. Sirakoulis.
\newblock Revisiting the cutting of the firing squad synchronization.
\newblock {\em Natural Computing}, 17(3):455--465, 2018.

\bibitem{Even_Litman_Winkler}
S.~Even, A.~Litman, and P.~Winkler.
\newblock Computing with snakes in deirected networks of automata.
\newblock {\em J. Algorithms}, 24:158--170, 1997.

\bibitem{Goldstein_Kobayashi_SIAM_2005}
D.~Goldstein and K.~Kobayashi.
\newblock On the complexity of network synchronization.
\newblock {\em SIAM J. Comput.}, 35(3):567--589, 2005.

\bibitem{Goldstein_Kobayashi_SIAM_2012}
D.~Goldstein and K.~Kobayashi.
\newblock On minimal-time solution of firing squad synchronization problem for
  networks.
\newblock {\em SIAM J. Comput.}, 41(3):618--669, 2012.

\bibitem{Goto_1962}
E.~Goto.
\newblock A minimal time solution of the firing squad problem.
\newblock {\em Course Notes for Applied Mathematics 298, Harvard University},
  pages 52--59, 1962.

\bibitem{Gruska_et_al_2004}
J.~Gruska, S.~La Torre, and M.~Parente.
\newblock Optimal time and communication solutions of firing squad
  synchronization problems on square arrays, toruses and rings.
\newblock In {\em DLT 2004}, volume 3340 of {\em Lecture Notes in Computer
  Science}, pages 200--211, 2004.

\bibitem{Kobayashi_Res_Rep_1976}
K.~Kobayashi.
\newblock A minimal time solution to the firing squad synchronization problem
  of rings with one-way information flow.
\newblock {\em Tokyo Institute of Technology, Department of Information
  Sciences, Research Reports on Information Sciences, No.C-8}, 1976.

\bibitem{Kobayashi_JCSS_1978}
K.~Kobayashi.
\newblock The firing squad synchronization problem for a class of polyautomata
  networks.
\newblock {\em Journal of Computer and System Sciences}, 17:300--318, 1978.

\bibitem{Kobayashi_TCS_2001}
K.~Kobayashi.
\newblock On time optimal solutions of the firing squad synchronization problem
  for two-dimensional paths.
\newblock {\em Theoretical Computer Science}, 259:129--143, 2001.

\bibitem{Kobayashi_TCS_2014}
K.~Kobayashi.
\newblock The minimum firing time of the generalized firing squad
  synchronization problem for squares.
\newblock {\em Theoretical Computer Science}, 547:46--69, 2014.

\bibitem{Kutrib_Vollmar_1995_Faulty}
M.~Kutrib and R.~Vollmar.
\newblock The firing squad synchronization problem in defective cellular
  automata.
\newblock {\em IEICE Trans. Inf. \& Syst.}, E78-D(7):895--900, 1995.

\bibitem{LaTorre_1996}
S.~LaTorre, M.~Napoli, and M.~Parente.
\newblock Synchronization of one-way connected processors.
\newblock {\em Complex Systems}, 10:239--255, 1996.

\bibitem{Mazoyer_1987}
J.~Mazoyer.
\newblock A six-state minimal time solution to the firing squad synchronization
  problem.
\newblock {\em Theoretical Computer Science}, 50:183--238, 1987.

\bibitem{Mazoyer_Survey}
J.~Mazoyer.
\newblock An overview of the firing synchronization problem.
\newblock In {\em Automata Networks, LITP Spring School on Theoretical Computer
  Science, Angel{\`e}s-Village, France, May 12-16, 1986, Proceedings}, volume
  316 of {\em Lecture Notes in Computer Science}, pages 82--94, 1988.

\bibitem{Moore}
E.~F. Moore.
\newblock {\em Sequential Machines, Selected Papers}.
\newblock Addison Wesley, Reading, MA, 1962.

\bibitem{Moore_Langdon_1968}
F.~R. Moore and G.~G. Langdon.
\newblock A generalized firing squad problem.
\newblock {\em Information and Control}, 12:212--220, 1968.

\bibitem{Napoli_Parente}
M.~Napoli and M.~Parente.
\newblock Minimum and non-minimum time solutions to the firing squad
  synchronization problem.
\newblock In {\em Gruska Restschrift}, volume 8808 of {\em Lecture Notes in
  Computer Science}, pages 114--128. Springer, 2014.

\bibitem{Nishitani_Honda}
Y.~Nishitani and N.~Honda.
\newblock The firing squad synchronization problem for graphs.
\newblock {\em Theoretical Computer Science}, 14:39--61, 1981.

\bibitem{Ostrovsky_Wilkerson}
R.~Ostrovsky and D.~Wilkerson.
\newblock Faster computation on directed networks of automata.
\newblock In {\em Proceedings of the 14th Annual ACM Symposium on Principles of
  Distributed Computing}, pages 38--46, 1995.

\bibitem{Roka_2000}
Z.~R\'oka.
\newblock The firing squad synchronization problem on {C}ayley graphs.
\newblock {\em Theoretical Computer Science}, 244:243--256, 2000.

\bibitem{Rosenstiehl_1966}
P.~Rosenstiehl.
\newblock Existence d'automates finis capables de s'accorder bien
  qu'arbitrairement connect\'{e}es nombreux.
\newblock {\em Internat. Comp. Centre Bull.}, 5:245--261, 1966.

\bibitem{Rosenstiehl_1972}
P.~Rosenstiehl, J.~R. Fiksel, and A.~Holliger.
\newblock Intelligent graphs: networks of finite automata capable of solving
  graph problems.
\newblock {\em ``Graph Theory and Computing'' (R. C. Read, Ed.), Academic
  Press}, pages 219--265, 1972.

\bibitem{Shinahr_1974}
I.~Shinahr.
\newblock Two- and three-dimensional firing-squad synchronization problem.
\newblock {\em Information and Control}, 24:163--180, 1974.

\bibitem{Szwerinski_1982}
H.~Szwerinski.
\newblock Time-optimal solution of the firing-squad-synchronization-problem for
  $n$-dimensional rectangles with the general at an arbitrary position.
\newblock {\em Theoretical Computer Science}, 19:305--320, 1982.

\bibitem{Umeo_2004_Faulty}
H.~Umeo.
\newblock A simple design of time-efficient firing squad synchronization
  algorithms with fault-tolerance.
\newblock {\em IEICE Trans. Inf. \& Syst.}, E87-D(3):733--739, 2004.

\bibitem{Umeo_2018_Goto_reconstruction}
H.~Umeo, M.~Hirota, Y.~Nozaki, K.~Imai, and T.~Sogabe.
\newblock A new reconstruction and the first implementation of {G}oto's fssp
  algorithm.
\newblock {\em Applied Mathematics and Computation}, 318:92--108, 2018.

\bibitem{Umeo_2005_Survey}
H.~Umeo, M.~Hisaoka, and T.~Sogabe.
\newblock A survey on optimum-time firing squad synchronization algorithms for
  one-dimensional cellular automata.
\newblock {\em Int. Journ. of Unconventional Computing}, 1:403--426, 2005.

\bibitem{Umeo_2017_PaCT}
H.~Umeo and N.~Kamikawa.
\newblock A new class of the smallest four-state partial fssp solutions for
  one-dimensional ring cellular automata.
\newblock In {\em Proceedings of 11th International Conference on Parallel
  Computing Technologies, PaCT 2017}, volume 10421 of {\em Lecture Notes in
  Computer Science}, pages 232--245, 2017.

\bibitem{Umeo_Kamikawa_Maeda_Fujita_2018_ACRI}
H.~Umeo, N.~Kamikawa, M.~Maeda, and G.~Fujita.
\newblock Implementations of {FSSP} algorithms on fault-tolerant cellular
  arrays.
\newblock In {\em Proceedings of ACRI 2018}, volume 11115 of {\em Lecture Notes
  in Computer Science}, pages 274--285, 2018.

\bibitem{Umeo_2012_ACRI}
H.~Umeo and K.~Kubo.
\newblock Recent developments in constructing square synchronizers.
\newblock In {\em Proceedings of 10th International Conference on Cellular
  Automata for Research and Industry, ACRI 2012}, volume 7495 of {\em Lecture
  Notes in Computer Science}, pages 171--183, 2012.

\bibitem{Umeo_Kubo_2015_CANDAR}
H.~Umeo and K.~Kubo.
\newblock An {FSSP} on torus.
\newblock In {\em Proceedings of CANDAR 2015}, pages 453--456, 2015.

\bibitem{Umeo_2015_Survey}
H.~Umeo, M.~Maeda, A.~Sousa, and K.~Taguchi.
\newblock A class of non-optimum-time $3n$-step fssp algorithms - a survey.
\newblock In {\em Proceedings of PaCT 2015}, volume 9251 of {\em Lecture Notes
  in Computer Science}, pages 231--245, 2015.

\bibitem{Waksman_1966}
A.~Waksman.
\newblock An optimum solution to the firing squad synchronization problem.
\newblock {\em Information and Control}, 9:66--78, 1966.

\bibitem{Yunes_2006_Faulty}
J.-B. Yun\`es.
\newblock Fault tolerant solutions to the firing squad synchronization problem
  in linear cellular automata.
\newblock {\em Journal of Cellular Automata}, 1(3):253--268, 2006.

\end{thebibliography}

\bigskip

\noindent
{\Large\bf Appendix}

\appendix

\section{Proof of Theorem \ref{theorem:thm011}}
\label{section:tedious_proof}

In this section we prove Theorem \ref{theorem:thm011}.
First we prove two lemmas.

\begin{lem}
\label{lemma:lem007}
Suppose that $w \geq 5$, 
$C$ is a configuration of $\abb{SH}[2]$, 
$v = (x, y)$, $v' = (x', y')$ are nodes of $C$ such that 
$v \in U \cup V$, 
either $|x - x'| \leq 1$ or $|y - y'| \leq 1$, 
and $\abb{d}_{C}(v, v') \leq \abb{d}_\abb{MH}(v, v') + 2$.
Then $\abb{d}_\abb{MH}(v_\abb{gen}, v) + \abb{d}_{C}(v, v') 
\leq 2w$.
\end{lem}

\begin{figure}[htbp]
\centering
\includegraphics[scale=1.0]{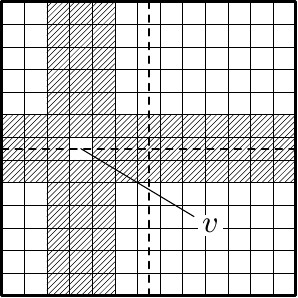}
\caption{Possible positions of $v'$ in Lemma \ref{lemma:lem007}.}
\label{figure:fig041}
\end{figure}

\noindent
{\it Proof}. 
In Fig. \ref{figure:fig041} we show possible positions of $v'$ 
by shadow.
We prove only the case $|x - x'| \leq 1$.
We have $y + |y' - y| \leq w$ because 
if $y \leq y'$ then 
$y + |y' - y| = y' \leq w$ and if 
$y' < y$ then 
$y + |y' - y| = 2y - y' \leq 2y \leq 2 \lfloor w / 2 \rfloor 
\leq w$.
Therefore 
$\abb{d}_\abb{MH}(v_\abb{gen}, v) + \abb{d}_{C}(v,  v') 
\leq \abb{d}_\abb{MH}(v_\abb{gen}, v) + \abb{d}_\abb{MH}(v, v') + 2 
= x + y + |x' - x| + |y' - y| + 2 
= (x + |x' - x|) + (y + |y' - y|) + 2
\leq \lfloor w / 2 \rfloor + 1 + w + 2 
= \lfloor w / 2 \rfloor + w + 3$.
If $w$ is odd then $w \geq 5$ and 
$\lfloor w / 2 \rfloor + w + 3 
= 2w - w / 2 + 5 / 2 \leq 2w$.
If $w$ is even then $w \geq 6$ and 
$\lfloor w / 2 \rfloor + w + 3 
= 2w - w / 2 + 3 \leq 2w$.
\hfill
$\Box$

\medskip

\begin{lem}
\label{lemma:lem008}
Suppose that the followings are true.
\begin{enumerate}
\item[{\rm (1)}] $C$ is a configuration of size $w$ of 
$\abb{SH}[2]$.
\item[{\rm (2)}] $v_{0} = (x_{0}, y_{0})$, 
$v_{1} = (x_{1}, y_{1})$ are nodes in $C$ 
such that $v_{0} \in U \cup V$.
\item[{\rm (3)}] There is a path in $C$ from $v_{0}$ to $v_{1}$ 
of the MH distance length.
\item[{\rm (4)}] $H$ is a set of positions in $S_{w}$ of the form of 
a rectangle 
{\rm (}that is, there are two positions $v_{2} = (x_{2}, y_{2})$, 
$v_{3} = (x_{3}, y_{3})$ 
in $S_{w}$ such that $x_{2} \leq x_{3}$, $y_{2} \leq y_{3}$, 
$H = \{ (x, y) ~|~ x_{2} \leq x \leq x_{3}, y_{2} \leq y 
\leq y_{3} \}${\rm )}.
\item[{\rm (5)}] There are no holes on the boundary of 
the rectangle $H$.
\item[{\rm (6)}] One of the followings is true:
\begin{enumerate}
\item[$\bullet$] $v_{0}$ is southwest of $v_{1}$ and 
$v_{1}$ is the southwest corner $(x_{2}, y_{2})$ of $H$.
\item[$\bullet$] $v_{0}$ is southeast of $v_{1}$ and 
$v_{1}$ is the southeast corner $(x_{3}, y_{2})$ of $H$.
\item[$\bullet$] $v_{0}$ is northeast of $v_{1}$ and 
$v_{1}$ is the northeast corner $(x_{3}, y_{3})$ of $H$.
\item[$\bullet$] $v_{0}$ is northwest of $v_{1}$ and 
$v_{1}$ is the northwest corner $(x_{2}, y_{3})$ of $H$.
\end{enumerate}
\end{enumerate}
Then $\abb{d}_\abb{MH}(v_\abb{gen}, v_{0}) + 
\abb{d}_{C}(v_{0}, v) \leq 2w$ for any node $v$ in $H$.
\end{lem}

\begin{figure}[htbp]
\centering
\includegraphics[scale=1.0]{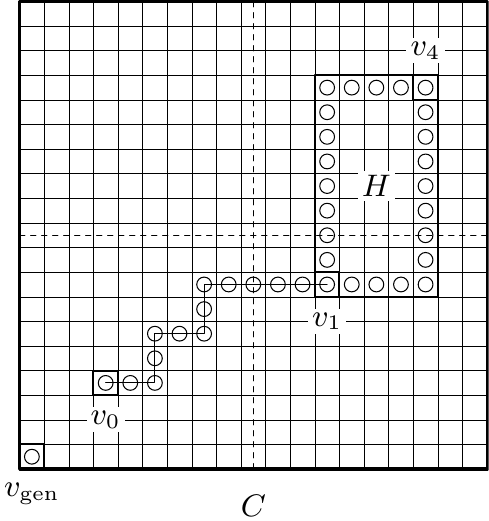}
\caption{An example of $C$ and so on in the proof of 
Lemma \ref{lemma:lem008}.}
\label{figure:fig044}
\end{figure}

\noindent
{\it Proof}. 
Let $v_{4} = (x_{4}, y_{4})$ be the corner of $H$ that is 
the opposite of $v_{1}$ 
(that is, $|x_{1} - x_{4}| = |x_{2} - x_{3}|$, 
$|y_{1} - y_{4}| = |y_{2} - y_{3}|$).
Fig. \ref{figure:fig044} shows an example of $C$, $v_{0}$, 
$v_{1}$, $v_{4}$, $H$ and a path from $v_{0}$ 
to $v_{1}$ in $C$ of the MH distance length (the bent line).
In this figure, a circle represents a node of $C$.
First we prove $\abb{d}_{C}(v_{1}, v) \leq 
\abb{d}_\abb{MH}(v_{1}, v_{4})$ for any node $v$ in $H$.

By (4), (5), $H$ is a configuration of $\abb{SH}[k]$ 
with $k \leq 2$ except that its shape is not necessarily 
a square.
We can naturally define maximal barriers of 
this configuration-like region $H$ and apply 
Theorem \ref{theorem:thm004} to them.

Suppose that $v$ is a node in $H$.
By Theorem \ref{theorem:thm004}, 
if $v$ is not in maximal barriers of $H$ 
then we have 
$\abb{d}_{H}(v_{1}, v) = \abb{d}_\abb{MH}(v_{1}, v) 
\leq \abb{d}_\abb{MH}(v_{1}, v_{4})$.
If $v$ is in a maximal barrier $R$ of $H$ 
then $v$ is adjacent to a node $v'$ that is 
in $H$ but is not in maximal barriers of $H$.
This is because the maximal barrier $R$ 
must be of the forms $S_{4}$ or $S_{5}$ 
shown in Fig. \ref{figure:fig010}.
Moreover this node $v'$ cannot be the corner $v_{4}$ by (5).
Therefore we have 
$\abb{d}_{H}(v_{1}, v) \leq \abb{d}_{H}(v_{1}, v') + 1 
= \abb{d}_\abb{MH}(v_{1}, v') + 1 
\leq \abb{d}_\abb{MH}(v_{1}, v_{4})$.
Therefore we have proved 
$\abb{d}_{C}(v_{1}, v) \leq \abb{d}_{H}(v_{1}, v) 
\leq \abb{d}_\abb{MH}(v_{1}, v_{4})$ for any 
node $v$ in $H$.

We have $x_{0} + |x_{4} - x_{0}| \leq w$ because 
if $x_{0} \leq x_{4}$ then $x_{0} + |x_{4} - x_{0}| 
= x_{4} \leq w$ 
and 
if $x_{4} < x_{0}$ then 
$x_{0} + |x_{4} - x_{0}| = 2x_{0} - x_{4} \leq 2x_{0} 
\leq 2 \lfloor w / 2 \rfloor \leq w$.
Similarly we have $y_{0} + |y_{4} - x_{0}| \leq w$.

Then, for any node $v$ in $H$, 
\begin{align*}
\abb{d}_\abb{MH}(v_\abb{gen}, v_{0}) + \abb{d}_{C}(v_{0}, v) 
& \leq \abb{d}_\abb{MH}(v_\abb{gen}, v_{0}) + 
\abb{d}_{C}(v_{0}, v_{1}) + \abb{d}_{C}(v_{1}, v) \\
& \leq \abb{d}_\abb{MH}(v_\abb{gen}, v_{0}) + \abb{d}_\abb{MH}(v_{0}, v_{1}) 
         + \abb{d}_\abb{MH}(v_{1}, v_{4}) \\
& = \abb{d}_\abb{MH}(v_\abb{gen}, v_{0}) + \abb{d}_\abb{MH}(v_{0}, v_{4}) 
\quad\quad \text{(by (6))} \\
& = (x_{0} + y_{0}) + (|x_{4} - x_{0}| + |y_{4} - y_{0}|) \\
& = (x_{0} + |x_{4} - x_{0}|) + (y_{0} + |y_{4} - y_{0}|) \\
& \leq 2w.
\end{align*}
\hfill $\Box$

\medskip

Now we prove Theorem \ref{theorem:thm011}.
Suppose that $w \geq 5$, $C$ is a configuration of size $w$ of 
$\abb{SH}[2]$, $v = (x, y) \in U \cup V$, 
and $v'  = (x', y') \in C$.
We denote the value $\abb{d}_\abb{MH}(v_\abb{gen}, v) + \abb{d}_{C}(v, v')$ 
by $F$.
Our goal is to prove $F \leq 2w$ except the four cases mentioned in the theorem.

We divide $S_{w}$ into four partially overlapping regions 
$\abb{Q}_\abb{I}$, \ldots, $\abb{Q}_\abb{IV}$ 
which we call \textit{quadrants} by analogy with quadrants in plane geometry:
$\abb{Q}_\abb{I} = \{ (x', y') \in S_{w} ~|~ x \leq x', y \leq y' \}$, 
$\abb{Q}_\abb{II} = \{ (x', y') \in S_{w} ~|~ x' \leq x, y \leq y' \}$, 
$\abb{Q}_\abb{III} = \{ (x', y') \in S_{w} ~|~ x' \leq x, y' \leq y \}$, 
$\abb{Q}_\abb{IV} = \{ (x', y') \in S_{w} ~|~ x \leq x', y' \leq y \}$.

First we prove the theorem for the case $v' \in \abb{Q}_\abb{I}$.
Later we explain how to modify the proof for other cases.

\begin{figure}[htbp]
\centering
\includegraphics[scale=1.0]{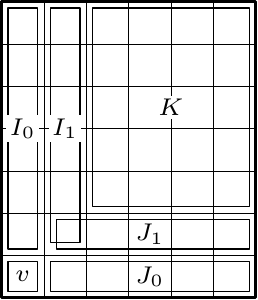}
\caption{Subsets $I_{0}, J_{0}, I_{1}, J_{1}, K$ of $\abb{Q}_\abb{I}$.}
\label{figure:fig043}
\end{figure}

We define five subsets $I_{0}$, $J_{0}$, $I_{1}$, $J_{1}$, $K$ 
of $\abb{Q}_\abb{I}$.
We show them in Fig. \ref{figure:fig043} 
assuming $\abb{Q}_\abb{I}$ has six columns and seven rows.
We assume that $x \leq w - 2$, $y \leq w - 2$ so that 
the set $\abb{Q}_\abb{I}$ has at least three columns and at least three rows 
and hence all of these five subsets are well-defined.
We can easily prove the theorem for the case $w - 1 \leq x \leq w$ and/or $w - 1 \leq y \leq w$ 
by modifying the ideas used in the following proof.

\begin{figure}[htbp] 
\centering
\includegraphics[scale=1.0]{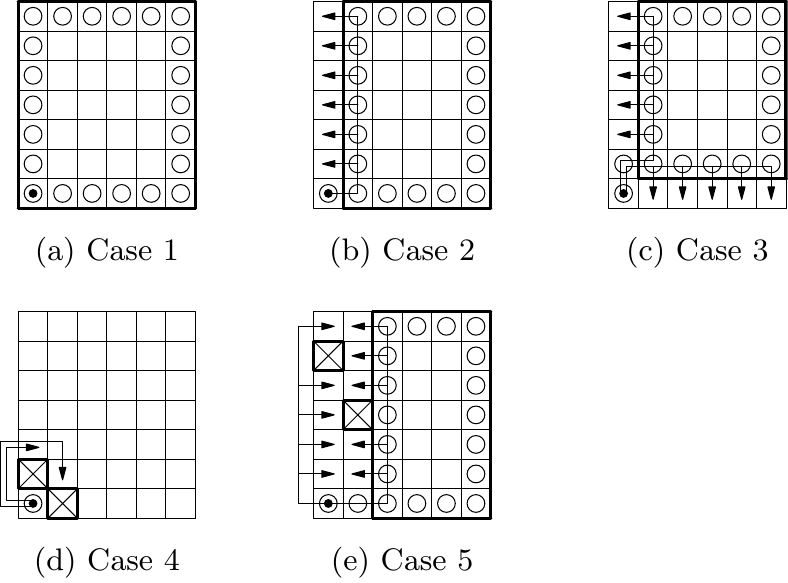}
\caption{Five cases for the proof of Theorem \ref{theorem:thm012}.}
\label{figure:fig045}
\end{figure}

We consider five cases.  
These cases are not necessarily disjoint.
By $\#I_{0}$ and so on we denote the number of holes in $I_{0}$ and so on.
In Fig. \ref{figure:fig045} (a) -- (e) we show situations for these five cases.
In these figures a bullet represents the position of $v$ 
and a circle represents a node.

\noindent
(Case 1) $\#I_{0} = \#J_{0} = 0$.
In this case $\abb{Q}_\abb{I}$ itself is a rectangle whose boundary has no holes.           
Therefore, if we select $H = \abb{Q}_\abb{I}$, $v_{0} = v_{1} = v$ then 
we can use Lemma \ref{lemma:lem008} to show 
$F \leq 2w$ for $v' \in \abb{Q}_\abb{I}$.

\medskip

\noindent
(Case 2)  Either $\#I_{1} = \#J_{0} = 0$ or $\#I_{0} = \#J_{1} = 0$.
The two sub cases are symmetric and we consider the first sub case.

In this sub case, if we select $H = I_{1} \cup J_{0} \cup J_{1} \cup K$, 
$v_{0} = v$, $v_{1} = v + (1, 0)$ then we can use 
Lemma \ref{lemma:lem008} to show 
$F \leq 2w$ 
for $v' \in H$.
For $v' = v + (0, s) \in I_{0}$ ($1 \leq s \leq w - y$), 
we have a path $v \rightarrow v + (1, 0) \rightarrow v + (1, s) \rightarrow v + (0, s) = v'$ 
from $v$ to $v'$  having length $\abb{d}_\abb{MH}(v, v') + 2$
(shown by arrow lines in Fig. \ref{figure:fig045} (b)) and hence 
$\abb{d}_{C}(v, v') \leq \abb{d}_\abb{MH}(v, v') + 2$.
Moreover we have $|x - x'| = 0$.
Therefore we can use Lemma \ref{lemma:lem007} to show 
$F \leq 2w$.

\medskip

\noindent
(Case 3) $\#I_{1} = \#J_{1} = 0$ and 
either $v + (0, 1)$ is a node or $v + (1, 0)$ is a node.
We consider the first sub case.

In this sub case, if we select $H = I_{1} \cup J_{1} \cup K$, 
$v_{0} = v$, $v_{1} = v + (1, 1)$ then we can use 
Lemma \ref{lemma:lem008} to show 
$F \leq 2w$ 
for $v' \in H$.
For $v' \in I_{0} \cup J_{0}$ we have a path from $v$ to $v'$ of 
length $\abb{d}_\abb{MH}(v, v') + 2$ as shown in Fig. \ref{figure:fig045} (c) and hence 
$\abb{d}_{C}(v, v') \leq \abb{d}_\abb{MH}(v, v') + 2$.
Moreover we have $|x - x'| = 0$ or $|y - y'| = 0$.
Therefore we can use Lemma \ref{lemma:lem007} to show 
$F \leq 2w$.

\medskip

\noindent
(Case 4)  The two positions $v + (0, 1)$, $v + (1, 0)$ have holes.
In this case these two holes are all the holes in $C$.

There is a column west of the set $I_{0}$ because $I_{0}$ contains a hole.
Therefore we have 
$\abb{d}_{C}(v, v + (0, 2)) \leq 4 = \abb{d}_\abb{MH}(v, v + (0, 2)) + 2$ 
(see the path from $v$ to $v + (0, 2)$ of length $4$ in Fig. \ref{figure:fig045} (d)).
From this we know that for any $v'$ such that $y + 2 \leq y'$ we have 
$F \leq \abb{d}_\abb{MH}(v_\abb{gen}, v) + d_\abb{MH}(v, v') + 2 
= \abb{d}_\abb{MH}(v_\abb{gen}, v') + 2$.
Similarly, for any $v'$ such that $x + 2 \leq x'$ we have 
$F \leq \abb{d}_\abb{MH}(v_\abb{gen}, v') + 2$.
This implies that for any $v'$ such that either $x + 2 \leq x'$ or 
$y + 2 \leq y'$ we have $F \leq \abb{d}_\abb{MH}(v_\abb{gen}, v') + 2 
\leq 2w$ except the three cases $v' = (w - 1, w), (w, w - 1), (w, w)$.

A node $v'$ such that $x' \leq x + 1$ and $y' \leq y + 1$ is 
$v' = v + (1, 1)$.
For this $v'$, $\abb{d}_{C}(v, v') = 6$ (see the path from $v$ to $v + (1, 1)$ 
of length $6$ shown in Fig. \ref{figure:fig045} (d)), and we have 
$F \leq \abb{d}_\abb{MH}(v_\abb{gen}, v) + 6 = (x + y) + 6 \leq 2\lfloor w / 2 \rfloor + 6 
\leq 2w$ using our assumption $w \geq 5$.

Summarizing, we have $F \leq 2w$ except the cases 
$v' = (w - 1, w), (w, w - 1), (w, w)$.

\medskip

We consider what cases remain here.
The cases ``$\#I_{0} = 2$'' and ``$\#J_{0} = 2$'' are  contained in Case 2.
The case ``$\#I_{0} = \#J_{0} = 1$'' is contained in Case 3 and Case 4.
The case ``$\#I_{0} = \#J_{0} = 0$'' is contained in Case 1.
The cases ``$\#I_{0} = 1$ and $\#J_{0} = \#I_{1} = 0$'' and 
``$\#J_{0} = 1$ and $\#I_{0} = \#J_{1} = 0$'' are contained in Case 2.
Therefore only the following case remains.

\medskip

\noindent
(Case 5)  Either $\#I_{0} = \#I_{1} = 1$ or $\#J_{0} = \#J_{1} = 1$.
We consider the first sub case.
The two holes in $I_{0}$, $I_{1}$ are all the holes in $C$.
Let $H$ be the rectangular subset of $\abb{Q}_\abb{I}$ 
having the nodes $v + (2, 0)$, $(w, w)$ as corners 
and let $v_{0}$ and $v_{1}$ be $v$ and $v + (2, 0)$ respectively.
Then we can use Lemma \ref{lemma:lem008} to show 
$F \leq 2w$ 
for $v' \in H$.

As for $v'$ in $I_{0} \cup I_{1} \cup \{ v + (1, 0) \}$, 
we can show $\abb{d}_{C}(v, v') \leq \abb{d}_\abb{MH}(v, v') + 2$ 
using arrow lines shown in Fig. \ref{figure:fig045} (e). 
(We use the fact that there is at least one column west of $I_{0}$.)
Moreover we have $|x - x'| \leq 1$.
Using Lemma \ref{lemma:lem007} we can show $F \leq 2w$.

This completes the proof for $\abb{Q}_\abb{I}$.
We explain how to modify the proof for other quadrants.
The two quadrants $\abb{Q}_\abb{II}$, $\abb{Q}_\abb{IV}$ are symmetric.
Therefore we consider only $\abb{Q}_\abb{II}$, $\abb{Q}_\abb{III}$.

We define the five subsets $I_{0}, \ldots, K$ 
similarly for $\abb{Q}_\abb{II}$, $\abb{Q}_\abb{III}$.
For example, we define $I_{1}$ to be 
the set $\{(x', y') ~|~ x' = x - 1, y + 1 \leq y' \leq w\}$ 
for $\abb{Q}_\abb{II}$ and the set 
$\{(x', y') ~|~ x' = x - 1, 0 \leq y' \leq y - 1 \}$ 
for $\abb{Q}_\abb{III}$.
With these modifications of the definitions of $I_{0}, \ldots, K$, 
all the proofs of Cases 1, 2, 3, 5 for $\abb{Q}_\abb{I}$ are valid also 
for $\abb{Q}_\abb{II}$, $\abb{Q}_\abb{III}$.
Only the proof for Case 4 needs essential modifications.

\begin{figure}[htbp]
\centering
\includegraphics[scale=1.0]{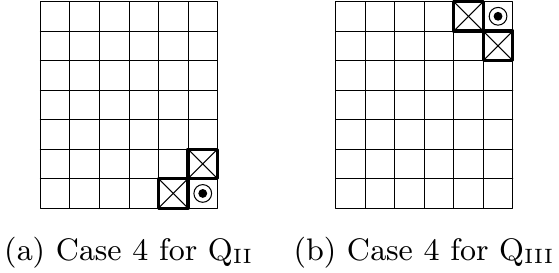}
\caption{The situations of Case 4 for $\abb{Q}_\abb{II}$ and $\abb{Q}_\abb{III}$.}
\label{figure:fig046}
\end{figure}

We show the situations of Case 4 for $\abb{Q}_\abb{II}$ and $\abb{Q}_\abb{III}$ 
in Fig. \ref{figure:fig046} (a), (b).
With the ideas used for $\abb{Q}_\abb{I}$ we can show the followings.
\begin{enumerate}
\item[$\bullet$] For $\abb{Q}_\abb{II}$ and $v' = v + (-1, 1)$, $F \leq x + y + 6$.
\item[$\bullet$] For $\abb{Q}_\abb{II}$ and $v' \not= v + (-1, 1)$, 
$F \leq (x + y) + \{(x - x') + (y' - y)\} + 2 = 2x - x' + y' + 2$.
\item[$\bullet$] For $\abb{Q}_\abb{III}$ and $v' = v + (-1, -1)$, $F \leq x + y + 6$.
\item[$\bullet$] For $\abb{Q}_\abb{III}$ and $v' \not= v + (-1, -1)$, 
$F \leq (x + y) + \{(x - x') + (y - y')\} + 2 = 2x + 2y - x' - y' + 2$.
\end{enumerate}
With these and our assumption $w \geq 5$ 
we can show that $F \leq 2w$ except the following cases 
by elementary calculation.
\begin{enumerate}
\item[$\bullet$] 
The quadrant is $\abb{Q}_\abb{II}$, 
$x = \lfloor w / 2 \rfloor$, $w$ is even, 
$v' = (0, w - 1), (1, w), (0, w)$.
\item[$\bullet$] The quadrant is $\abb{Q}_\abb{II}$, $x = \lfloor w / 2 \rfloor$, $w$ is odd, 
$v' = (0, w)$.
\item[$\bullet$] The quadrant is $\abb{Q}_\abb{III}$, $x = y = \lfloor w / 2 \rfloor$, $w$ is even, 
$v' = (0, 1), (1, 0), (0, 0)$.
\end{enumerate}

This, together with our previous result for $\abb{Q}_{\abb{I}}$, 
completes the proof of Theorem \ref{theorem:thm011}.

\medskip


\end{document}